\documentclass[
showpacs,  
showkeys,  
aps,       
amsthm,    
amsmath,   
amsfonts,  
amssymb    
]{revtex4-1}          

\usepackage{amsthm}   
\usepackage{amsmath}  
\usepackage{amsfonts} 
\usepackage{amssymb}  
\usepackage{graphicx} 
\usepackage{dcolumn}  
\usepackage{bm}       
\usepackage{extarrows}
\usepackage[]{hyperref} 
\usepackage{float}    
\usepackage{subfigure}
\usepackage{epsfig}   
\usepackage{color}    
\usepackage[roman]{complexity}

\begin{document}

\title{Non-Markovian Dynamics of Time-Fractional Open Quantum Systems}

\author{Dongmei Wei}

\author{Hailing Liu}

\author{Yongmei Li}

\author{Sujuan Qin}
\email{qsujuan@bupt.edu.cn}

\author{Qiaoyan Wen}

\author{Fei Gao}
\email{gaof@bupt.edu.cn}

\affiliation{State Key Laboratory of Networking and Switching Technology, Beijing University of Posts and Telecommunications, Beijing 100876, China}

\begin{abstract}
Applications of Time-Fractional Schr\"{o}dinger Equations (TFSEs) to quantum processes are instructive for understanding and describing the time behavior of real physical systems. By applying three popular TFSEs, namely Naber's TFSE I, Naber's TFSE II, and XGF's TFSE, to a basic open system model of a two-level system (qubit) coupled resonantly to a dissipative environment, we solve exactly for Time-Fractional Single Qubit Open Systems (TFSQOSs). However, we find that the three TFSEs perform badly for the following reasons. On the other hand, in the respective frameworks of the three TFSEs the total probability for obtaining the system in a single-qubit state is not equal to one over time at fractional order $\beta\in\left({0,1}\right)$, implying that time-fractional quantum mechanics violates quantum mechanical probability conservation. On the other hand, neither of the latter two TFSEs are capable of describing the non-Markovian dynamics of the system at all $\beta\in\left({0,1}\right]$, only at some $\beta$. To address this, we introduce a well-performed TFSE by constructing a new analytic continuation of time combined with the conformable fractional derivative, in which for all $\beta$, not only does the total probability for the system equal one at all times but also the non-Markovian features can be observed throughout the time evolution of the system. Furthermore, we study the performances of the four TFSEs applying to an open system model of two isolated qubits each locally interacting with its own dissipative environment. By deriving the exact solutions for time-fractional two qubits open systems, we show that our TFSE still possesses the above two advantages compared with the other three TFSEs.
\end{abstract}


\maketitle

\section{Introduction}
\label{intro}
All quantum systems are open since they inevitably interact with an environment \cite{Breuer2002}, where dissipation and/or decoherence take place as energy and/or information stream from the open system to the environment \cite{Breuer2002,Weiss2008}. The backaction mechanisms arising from the interaction induces the non-Markovian memory effects of the environment \cite{Caruso2014}, this is the case: when energy and/or information stream back from the environment to the open system, the future state of the open system depends not only on its last state but also on other previous states. This implies that the nontrivial time correlation exists among the states of the open system at different times in the dynamical exchanges of energy and/or information between the open system and the environment \cite{Fanchini2014,Addis2016,Mortezapour2018}. Such dynamical behavior can serve as a resource for different quantum information tasks, such as creating steady state entanglement \cite{Huelga2012}, accelerating quantum evolution \cite{Deffner2013,Wei2023}, achieving entangling protocols \cite{1Mirkin2019}, extracting work from an Otto cycle \cite{Thomas2018}.

The fact that two-level systems (qubits) serve as fundamental elements in the fields of quantum communication, computing, sensing and materials \cite{Stas2022,Noiri2022,Arunkumar2023,LiuXL2019} has stimulated much interest in their theoretical analyses and actual implementations. There are several methods to create qubits by exploiting current quantum technologies, and each of them is based on a different quantum system, including quantum optics, microscopic quantum objects (electrons, ions, atoms) in traps, quantum dots, and quantum circuits \cite{Majer2005,Berkley2003,Pashkin2003,Bellomo2007}. Unfortunately, it is inevitable that the different implementations of qubits will be affected by the environmental noise. In consequence, the dissipative dynamics of two-level open systems represents a vital issue in the study of open quantum system theory in recent years \cite{Ferraro2009,Laine2012,2Mirkin2019,Roy2023}.

Most of the results on open system dynamics are based on the Standard Schr\"{o}dinger Equation (SSE). For comparison, the dynamical evolution of open systems in the framework of the time-fractional formalism is more complicated by employing different definitions of the time-fractional derivative and various analytic continuations of time in the SSE. In fact, the time-fractional quantum dynamics is a generalization of the standard one, which aims to explore nonlocal phenomena arising from the interaction of a quantum system with its environment \cite{Naber2004,Iomin2009,Huang2011}. There are several methods available to model such dynamics as follows:

(I) The Time-Fractional Schr\"{o}dinger Equation (TFSE) was originally constructed by Naber \cite{Naber2004}, which adopted the Caputo Fractional Derivative (Ca-FD) in the form of convolution with a power law memory kernel \cite{Mainardi1996} and a Analytic Continuation of Time (ACT) $t\to t/i\hbar_\beta$ with order of the imaginary unit $i$ identical to that of the time derivative, and we denote as Naber's TFSE I. It has been showed to describe the non-Markovian evolution of time-fractional open quantum systems \cite{Iomin2009}. Its extensions into the space-time-fractional quantum dynamics have been executed \cite{Dong2008}. Its unitary evolution for a traceless non-Hermitian two-level system was shown to be possible \cite{Cius2022}. But in its framework, the total probability for a free particle in a potential well was found to be greater than one, which means that time-fractional quantum mechanics does not support quantum mechanical probability conservation \cite{Naber2004,Laskin2017}.

(II) As Naber stressed in Ref. \cite{Naber2004}, an alternative way to fractionalize the time derivative is to consider another interesting ACT $t\to t/{(i\hbar_\beta)^{1/\beta}}$ without raising $i$ to the same order as that of the time derivative, which can be treated as a possible case for studying the TFSE, and we call Naber's TFSE II. A lot of subsequent work based on Naber's TFSE II has been completed, for instance, the solutions to it for different examples were found \cite{Muslih2010}, it was generalized to the time-fractional Pauli equation \cite{Nasrolahpour2011}, it was discovered that the dynamical evolution of its description involves the oscillating and decaying parts \cite{Iomin2019}. Notably, the total probability for a free particle in a potential well within its framework is found to go to zero over time by means of the Feynman path integral method \cite{Achar2013}, which is not conserved. In addition, there is no research on how it differs from Naber's TFSE I in describing the non-Markovian evolution.

(III) Recently, a simple Conformable Fractional Derivative (Co-FD) which depends just on the basic limit definition of the derivative introduced by Khalil et al. \cite{Khalil2014} has attracted much concern. For one thing, other theories and features related to this definition were further developed in Refs. \cite{Abdeljawad2015,Abdelhakim2018}, many applications and phenomena could be modeled based on this definition \cite{Hammad2014,Ayati2017,Kareem2017}. For another, different forms of fractional differential equations in the sense of this definition has been raised and solved with the help of many complex tools in Refs. \cite{ElAjou2019,Oqielat2020,Liaqat2022}. Particularly, Xiang et al. \cite{Xiang2019} proposed a TFSE by constructing a ACT $t\to t/i\hbar_\beta {\beta^{1/\beta}}$ with order of $i$ identical to that of the time derivative and employing the Co-FD, and we refer to such a TFSE as XGF's TFSE. And for it, they found that the total probability for a free particle in potential well is a constant over time. Such result reveals that time-fractional quantum mechanics violates conservation of quantum mechanical probability. However, no attention has been paid to whether it holds the ability to describe the non-Markovian evolution.

In this paper, we present applications of Naber's TFSE I, Naber's TFSE II, and XGF's TFSE, to different open quantum system models and further construct a new TFSE to better describe the non-Markovian dynamics of time-fractional open quantum systems. By applying the three TFSEs to a basic open system model of a qubit coupled resonantly to a dissipative environment, we derive the exact solutions for Time-Fractional Single Qubit Open Systems (TFSQOSs). But from the numerical results we find that in the respective frameworks of the three TFSEs, the total probability for obtaining the system in a single-qubit state is not one as time passes at $\beta\in\left({0,1}\right)$ from the numerical results, which verifies the previous results \cite{Naber2004,Achar2013,Xiang2019}. In this case, time-fractional quantum mechanics does not support quantum mechanical probability conservation. We also discover that the latter two TFSEs are not capable of describing the non-Markovian dynamics of the system at some $\beta$ in $\beta\in\left({0,1}\right]$. To tackle them, we propose a well-performed TFSE by combing a new ACT with the Co-FD, where for all $\beta$ the total probability for the system is one at all times, and the system evolves in a non-Markovian manner. Furthermore, by applying the four TFSEs to an open system model of two isolated qubits each locally interacting with its own dissipative environment, we solve exactly for Time-Fractional Two Qubits Open Systems (TFTQOSs). The results demonstrate that in the case of the TFTQOS, our TFSE still has the above two advantages over the other three TFSEs.

The remainder of the paper is structured as follows. In Sec. \ref{sec:2}, we present a procedure for obtaining the matrix forms of the Hamiltonian for multiqubit open system models. In Sec. \ref{sec:3}, we derive the exact solutions to Naber's TFSE I, Naber's TFSE II, and XGF's TFSE for TFSQOSs and TFTQOSs and further introduce a new well-performed TFSE to describe the time-fractional open quantum dynamics more effectively. In Sec. \ref{sec:4}, we provide extensive comparisons of the four TFSEs with respect to the exact dynamics of the total probabilities for TFSQOSs and TFTQOSs, as well as that of the probabilities of being in two excited states. Finally, in Sec. \ref{sec:5}, we give some conclusive remarks.

\section{Open quantum system model}
\label{sec:2}
In this section, a microscopic open quantum system model with $l$ ($l\in{N^+}$) qubits coupled to $l$ independent dissipative environments is introduced in Sec. \ref{subsec:21}. The procedure for obtaining the matrix forms of the Hamiltonian for multiqubit open system models is presented in Sec. \ref{subsec:22}.

\subsection{The microscopic model}
\label{subsec:21}
We now consider a composite quantum system consisting of $l$ separated two-level atoms ($A_1, ..., A_l$) each independently interacting with its own leaky cavity ($B_1, ..., B_l$) at zero temperature \cite{Breuer2002,Bellomo2007}, which has been studied extensively in the literature since it is exploitable to model effectively many real situations \cite{Ferraro2009}. Following the rotating wave approximation, the Hamiltonian for the total system (system plus environment) takes the following form ($\hbar=1$)
\begin{equation}
\label{e1}
{H_{total}} = {H_S} + {H_E} + {H_I}.
\end{equation}
Here
\begin{equation}
\label{e2}
{H_S} = \sum\limits_{l \in N^+}{\omega_0^{{A_l}}\sigma_+ ^{{A_l}}\sigma_- ^{{A_l}}},\,\,\,\,\,\,
{H_E} = \sum\limits_{l\in N^+} {\sum\limits_{j \in N^+} {\omega_j^{{B_l}}(} b_j^{{B_l}}{)^\dag }b_j^{{B_l}}}
\end{equation}
contain the free Hamiltonian of $l$ qubits and that of $l$ independent environments. The transition frequency of the two-level atom over the excited and ground states is denoted by ${\omega _0}$, whereas ${\sigma _ \pm } \equiv ({\sigma _x} \pm i{\sigma _y})/2$ are usually the raising and lowering Pauli operators. The $l$ independent cavities are described by ${\omega _j}$ and $b_j^\dag$ (${b_j}$) ($j\in{N^+}$) which are correspondingly the angular frequency and the creation (annihilation) operators of the $j$th mode of the cavity.

The interaction Hamiltonian is
\begin{equation}
\label{e3}
{H_I} = \sum\limits_{l \in N^+} {\sum\limits_{j \in N^+} {\lambda_j^{{A_l}{B_l}}(\sigma_ - ^{{A_l}}} {{(b_j^{{B_l}})}^\dag} + \sigma_+ ^{{A_l}}b_j^{{B_l}})}
\end{equation}
between each two-level atom and its own cavity with the coupling strength ${\lambda _j}\in\left[{0,1}\right]$. Starting now, we will concentrate on the interaction picture defined by ${H_S} + {H_E}$, where the interaction Hamiltonian is rewritten to
\begin{equation}
\label{e4}
{H^l_I}(t) = \sum\limits_{l \in N^+} {\sum\limits_{j \in N^+} {\lambda_j^{{A_l}{B_l}}(\sigma_- ^{{A_l}}} {{(b_j^{{B_l}})}^\dag }{e^{ - i(\omega_0^{{A_l}} - \omega_j^{{B_l}})t}} + \sigma_ + ^{{A_l}}b_j^{{B_l}}{e^{i(\omega_0^{{A_l}} - \omega_j^{{B_l}})t}})}.
\end{equation}
For simplicity, the settings of parameters are $\omega_0^{A_l}={\omega_0}$, $\omega_j^{B_l}={\omega_j}$ and $\,\lambda_j^{{A_l}{B_l}} = {\lambda_j}$, the interaction Hamiltonian can thus be simplified to
\begin{equation}
\label{e5}
{H^l_I}(t) = \sum\limits_{l \in N^+} {\sum\limits_{j \in N^+} {{\lambda_j}(\sigma_- ^{{A_l}}} {{(b_j^{{B_l}})}^\dag }{e^{- i({\omega_0} - {\omega_j})t}} + \sigma_ + ^{{A_l}}b_j^{{B_l}}{e^{i({\omega_0} - {\omega_j})t}})}.
\end{equation}

\subsection{The matrix form for the model Hamiltonian}
\label{subsec:22}
To study open system dynamics, we will proceed from the matrix form of the model Hamiltonian. One excitation is supposed to exist in the composite atom-cavity system, where the excited and ground states of each atom are provided individually as $\left| e \right\rangle$ and $\left| g \right\rangle$, and each cavity in the form of single mode (i.e., $j=1$) is initially at the Fock state denoted by $\left| n \right\rangle$ ($n\in{N}$). For $l=1$, ${H^1_I}(t)$ can lead to a transition that dominates the atom-cavity dynamics to be $\{$ $|e,n\rangle$ $\leftrightarrow$ $|g,n+1\rangle$ $\}$. Therefore, the matrix form of ${H^1_I}(t)$ in the composite computational basis $\{$ $|g,n+1\rangle$, $|e,n\rangle$ $\}$ is
\begin{equation}
\label{e6}
{H^1_I}(t) = \left( {\begin{array}{*{20}{c}}
0&{\lambda \sqrt {n + 1} {e^{ - i\Delta t}}}\\
{\lambda \sqrt {n + 1} {e^{i\Delta t}}}&0
\end{array}} \right).
\end{equation}
When $l=2$, the transition arising from ${H^2_I}(t)$ can be $\{$ $|e,e,n,n\rangle$ $\leftrightarrow$ $|g,g,n+1,n+1\rangle$, $|e,g,n,n+1\rangle$ $\leftrightarrow$ $|g,e,n+1,n\rangle$ $\}$. And so, in the composite computational basis $\{$ $|g,g,n+1,n+1\rangle$, $|g,e,n+1,n\rangle$, $|e,g,n,n+1\rangle$, $|e,e,n,n\rangle$ $\}$, the matrix form of ${H^2_I}(t)$ is
\begin{equation}
\label{e7}
{H^2_I}(t) = \left( {\begin{array}{*{20}{c}}
0&{\lambda \sqrt {n + 1} {e^{ - i\Delta t}}}&{\lambda \sqrt {n + 1} {e^{ - i\Delta t}}}&0\\
{\lambda \sqrt {n + 1} {e^{i\Delta t}}}&0&0&{\lambda \sqrt {n + 1} {e^{ - i\Delta t}}}\\
{\lambda \sqrt {n + 1} {e^{i\Delta t}}}&0&0&{\lambda \sqrt {n + 1} {e^{ - i\Delta t}}}\\
0&{\lambda \sqrt {n + 1} {e^{i\Delta t}}}&{\lambda \sqrt {n + 1} {e^{i\Delta t}}}&0
\end{array}} \right).
\end{equation}

Importantly, $\Delta  = {\omega _0} - {\omega _j}$ means the detuning. If $\Delta>0$, the Hamiltonian will be non-Hermitian in time. While the atomic transition resonates with the cavity mode, that is, $\Delta=0$, ${H^1_I}(t)$ and ${H^2_I}(t)$ can further be represented by
\begin{equation}
\label{e8}
{H^1_I} = \left( {\begin{array}{*{20}{c}}
0&{\lambda \sqrt {n + 1} }\\
{\lambda \sqrt {n + 1} }&0
\end{array}} \right), \,\,\,\,\,\,\,
{H^2_I} = \left( {\begin{array}{*{20}{c}}
0&{\lambda \sqrt {n + 1} }&{\lambda \sqrt {n + 1} }&0\\
{\lambda \sqrt {n + 1} }&0&0&{\lambda \sqrt {n + 1} }\\
{\lambda \sqrt {n + 1} }&0&0&{\lambda \sqrt {n + 1} }\\
0&{\lambda \sqrt {n + 1} }&{\lambda \sqrt {n + 1} }&0
\end{array}} \right).
\end{equation}
In this case, ${H^1_I}$ and ${H^2_I}$ are Hermitian operators, hence they have spectral decompositions, which is expressed by ${H^l_I} = \sum\limits_{q \in N^+} {{\alpha _q}} \left| {{u_q}} \right\rangle \left\langle {{u_q}} \right|$ with the eigenvalues ${{\alpha _q}}$ and the corresponding normalized eigenvectors $\left| {{u_q}} \right\rangle$.

\section{Exact dynamics of Time-Fractional Open Systems}
\label{sec:3}
In this section, we derive the exact solutions to Naber's TFSE I in Sec. \ref{subsec:31}, Naber's TFSE II in Sec. \ref{subsec:32}, and XGF's TFSE in Sec. \ref{subsec:33} for TFSQOSs and TFTQOSs. To better describe the time-fractional open quantum dynamics, we propose a new well-performed TFSE in Sec. \ref{subsec:34}.

\subsection{Naber's TFSE I}
\label{subsec:31}
To describe the time-fractional quantum dynamics, Naber introduced Naber's TFSE I \cite{Naber2004}
\begin{equation}
\label{e9}
{(i\hbar_\beta)^\beta}\frac{{{\partial ^\beta }\left| {\psi (\boldsymbol{x},t)} \right\rangle }}{{\partial {t^\beta }}} = {H_\beta }\left| {\psi (\boldsymbol{x},t)} \right\rangle,
\end{equation}
where $\beta\in\left({0,1}\right]$ denotes the fractional order of the time derivative, ${H_\beta}$ is called the pseudo-Hamiltonian \cite{Dong2008,Laskin2017}. To ensure the dimensional consistency in the time-fractional quantum dynamics, each variable and parameter is treated as dimensionless in Eq. (\ref{e9}), and thus the scaled Planck constant $\hbar_\beta$ is taken to be dimensionless \cite{Naber2004,Dong2008}. In this case, the ACT $t\to t/i\hbar_\beta$ with order of $i$ identical to that of the time derivative is constructed in Eq. (\ref{e9}).

For $\beta=1$, Eq. (\ref{e9}) is the SSE with a first-order time derivative $\frac{\partial }{{\partial t}}$ to describe the standard quantum dynamics. For $0<\beta<1$, the fractional-order time derivative $\frac{{{\partial ^\beta }}}{{\partial {t^\beta }}}$ is formed by a convoluted integration with a power law memory kernel, which is defined as the Ca-FD \cite{Mainardi1996}
\begin{equation}
\label{e10}
\frac{{{\partial ^\beta }\left| {\psi (t)} \right\rangle }}{{\partial {t^\beta }}} \equiv D_t^\beta \left| {\psi (t)} \right\rangle  = \int_0^t {\frac{{{{(t - \tau )}^{ - \beta }}}}{{\Gamma (1 - \beta )}}} \frac{{\partial \left| {\psi (t)} \right\rangle }}{{\partial t}}d\tau  = k(t) * \left| {\dot \psi (t)} \right\rangle,
\end{equation}
with the gamma function $\Gamma(z)=\int_0^{ + \infty}{{t^{z - 1}}}{e^{ - t}}dt\,(z\in C)$ and the memory kernel function $k(t)=\frac{{{{(t-\tau)}^{-\beta }}}}{{\Gamma(1-\beta)}}$. The Ca-FD is of great importance for solving TFSE since it both allows for the application of the Laplace transform to the time-fractional derivative and provides a time-fractional evolution supported by the Mittag-Leffler function ${E_{\beta ,1}}(z) \equiv {E_\beta }(z) = \int_{j = 0}^{ + \infty } {\frac{{{z^j}}}{{\Gamma (\beta j + 1)}}}$. When ${H_\beta}$ is time-independent, the wave function can be expressed in the following operator form \cite{Iomin2009}
\begin{equation}
\label{e11}
\left| {\psi (t)} \right\rangle  = {E_\beta }\left[ {{{(\frac{{ - it}}{\hbar_\beta })}^\beta }{H_\beta }} \right]\left| {\psi (0)} \right\rangle,
\end{equation}
Notably, the Time Evolution Operator (TEO) ${E_\beta}\left[{{{(\frac{{-it}}{\hbar_\beta})}^\beta}{H_\beta}}\right]$ can be split into the oscillation and decay part in time \cite{Naber2004,Iomin2009}, which suggests that Naber's TFSE I can depict the dynamics of a quantum system through the history information depending on the changes of time intervals, rather than depicting that according to the changes of point times.

In this work, we consider the initial states $\left|{\psi(0)}\right\rangle_{SE}^{l=1, 2}$ for the total systems to be
\begin{equation}
\label{e12}
{\left|{\psi(0)}\right\rangle_{SE}^{1}} = {\left|{e,n}\right\rangle_{{A_1}{B_1}}},
\end{equation}
\begin{equation}
\label{e13}
\begin{aligned}
{\left|{\psi(0)}\right\rangle_{SE}^{2}}=a{\left|{g,g,n,n}\right\rangle_{{A_1}{A_2}{B_1}{B_2}}} + b{\left|{e,e,n,n} \right\rangle_{{A_1}{A_2}{B_1}{B_2}}}
\end{aligned}
\end{equation}
with $a,b \in \left[ {0,1} \right]$ and $a^2+b^2=1$. By substituting ${H^1_I}$ and Eq. (\ref{e12}),
${H^2_I}$ and Eq. (\ref{e13}) into Eq. (\ref{e11}) ($\hbar = {\hbar_\beta }=1$), respectively, we can obtain the exact solutions to Naber's TFSE I for the total systems displayed by
\begin{equation}
\label{e14}
\begin{aligned}
{\left|{\psi(t)} \right\rangle_{SE1}^{1}}=a'b'({E_{\beta_1^ + }^1}-{E_{\beta_1^ - }^1}){\left| {g,n + 1} \right\rangle_{{A_1}{B_1}}}+{{b'}^2}({E_{\beta_1^ + }^1}+{E_{\beta_1^ - }^1}){\left|{e,n} \right\rangle_{{A_1}{B_1}}}
\end{aligned}
\end{equation}
and
\begin{equation}
\label{e15}
\begin{aligned}
{\left|{\psi(t)} \right\rangle_{SE1}^{2}} &= a{\left| {g,g,n,n} \right\rangle_{{A_1}{A_2}{B_1}{B_2}}}\\
&+(({E_{\beta_1^ + }^2}+{E_{\beta_1^ - }^2})chb - mkb){\left| {g,g,n + 1,n + 1} \right\rangle _{{A_1}{A_2}{B_1}{B_2}}}\\
&+ ({E_{\beta_1^ + }^2}-{E_{\beta_1^ - }^2})dhb{\left| {g,e,n + 1,n} \right\rangle _{{A_1}{A_2}{B_1}{B_2}}}\\
&+ ({E_{\beta_1^ + }^2}-{E_{\beta_1^ - }^2})fhb{\left| {e,g,n,n + 1} \right\rangle _{{A_1}{A_2}{B_1}{B_2}}}\,\\
&+ (({E_{\beta_1^ + }^2}+{E_{\beta_1^ - }^2}){h^2}b\, + {k^2}b){\left| {e,e,n,n} \right\rangle _{{A_1}{A_2}{B_1}{B_2}}},
\end{aligned}
\end{equation}
where
\begin{equation}
\label{e16}
\begin{aligned}
E_{\beta_1^\pm}^1 = {E_\beta }\left[{\pm {{(-it)}^\beta}\lambda\sqrt {n+1}}\right],\,\,\,\,\,\,
E_{\beta_1^\pm}^2 = {E_\beta}\left[{\pm 2{{(-it)}^\beta}\lambda \sqrt {n+1}}\right],
\end{aligned}
\end{equation}
$a',b',c,d,f,h,m,k\in[0,1]$, ${a'}^2+{b'}^2=1$, ${c^2}+{d^2}+{f^2}+{h^2}=1$, ${m^2}+{k^2}=1$ are the normalized conditions for the eigenvectors of ${H^1_I}$ and ${H^2_I}$.

The information about TFSQOSs or TFTQOSs is included in the reduced density matrix for the atom(s). By tracing out the cavity parts of the total systems from Eqs. (\ref{e14}) and (\ref{e15}), the elements of ${\rho^1_{S1}(t)}$ in the standard computational basis $\left\{{\left|g\right\rangle,\left|e\right\rangle}\right\}$ and ${\rho^2_{S1}(t)}$ in the standard computational basis $\left\{ {\left|{gg}\right\rangle,\left|{ge}\right\rangle,\left|{eg} \right\rangle,\left|{ee}\right\rangle}\right\}$ are
\begin{equation}
\label{e17}
\begin{aligned}
{\rho_{S1}^{1(11)}} &= (E_{\beta_1^ + }^1 - E_{\beta_1^ - }^1)({E{_{\beta _1^ + }^1}^*} - {E{_{\beta _1^ - }^1}^*}){a'}^2{b'}^2,\\
{\rho_{S1}^{1(22)}} &= (E_{\beta_1^ + }^1 + E_{\beta_1^ - }^1)({E{_{\beta _1^ + }^1}^*} + {E{_{\beta _1^ - }^1}^*}){b'}^4,
\end{aligned}
\end{equation}
and
\begin{equation}
\label{e18}
\begin{aligned}
{\rho_{S1}^{2(11)}}&=(({E_{\beta_1^ +}^2}+{E_{\beta_1^ -}^2})chb-mkb)(({{E_{\beta_1^ + }^2}^*} + {{E_{\beta_1^ - }^2}^*})chb-mkb)+ {a^2},\\
{\rho_{S1}^{2(22)}}&=({E_{\beta_1^ +}^2}-{E_{\beta_1^ -}^2})({{E_{\beta_1^ + }^2}^*} - {{E_{\beta_1^ - }^2}^*}){d^2}{h^2}{b^2},\\
{\rho_{S1}^{2(33)}}&=({E_{\beta_1^ +}^2}-{E_{\beta_1^ -}^2})({{E_{\beta_1^ + }^2}^*} - {{E_{\beta_1^ - }^2}^*}){f^2}{h^2}{b^2},\\
{\rho_{S1}^{2(44)}}&=(({E_{\beta_1^ +}^2}+{E_{\beta_1^ -}^2}){h^2}b+{k^2}b)(({{E_{\beta_1^ + }^2}^*} + {{E_{\beta_1^ - }^2}^*}){h^2}b +{k^2}b),\\
{\rho_{S1}^{2(14)}}&=({{E_{\beta_1^ + }^2}^*} + {{E_{\beta_1^ - }^2}^*}){h^2}ab+{k^2}ab,\\
{\rho_{S1}^{2(41)}}&=({E_{\beta_1^ +}^2}+{E_{\beta_1^ -}^2}){h^2}ab+{k^2}ab.
\end{aligned}
\end{equation}

\subsection{Naber's TFSE II}
\label{subsec:32}
As a possible case for studying TFSE, Naber's TFSE II proposed by Naber \cite{Naber2004} has been widely considered by replacing the ACT $t\to t/i\hbar_\beta$ in Eq. (\ref{e9}) with $t\to t/{(i\hbar_\beta)^{1/\beta}}$ that the equation is stated to be
\begin{equation}
\label{e19}
{i\hbar_\beta}D_t^\beta \left| {\psi (\boldsymbol{x},t)} \right\rangle= {H_\beta }\left| {\psi (\boldsymbol{x},t)} \right\rangle.
\end{equation}
The infeasibility of raising $i$ to the same order as that of the time derivative was analyzed in detail in Ref. \cite{Achar2013}, which indicated that $i$ can not be altered arbitrarily due to its relation to the phase factor in the Feynman propagator.

It should be remarked that like Eq. (\ref{e9}), Eq. (\ref{e19}) is also used to describe the non-Markovian evolution of time-fractional open quantum systems since it contains the Ca-FD \cite{Iomin2019}. For the time-independent ${H_\beta}$, the wave function can be formulated with
\begin{equation}
\label{e20}
\left| {\psi (t)} \right\rangle  = {E_\beta }\left[ {\frac{{ - i{t^\beta }}}{{{\hbar _\beta }}}{H_\beta }} \right]\left| {\psi (0)} \right\rangle.
\end{equation}

Similar to solving Naber's TFSE I, substituting ${H^1_I}$ and Eq. (\ref{e12}), ${H^2_I}$ and Eq. (\ref{e13}) into Eq. (\ref{e20}), the exact solutions to Naber's TFSE II for the total systems can be obtained by
\begin{equation}
\label{e21}
\begin{aligned}
{\left|{\psi(t)} \right\rangle_{SE2}^{1}}=a'b'({E_{\beta_2^ - }^1}-{E_{\beta_2^ + }^1}){\left| {g,n + 1} \right\rangle_{{A_1}{B_1}}}+{{b'}^2}({E_{\beta_2^ - }^1}+{E_{\beta_2^ + }^1}){\left|{e,n} \right\rangle_{{A_1}{B_1}}}
\end{aligned}
\end{equation}
and
\begin{equation}
\label{e22}
\begin{aligned}
{\left|{\psi(t)} \right\rangle_{SE2}^{2}} &= a{\left| {g,g,n,n} \right\rangle_{{A_1}{A_2}{B_1}{B_2}}}\\
&+(({E_{\beta_2^ - }^2}+{E_{\beta_2^ + }^2})chb - mkb){\left| {g,g,n + 1,n + 1} \right\rangle _{{A_1}{A_2}{B_1}{B_2}}}\\
&+ ({E_{\beta_2^ - }^2}-{E_{\beta_2^ + }^2})dhb{\left| {g,e,n + 1,n} \right\rangle _{{A_1}{A_2}{B_1}{B_2}}}\\
&+ ({E_{\beta_2^ - }^2}-{E_{\beta_2^ + }^2})fhb{\left| {e,g,n,n + 1} \right\rangle _{{A_1}{A_2}{B_1}{B_2}}}\,\\
&+ (({E_{\beta_2^ -}^2}+{E_{\beta_2^ + }^2}){h^2}b\, + {k^2}b){\left| {e,e,n,n} \right\rangle _{{A_1}{A_2}{B_1}{B_2}}},
\end{aligned}
\end{equation}
where
\begin{equation}
\label{e23}
\begin{aligned}
E_{\beta_2^\pm}^1 = {E_\beta }\left[{\pm {i{t^\beta}}\lambda\sqrt {n+1}}\right],\,\,\,\,\,\,
E_{\beta_2^\pm}^2 = {E_\beta}\left[{\pm 2i{t^\beta}\lambda \sqrt {n+1}}\right].
\end{aligned}
\end{equation}
After performing ${\rho_{S2}^{l}(t)} = t{r_E}({\left| {\psi (t)} \right\rangle _{SE2}^{l}}\left\langle {\psi (t)} \right|)$, the density matrix ${\rho_{S2}^{1}(t)}$ for TFSQOSs and the density matrix ${\rho_{S2}^{2}(t)}$ for TFTQOSs will be obtained easily.

\subsection{XGF's TFSE}
\label{subsec:33}
The equation that is formed by replacing the Ca-FD \cite{Mainardi1996} in Eq. (\ref{e9}) with the Co-FD \cite{Khalil2014} is what we call XGF's TFSE \cite{Xiang2019}, which constructs the ACT $t\to t/i\hbar_\beta {\beta^{1/\beta}}$ with order of $i$ identical to that of the time derivative, as indicated below
\begin{equation}
\label{e24}
{(i\hbar_\beta)^\beta}T_t^\beta \left| {\psi (\boldsymbol{x},t)} \right\rangle= {H_\beta }\left| {\psi (\boldsymbol{x},t)} \right\rangle.
\end{equation}
It is important to emphasize that the Co-FD is a new simple fractional derivative in the sense that it depends just on the basic limit definition of the derivative, defined by
\begin{equation}
\label{e25}
T_t^\beta \left| {\psi (t)} \right\rangle  = \mathop {\lim }\limits_{\varepsilon  \to 0} \frac{{\left| {\psi (t + \varepsilon {t^{1 - \beta }})} \right\rangle - \left| {\psi (t)} \right\rangle }}{t}
\end{equation}
with $\beta\in\left({0,1}\right]$. For $\beta=1$, Eq. (\ref{e24}) is reduced to the SSE. For $0<\beta<1$, the Co-FD was proved to satisfy all the properties of the standard derivative. It can be observed that the Co-FD is local in time. In the case of ${H_\beta}$ being time-independent, the wave function is derived to be
\begin{equation}
\label{e26}
\left| {\psi(t)} \right\rangle = {e^{\frac{{{{(\frac{{ - i}}{{{\hbar _\beta}}})}^\beta}{t^\beta}{H_\beta }}}{\beta}}}\left| {\psi(0)} \right\rangle.
\end{equation}

Similarly, the exact solutions to XGF's TFSE for the total systems are given by
\begin{equation}
\label{e27}
\begin{aligned}
{\left|{\psi(t)} \right\rangle_{SE3}^{1}}=a'b'({e_{\beta_1^ + }^1}-{e_{\beta_1^ - }^1}){\left| {g,n + 1} \right\rangle_{{A_1}{B_1}}}+{{b'}^2}({e_{\beta_1^ + }^1}+{e_{\beta_1^ - }^1}){\left|{e,n} \right\rangle_{{A_1}{B_1}}}
\end{aligned}
\end{equation}
and
\begin{equation}
\label{e28}
\begin{aligned}
{\left|{\psi(t)} \right\rangle_{SE3}^{2}} &= a{\left| {g,g,n,n} \right\rangle_{{A_1}{A_2}{B_1}{B_2}}}\\
&+(({e_{\beta_1^ + }^2}+{e_{\beta_1^ - }^2})chb - mkb){\left| {g,g,n + 1,n + 1} \right\rangle _{{A_1}{A_2}{B_1}{B_2}}}\\
&+ ({e_{\beta_1^ + }^2}-{e_{\beta_1^ - }^2})dhb{\left| {g,e,n + 1,n} \right\rangle _{{A_1}{A_2}{B_1}{B_2}}}\\
&+ ({e_{\beta_1^ + }^2}-{e_{\beta_1^ - }^2})fhb{\left| {e,g,n,n + 1} \right\rangle _{{A_1}{A_2}{B_1}{B_2}}}\,\\
&+ (({e_{\beta_1^ + }^2}+{e_{\beta_1^ - }^2}){h^2}b\, + {k^2}b){\left| {e,e,n,n} \right\rangle _{{A_1}{A_2}{B_1}{B_2}}},
\end{aligned}
\end{equation}
where
\begin{equation}
\label{e29}
\begin{aligned}
e_{\beta_1^\pm}^1 &= {e^{\frac{{\pm{{(-it)}^\beta}\lambda \sqrt {n + 1} }}{\beta}}},\,\,\,\,\,\,\,\,
e_{\beta_1^\pm}^2 &= {e^{\frac{{\pm2{{(-it)}^\beta}\lambda \sqrt {n + 1} }}{\beta}}}.
\end{aligned}
\end{equation}
Then, it is easy to yield ${\rho_{S3}^{1}(t)}$ for TFSQOSs and ${\rho_{S3}^{2}(t)}$ for TFTQOSs by performing ${\rho_{S3}^{l}(t)} = t{r_E}({\left| {\psi (t)} \right\rangle _{SE3}^{l}}\left\langle {\psi (t)} \right|)$.

\subsection{Our TFSE}
\label{subsec:34}
To describe the non-Markovian evolution of time-fractional open quantum systems more effectively, we propose a well-behaved TFSE by replacing the ACT $t\to t/i\hbar_\beta {\beta^{1/\beta}}$ in Eq. (\ref{e24}) with a new ACT $\,t \to t/{(i\hbar_\beta \beta)^{1/\beta}}$, where the new one does not raise $i$ to the same order as the time derivative. As a consequence, such a TFSE is denoted for
\begin{equation}
\label{e30}
{i\hbar_\beta}T_t^\beta \left| {\psi (\boldsymbol{x},t)} \right\rangle = {H_\beta }\left| {\psi (\boldsymbol{x},t)} \right\rangle.
\end{equation}
With ${H_\beta}$ independent of time, the wave function takes the following form
\begin{equation}
\label{e31}
\left| {\psi (t)} \right\rangle = {e^{\frac{{\frac{{ - i}}{{{\hbar _\beta }}}{t^\beta }{H_\beta }}}{\beta }}}\left| {\psi (0)} \right\rangle.
\end{equation}

Similarly, we solve the TFSE exactly for the total systems
\begin{equation}
\label{e32}
\begin{aligned}
{\left|{\psi(t)} \right\rangle_{SE4}^{1}}=a'b'({e_{\beta_2^ - }^1}-{e_{\beta_2^ + }^1}){\left| {g,n + 1} \right\rangle_{{A_1}{B_1}}}+{{b'}^2}({e_{\beta_2^ - }^1}+{e_{\beta_2^ + }^1}){\left|{e,n} \right\rangle_{{A_1}{B_1}}}
\end{aligned}
\end{equation}
and
\begin{equation}
\label{e33}
\begin{aligned}
{\left|{\psi(t)} \right\rangle_{SE4}^{2}} &= a{\left| {g,g,n,n} \right\rangle_{{A_1}{A_2}{B_1}{B_2}}}\\
&+(({e_{\beta_2^ - }^2}+{e_{\beta_2^ + }^2})chb - mkb){\left| {g,g,n + 1,n + 1} \right\rangle _{{A_1}{A_2}{B_1}{B_2}}}\\
&+ ({e_{\beta_2^ - }^2}-{e_{\beta_2^ + }^2})dhb{\left| {g,e,n + 1,n} \right\rangle _{{A_1}{A_2}{B_1}{B_2}}}\\
&+ ({e_{\beta_2^ - }^2}-{e_{\beta_2^ + }^2})fhb{\left| {e,g,n,n + 1} \right\rangle _{{A_1}{A_2}{B_1}{B_2}}}\,\\
&+ (({e_{\beta_2^ - }^2}+{e_{\beta_2^ + }^2}){h^2}b\, + {k^2}b){\left| {e,e,n,n} \right\rangle _{{A_1}{A_2}{B_1}{B_2}}},
\end{aligned}
\end{equation}
where
\begin{equation}
\label{e34}
\begin{aligned}
e_{\beta_2^\pm}^1 &= {e^{\frac{{\pm i{{t}^\beta}\lambda \sqrt {n + 1} }}{\beta}}},\,\,\,\,\,\,\,\,
e_{\beta_2^\pm}^2 &= {e^{\frac{{\pm2i{{t}^\beta}\lambda \sqrt {n + 1} }}{\beta}}}.
\end{aligned}
\end{equation}
In the next step, by performing ${\rho_{S4}^{l}(t)} = t{r_E}({\left| {\psi (t)} \right\rangle _{SE4}^{l}}\left\langle {\psi (t)} \right|)$, we can get ${\rho_{S4}^{1}(t)}$ for TFSQOSs and ${\rho_{S4}^{2}(t)}$ for TFTQOSs.

See Table \ref{tab1} for the comparisons of the significant features among Naber's TFSE I, Naber's TFSE II, XGF's TFSE, and our TFSE.
\begin{table*}
\caption{\label{tab1}
Comparisons of the significant features of three popular TFSEs with our TFSE.}
\begin{ruledtabular}
\begin{tabular}{ccccc}
TFSEs &FD &ACT  &TEO &In Time \\ \hline
Naber's TFSE I~\cite{Naber2004} &Ca-FD &$t\to t/i\hbar_\beta$ &${E_\beta}\left[{{{(\frac{{-it}}{\hbar_\beta})}^\beta }{H_\beta}}\right]$ &non-local  \\
Naber's TFSE II~\cite{Naber2004} &Ca-FD &$t\to t/{(i\hbar_\beta)^{1/\beta}}$ &${E_\beta}\left[{\frac{{- i{t^\beta}}}{{{\hbar_\beta}}}{H_\beta}}\right]$ &non-local \\
XGF's TFSE~\cite{Xiang2019} &Co-FD  &$t\to t/i\hbar_\beta {\beta^{1/\beta}}$  &${e^{\frac{{{{(\frac{{-i}}{{{\hbar_\beta}}})}^\beta }{t^\beta }{H_\beta }}}{\beta }}}$  &local \\
Our TFSE  &Co-FD  &$\,t \to t/{(i\hbar_\beta \beta)^{1/\beta}}$  &${e^{\frac{{\frac{{- i}}{{{\hbar_\beta }}}{t^\beta}{H_\beta}}}{\beta}}}$   &local \\
\end{tabular}
\end{ruledtabular}
\end{table*}

\section{Comparisons of the dynamics described by Naber's TFSE I, Naber's TFSE II, XGF's TFSE and our TFSE}
\label{sec:4}
In this section, to establish whether one TFSE describes the time-fractional open quantum dynamics better than the others, we make detailed comparisons of the time behavior of the total probabilities for the TFSQOS in a single-qubit state and the TFTQOS in a two-qubit state in Sec. \ref{subsec:41} and that of the probabilities for the TFSQOS in the excited state $\left|{1}\right\rangle$ and the TFTQOS in the excited state $\left|{11}\right\rangle$ in Sec. \ref{subsec:42} by employing the results obtained above.

\subsection{Exact dynamics of the total probabilities for the TFSQOS in a single-qubit sate and the TFTQOS in a two-qubit sate}
\label{subsec:41}
The total probabilities $P_{\gamma=1,2,3,4}^{total\,1}(t)$ that the TFSQOS is in a single-qubit state and $P_{\gamma=1,2,3,4}^{total\,2}(t)$ that the TFTQOS is in a two-qubit state can be given by
\begin{equation}
\label{e35}
\begin{aligned}
P_{\gamma=1,2,3,4}^{total\,1}(t)=\rho_{S{\gamma}}^{1(11)} + \rho_{S{\gamma}}^{1(22)},\,\,\,\,\,\,\,\,\,
P_{\gamma=1,2,3,4}^{total\,2}(t)=\rho_{S{\gamma}}^{2(11)} + \rho_{S{\gamma}}^{2(22)} + \rho_{S{\gamma}}^{2(33)} + \rho_{S{\gamma}}^{2(44)}.
\end{aligned}
\end{equation}
For simplicity, $a'=b'=m=k=\frac{{\sqrt 2 }}{2}$, $c=d=f=h=\frac{1}{2}$ are set. It is notable that $P_{\gamma}^{total\,2}(t)$ differs from $P_{\gamma}^{total\,1}(t)$ in that $P_{\gamma}^{total\,2}(t)$ is related to the initial entanglement $(a,b)$. The entanglement of the bipartite system can be characterized by Wootter's concurrence \cite{Wootter1998}. Thus for the initial two-qubit state ${\rho^2 _S (0)}$, the concurrence can be calculated as $C_0=2ab$.

\begin{figure}[htbp]
\centering
    \includegraphics[width=0.32\linewidth]{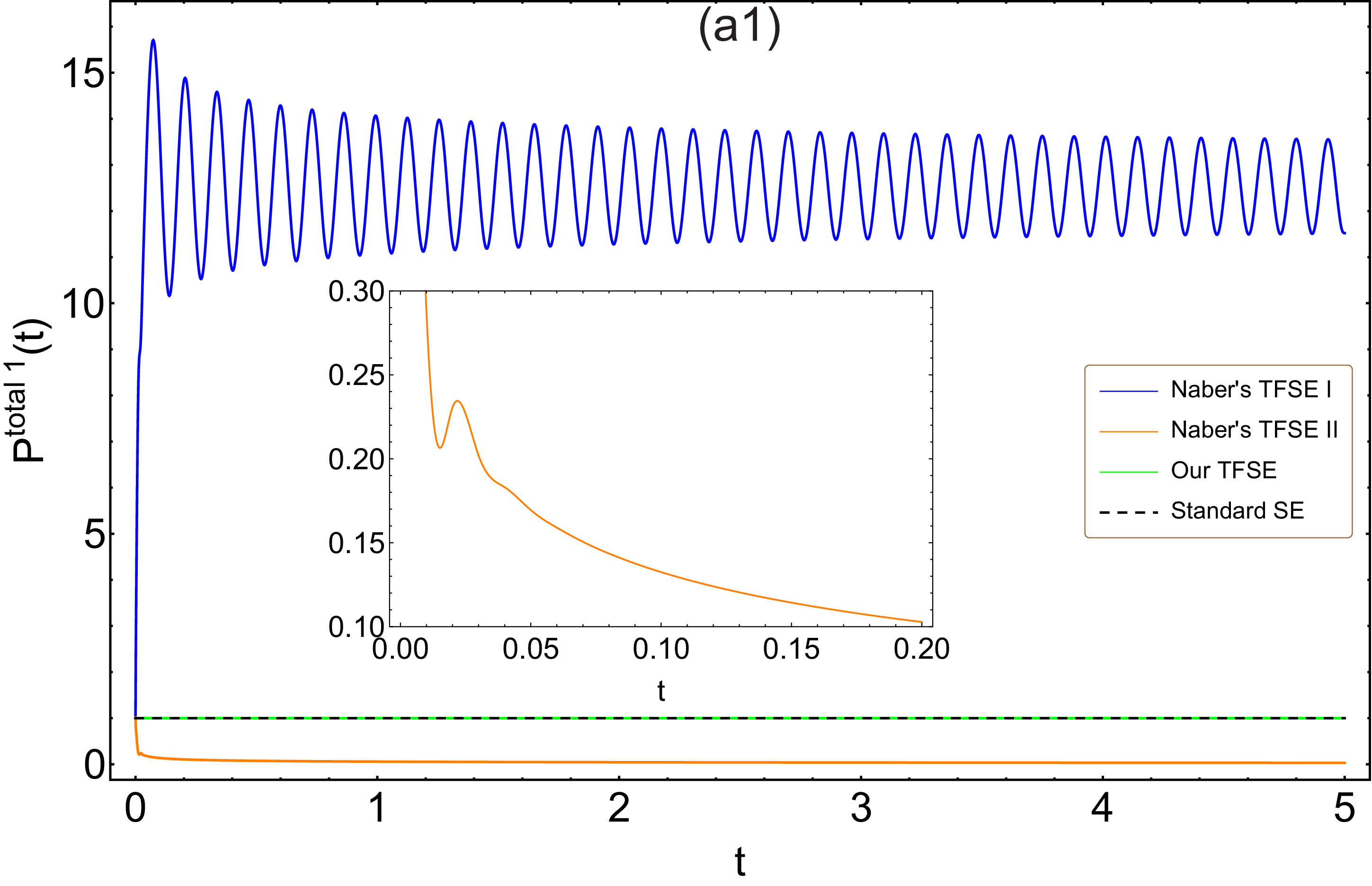}
    \includegraphics[width=0.32\linewidth]{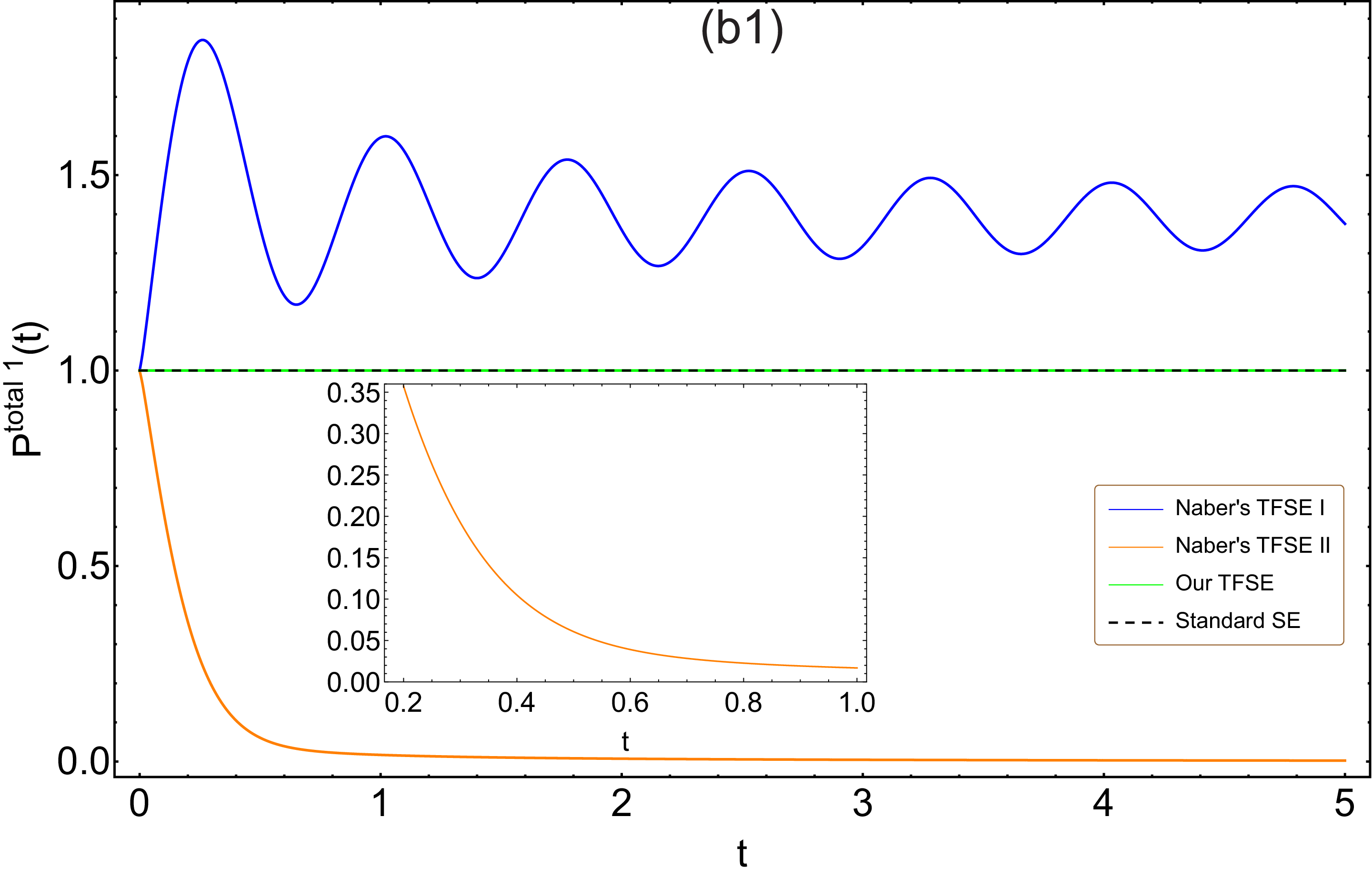}
    \includegraphics[width=0.32\linewidth]{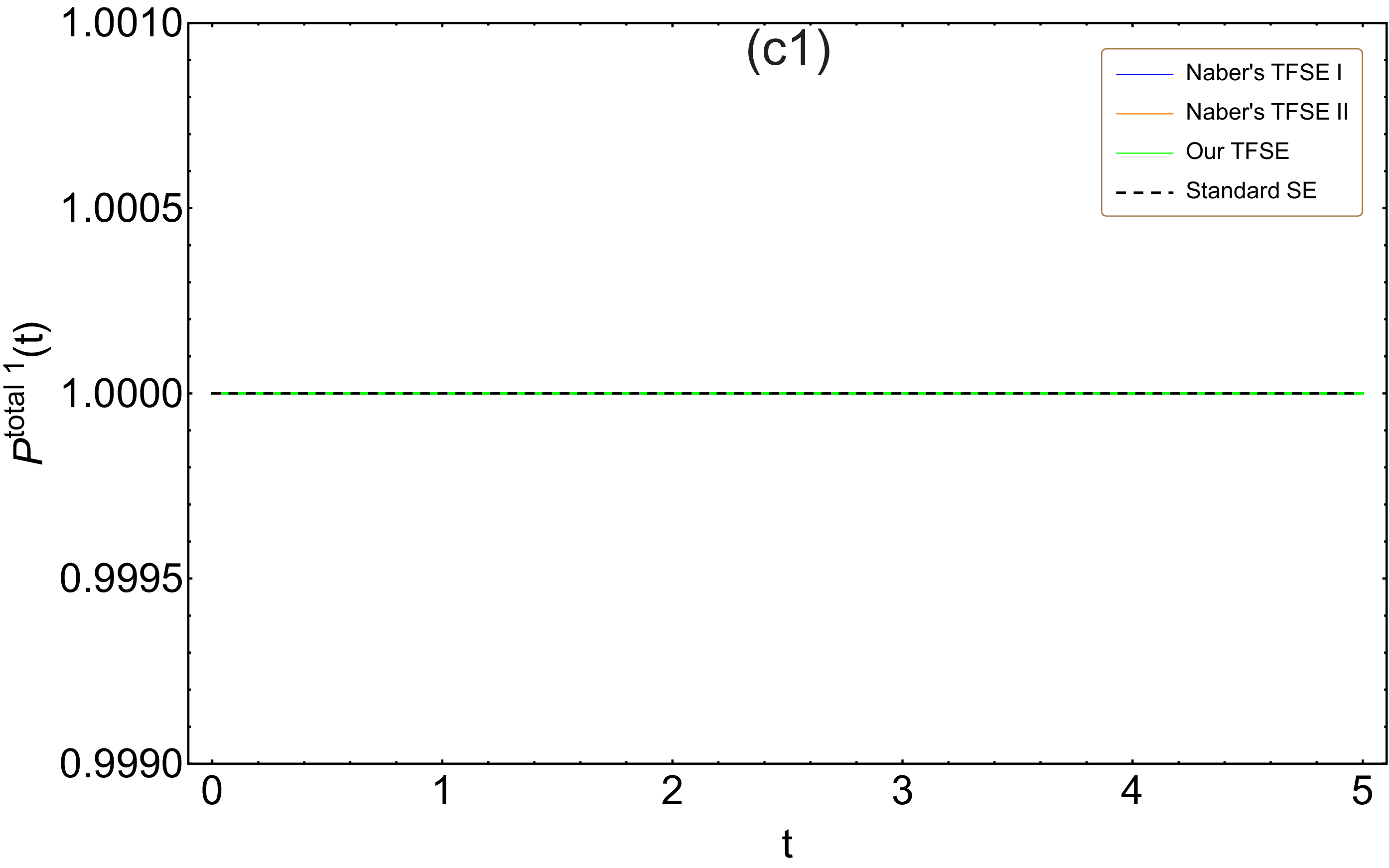}\\
    \includegraphics[width=0.32\linewidth]{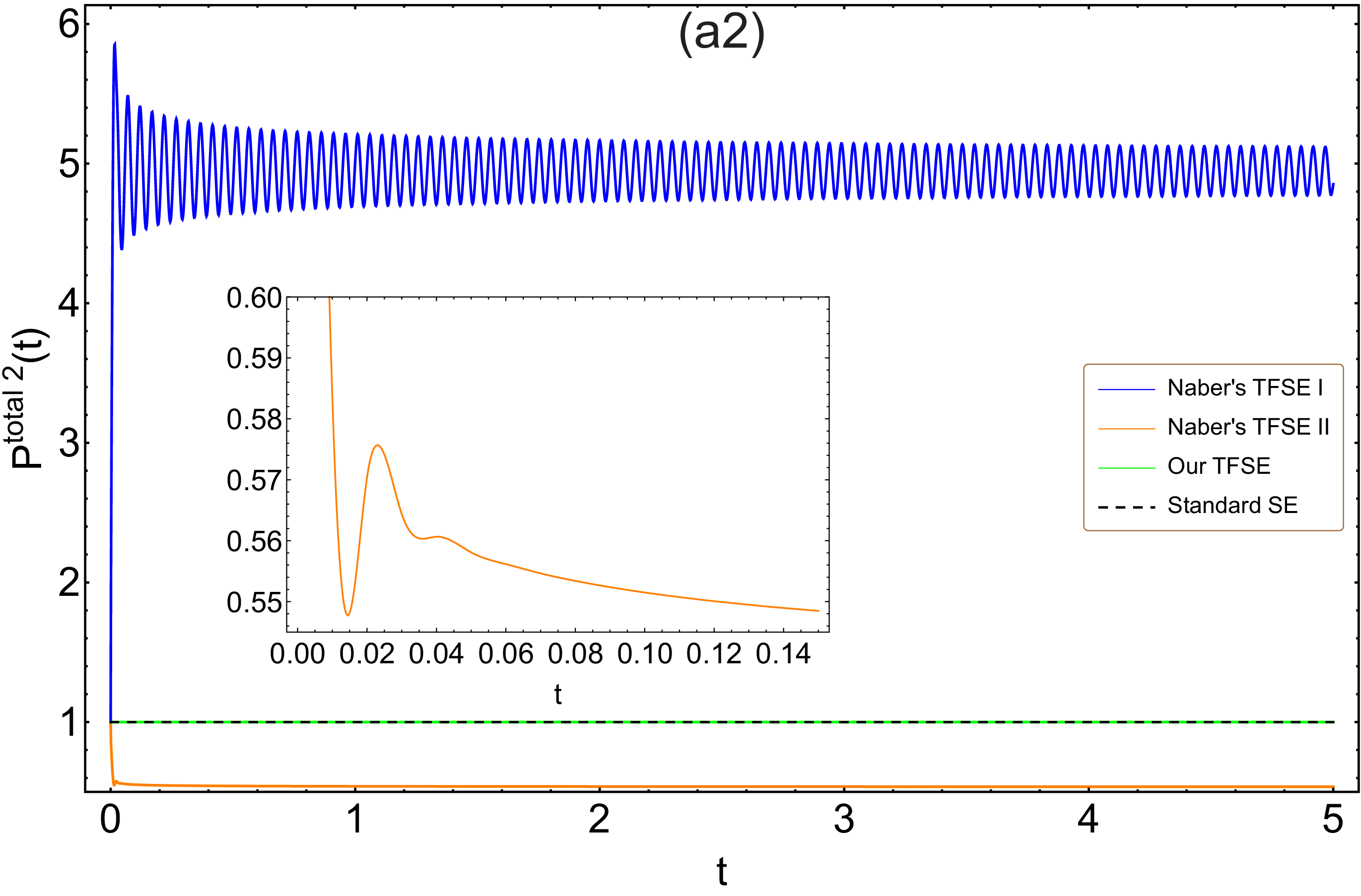}
    \includegraphics[width=0.32\linewidth]{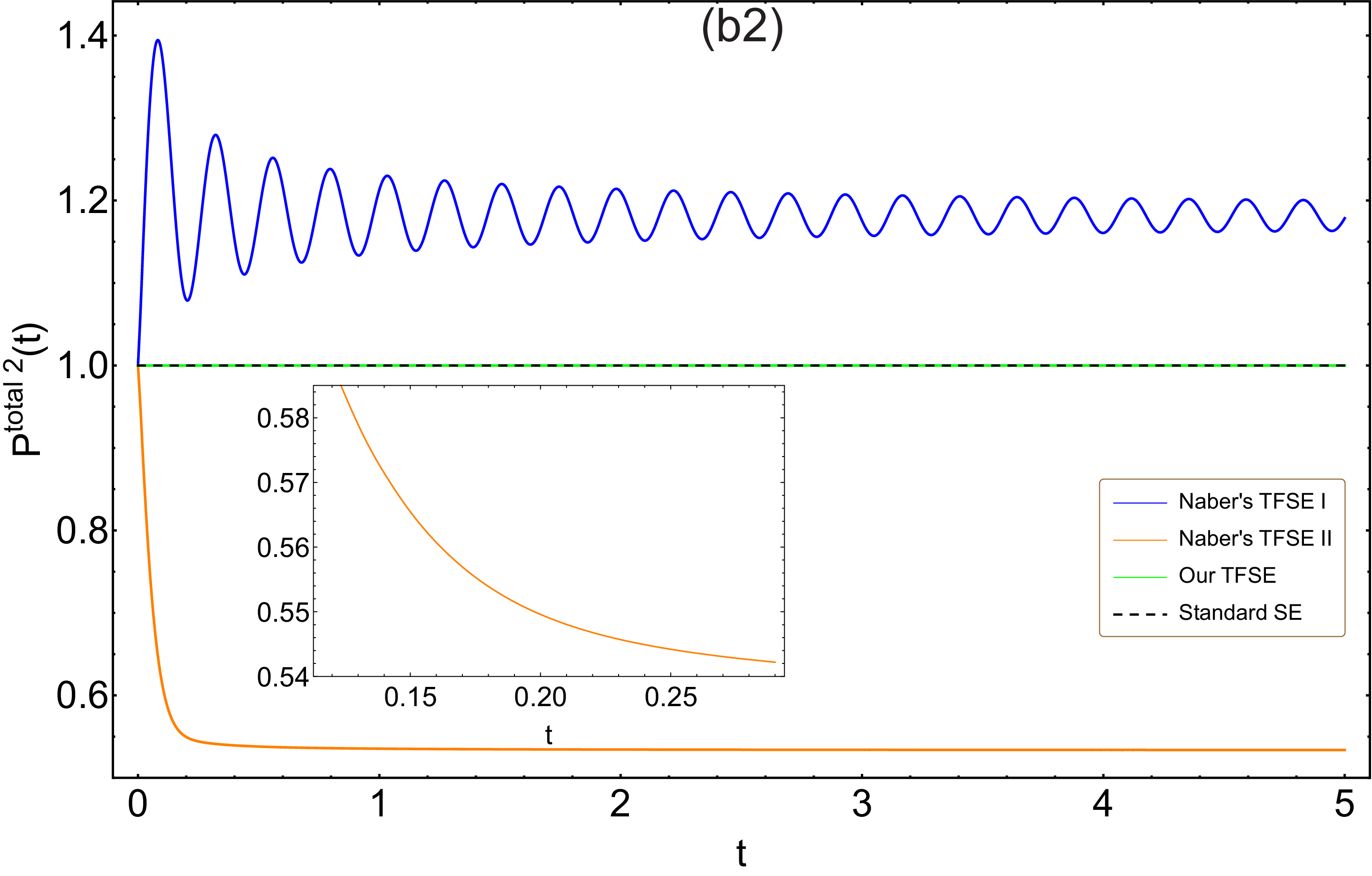}
    \includegraphics[width=0.32\linewidth]{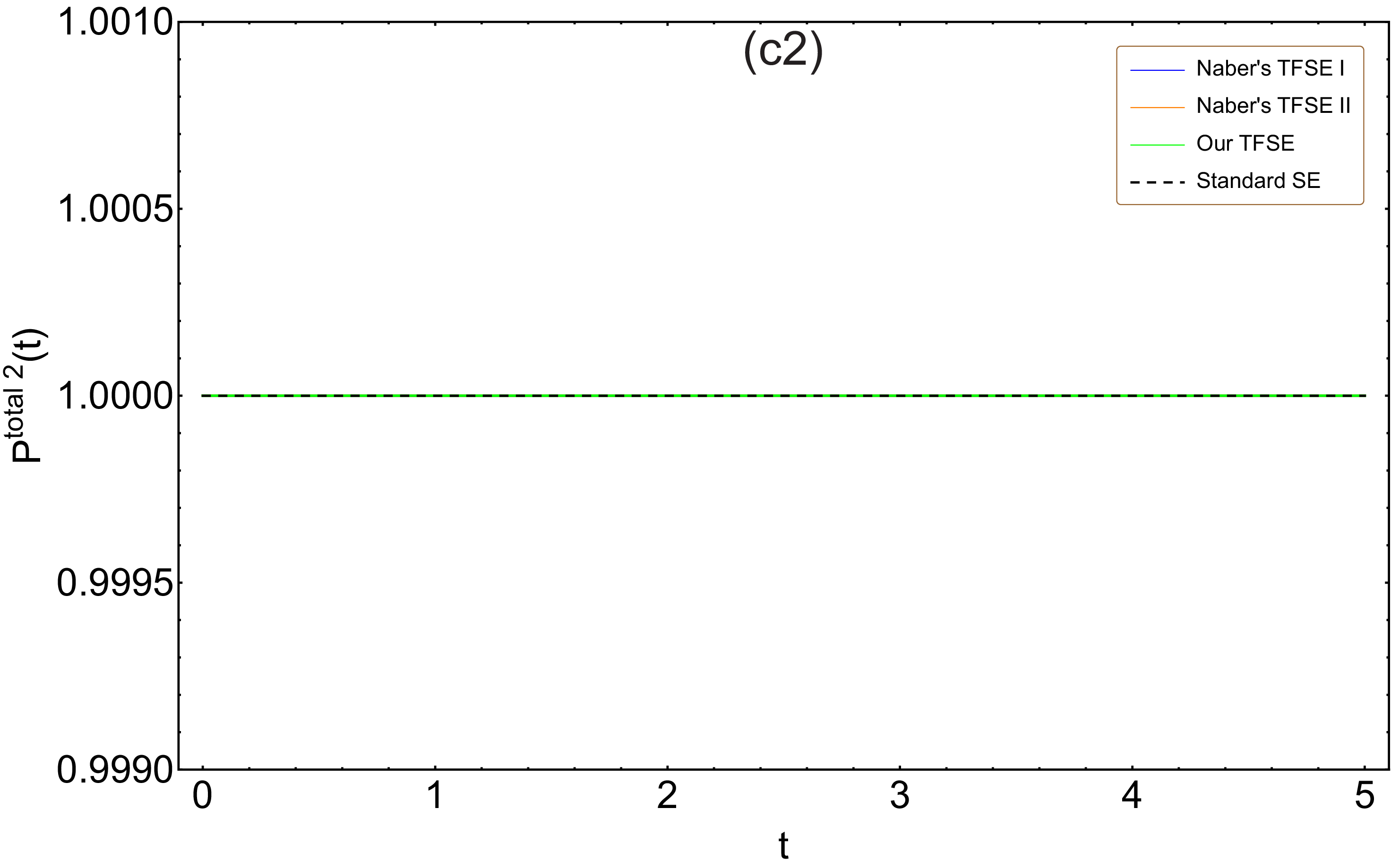}\\
\caption{The time evolution of $P_{\gamma=1,2,4}^{total\,1}(t)$ and $P_{\gamma=1,2,4}^{total\,2}(t)$ is plotted for $\beta=0.2, 0.6, 1$. The other parameters are set as $\lambda=0.5$, $n=50$, and $C_0=0.5$. The insets show how $P_2^{total\,1}(t)$ and $P_2^{total\,2}(t)$ change in a very short time scale.}
\label{Fig1}
\end{figure}

Fig. \ref{Fig1} shows the time behavior of $P_{\gamma=1,2,4}^{total\,1}(t)$ and $P_{\gamma=1,2,4}^{total\,2}(t)$ described by Naber's TFSE I, Naber's TFSE II and our TFSE by considering $\beta=0.2,0.6,1$, where we have chosen $\lambda=0.5$, $n=50$, and $C_0=0.5$. We find that $P_1^{total\,1}(t)$ and $P_1^{total\,2}(t)$ are always greater than or equal to one over time. Moreover, their dynamics behave the non-Markovian features when $\beta\in\left({0,1}\right)$, where the larger $\beta$ leads to the weaker frequencies and amplitudes of the non-Markovian oscillations, these are also displayed in $P_2^{total\,1}(t)$ and $P_2^{total\,2}(t)$. Differently, $P_2^{total\,1}(t)$ and $P_2^{total\,2}(t)$ are less than or equal to one for any $t$, and decay monotonically for large $\beta$. Note that $P_4^{total\,1}(t)$ and $P_4^{total\,2}(t)$ are conserved for all $\beta$ at all $t$, contrary to $P_{\gamma=1,2}^{total\,1}(t)$ and $P_{\gamma=1,2}^{total\,2}(t)$ which are conserved only for $\beta=1$.

\begin{figure}[htbp]
\centering
    \includegraphics[width=0.32\linewidth]{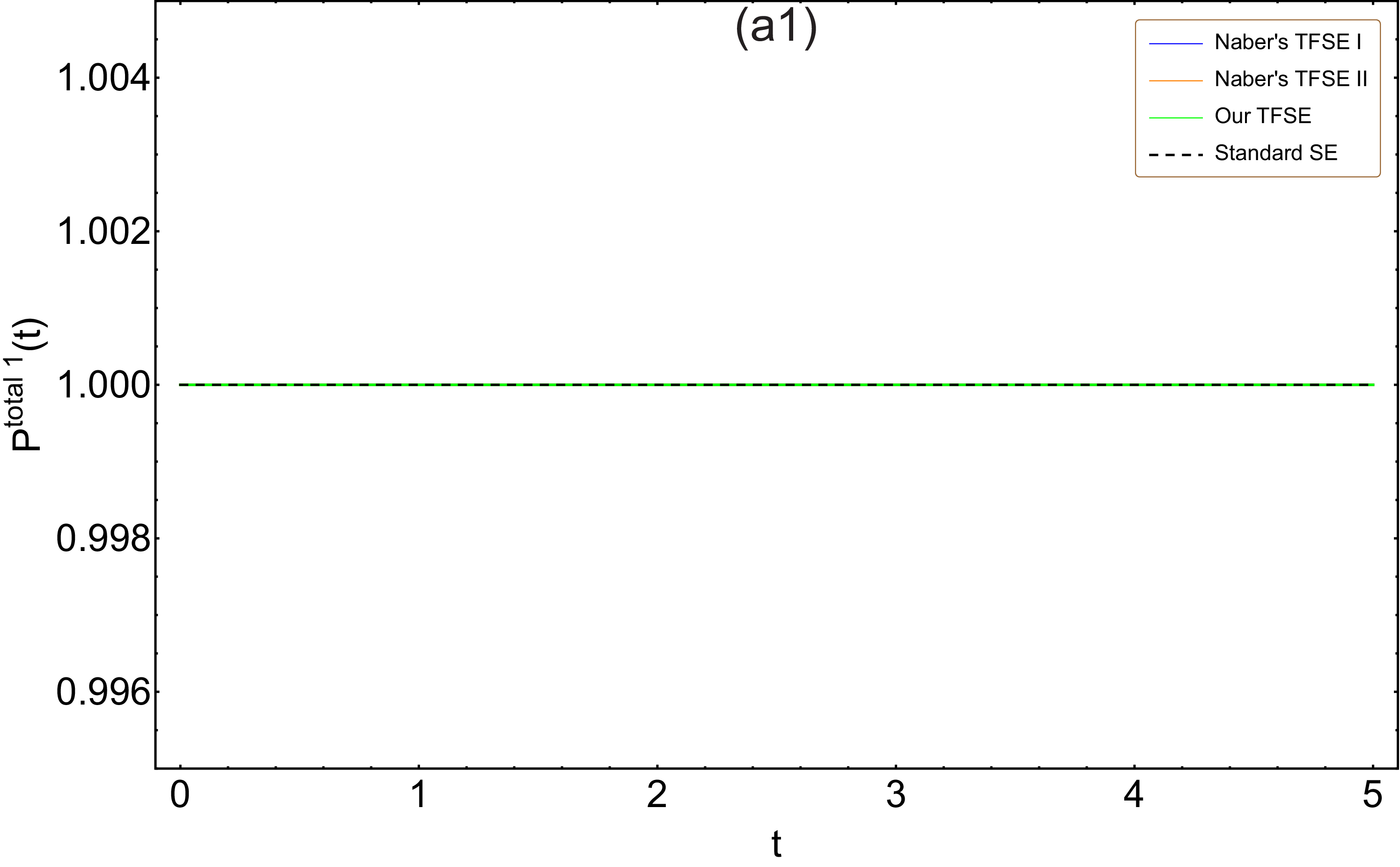}
    \includegraphics[width=0.32\linewidth]{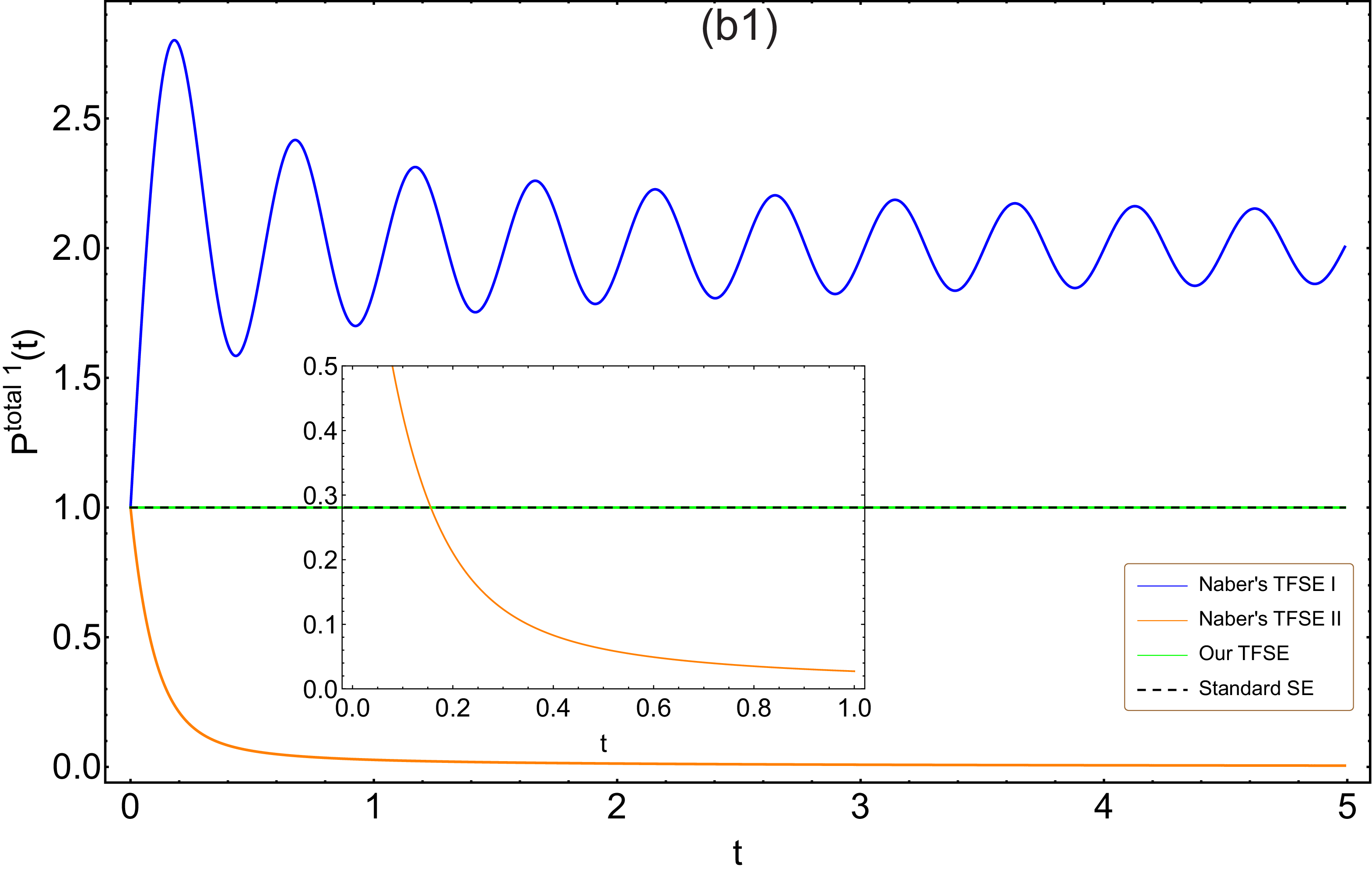}
    \includegraphics[width=0.32\linewidth]{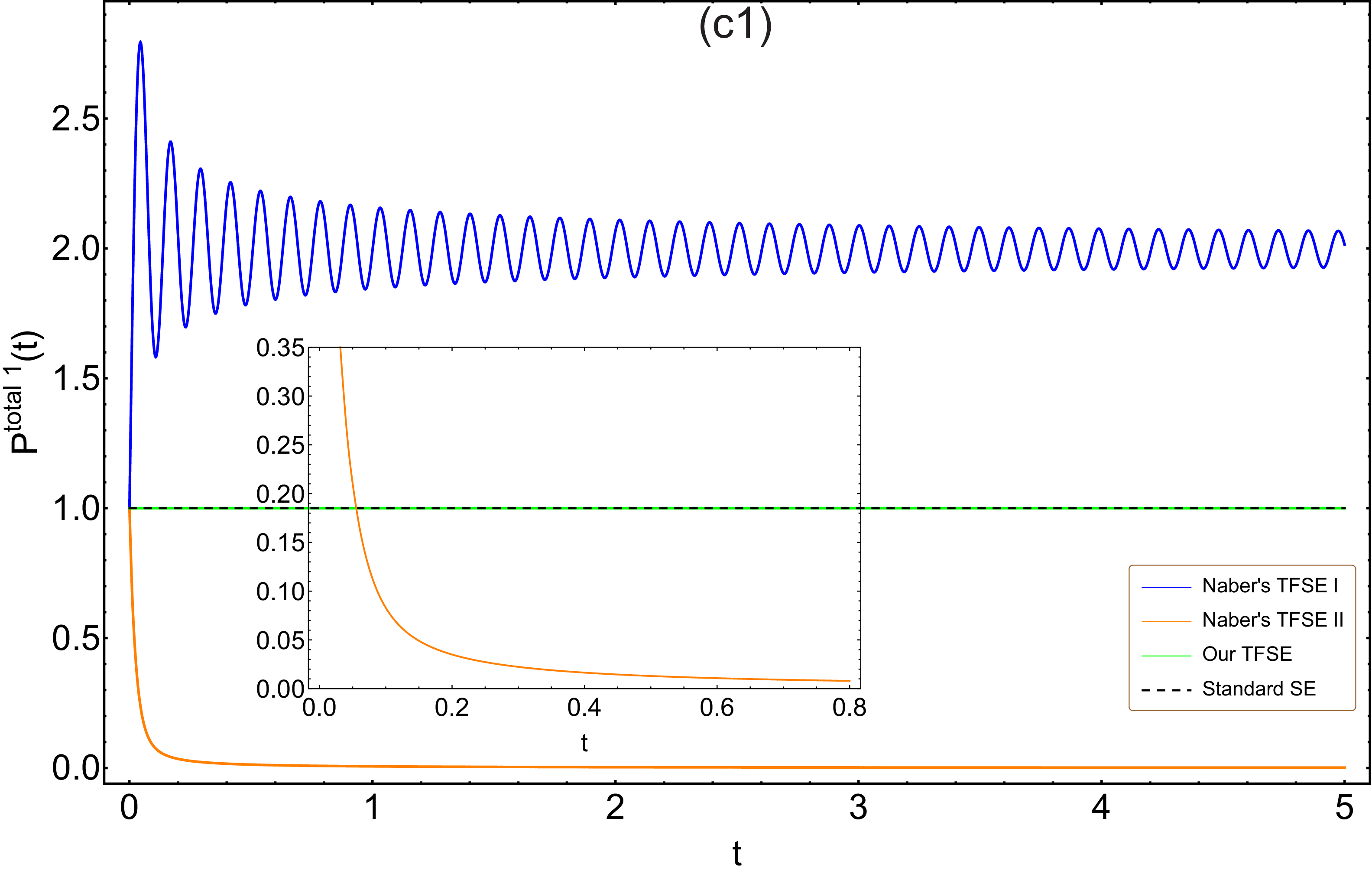}\\
    \includegraphics[width=0.32\linewidth]{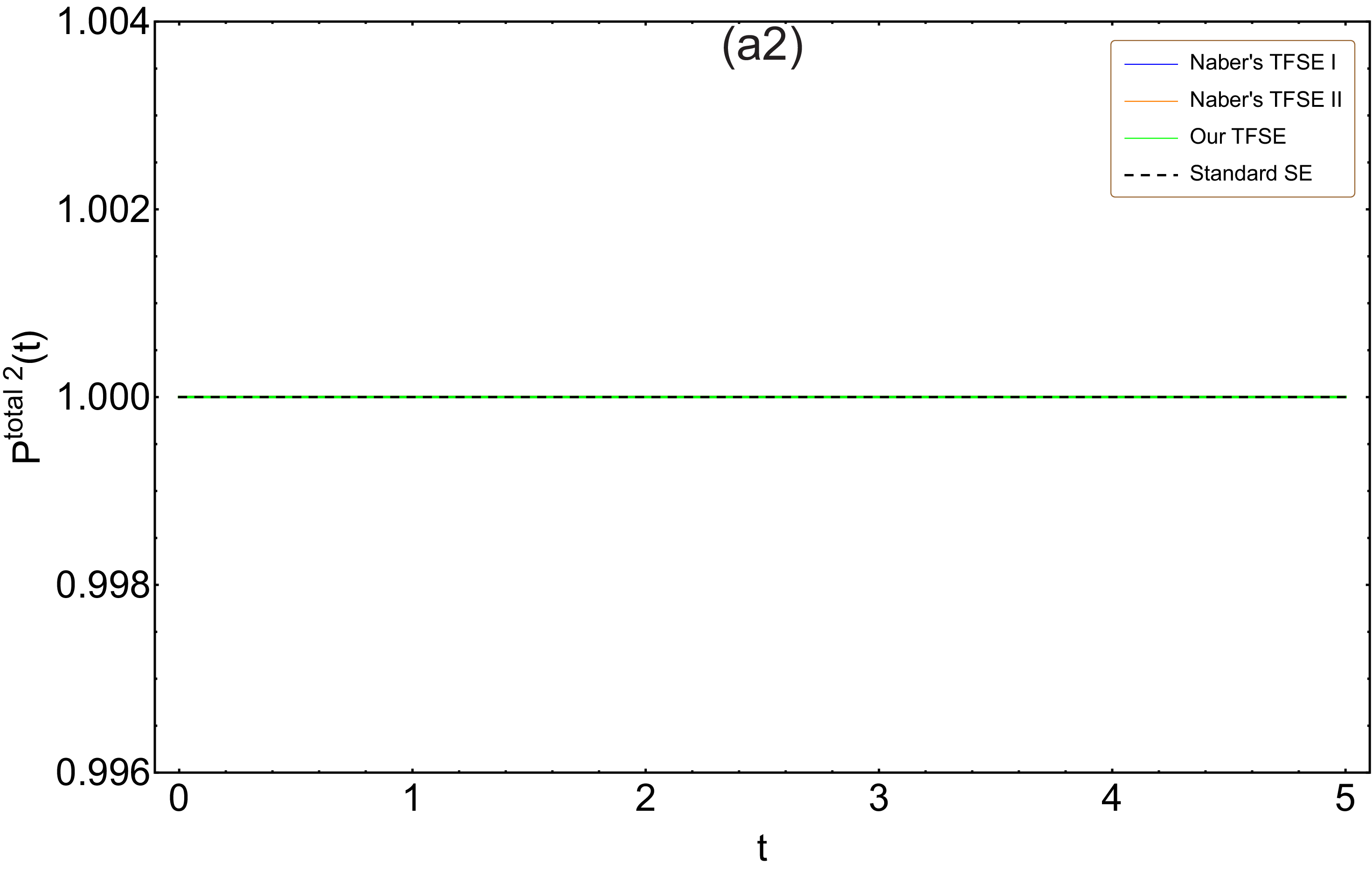}
    \includegraphics[width=0.32\linewidth]{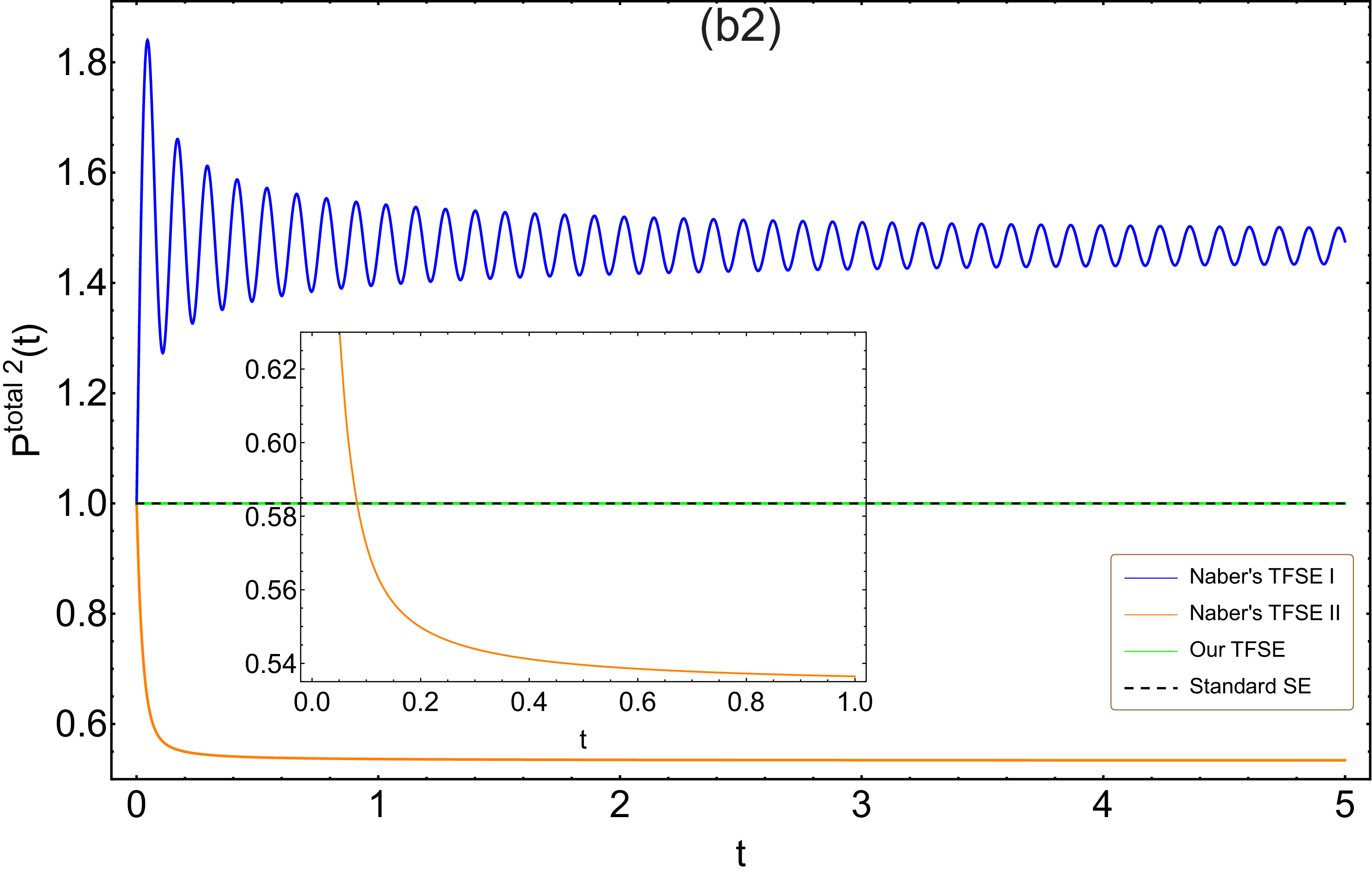}
    \includegraphics[width=0.32\linewidth]{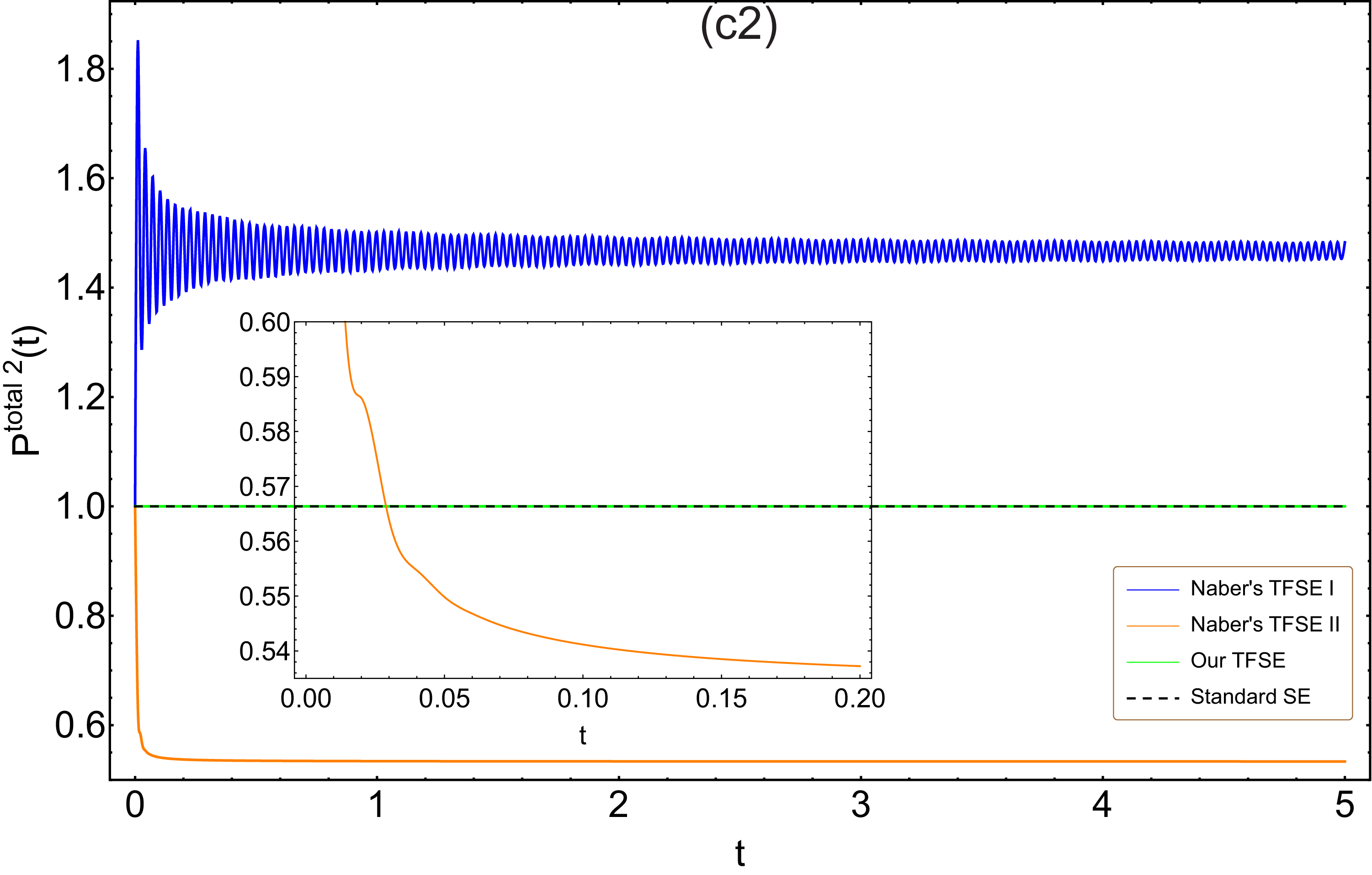}\\
\caption{The time evolution of $P_{\gamma=1,2,4}^{total\,1}(t)$ and $P_{\gamma=1,2,4}^{total\,2}(t)$ is plotted for $\lambda=0, 0.5, 1$. The settings of other parameters are $\beta=0.5$, $n=50$, and $C_0=0.5$. The insets show a very-short-time-scale evolution of $P_2^{total\,1}(t)$ and $P_2^{total\,2}(t)$.}
\label{Fig2}
\end{figure}

In Fig. \ref{Fig2} the time behavior of $P_{\gamma=1,2,4}^{total\,1}(t)$ and $P_{\gamma=1,2,4}^{total\,2}(t)$ is obtained from $\lambda=0, 0.5, 1$ by setting $\beta=0.5$, $n=50$, and $C_0=0.5$. For $\lambda=0$, $P_{\gamma=1,2,4}^{total\,1}(t)$ and $P_{\gamma=1,2,4}^{total\,2}(t)$ are all equal to one with time. In this case, the three TFSEs all perform well. When $\lambda\in(0,1]$, $P_1^{total\,1}(t)$ and $P_1^{total\,2}(t)$ are greater than or equal to one, while $P_2^{total\,1}(t)$ and $P_2^{total\,2}(t)$ are less than or equal to one. Moreover, as small $\lambda$ starts to increase and exceeds a certain critical value, the non-Markovian oscillations of $P_1^{total\,1}(t)$ and $P_1^{total\,2}(t)$ emerge and enhance. For small and intermediate $\lambda$, $P_2^{total\,1}(t)$ and $P_2^{total\,2}(t)$ become the monotonically decreasing functions of $t$. While for large $\lambda$, the time evolution of $P_2^{total\,1}(t)$ and $P_2^{total\,2}(t)$ may display oscillations. Clearly, for any $t$ and $\lambda$, $P_4^{total\,1}(t)$ and $P_4^{total\,2}(t)$ are conserved.

\begin{figure}[htbp]
\centering
    \includegraphics[width=0.32\linewidth]{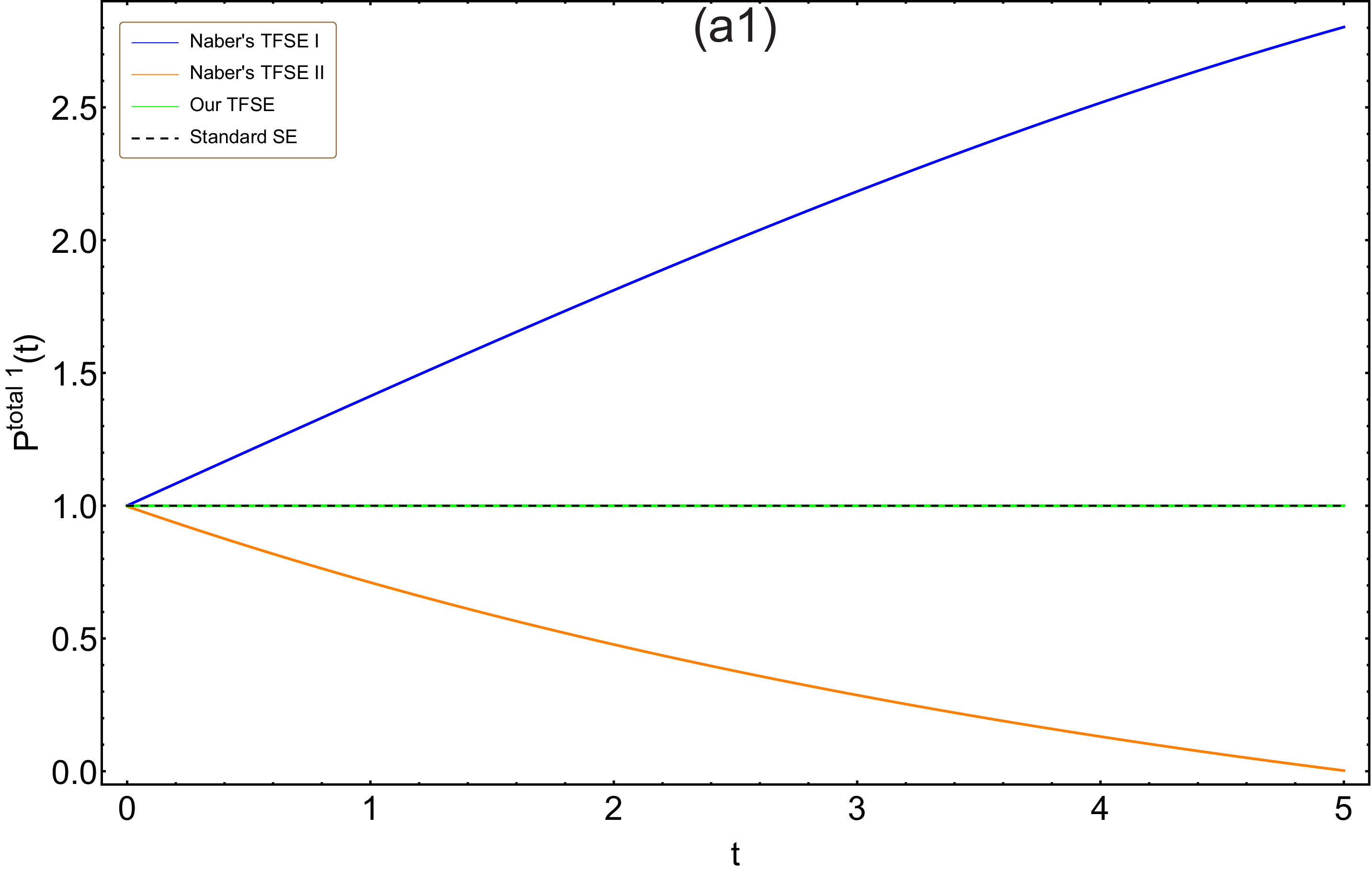}
    \includegraphics[width=0.32\linewidth]{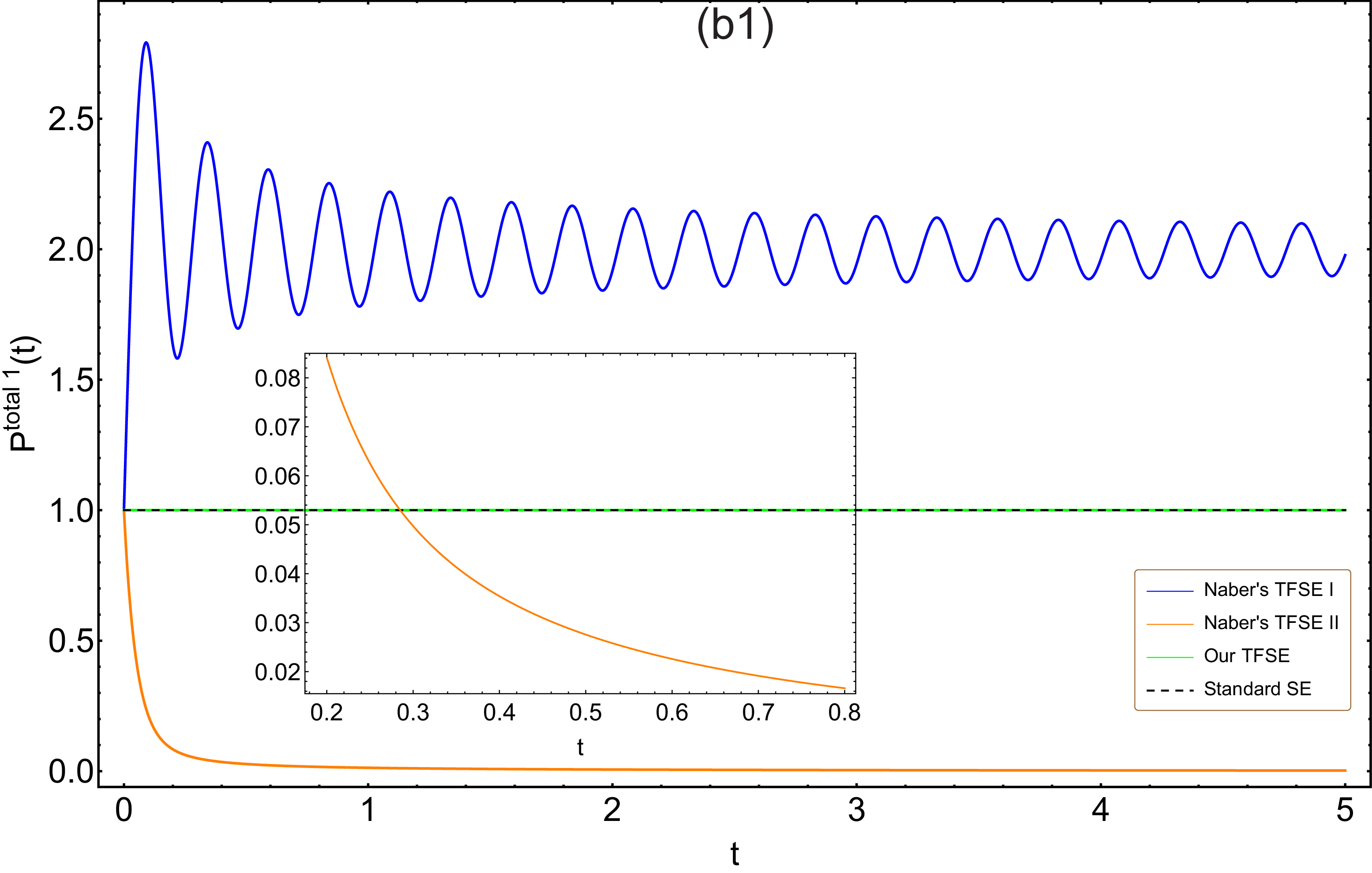}
    \includegraphics[width=0.32\linewidth]{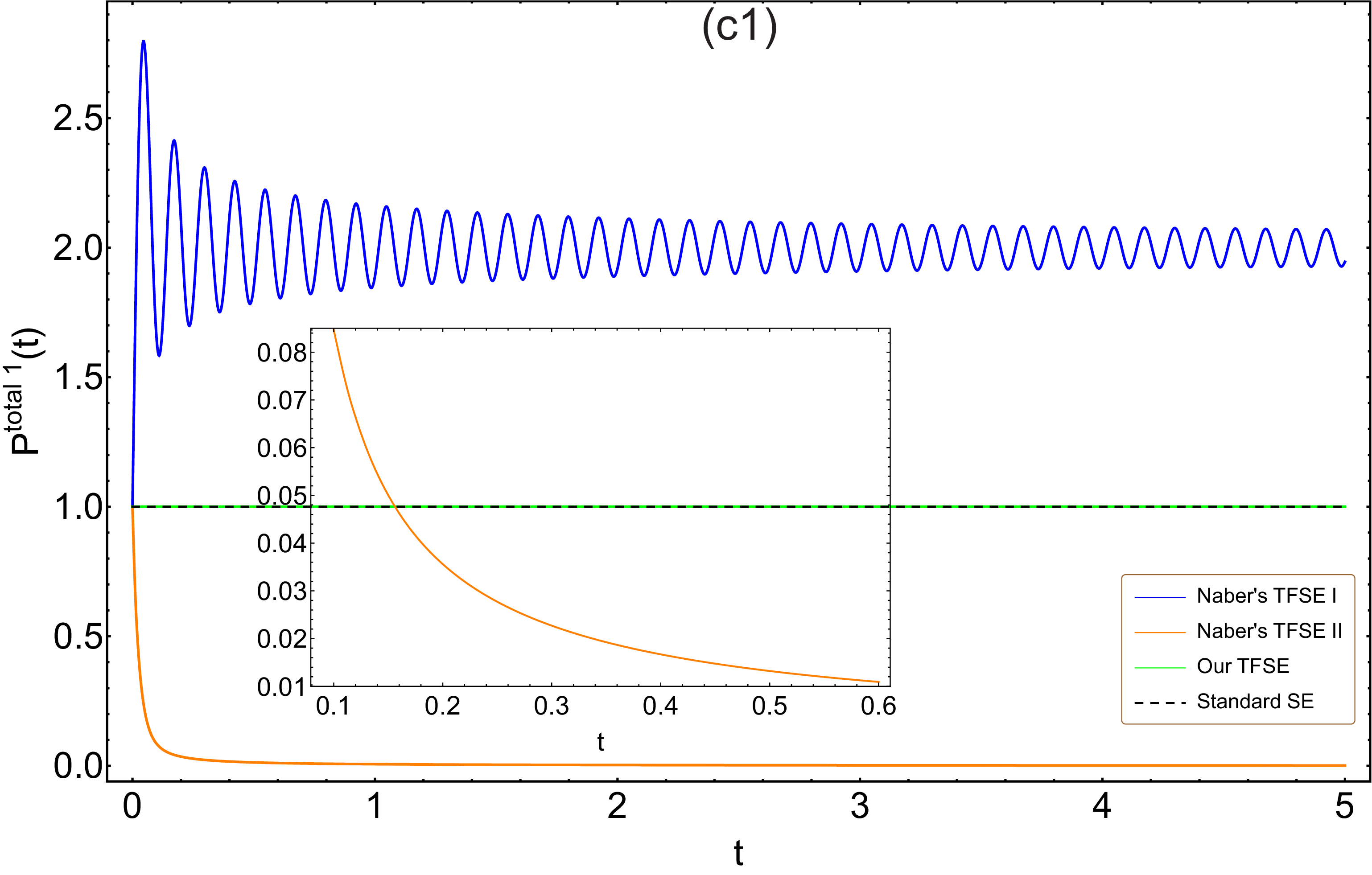}\\
    \includegraphics[width=0.32\linewidth]{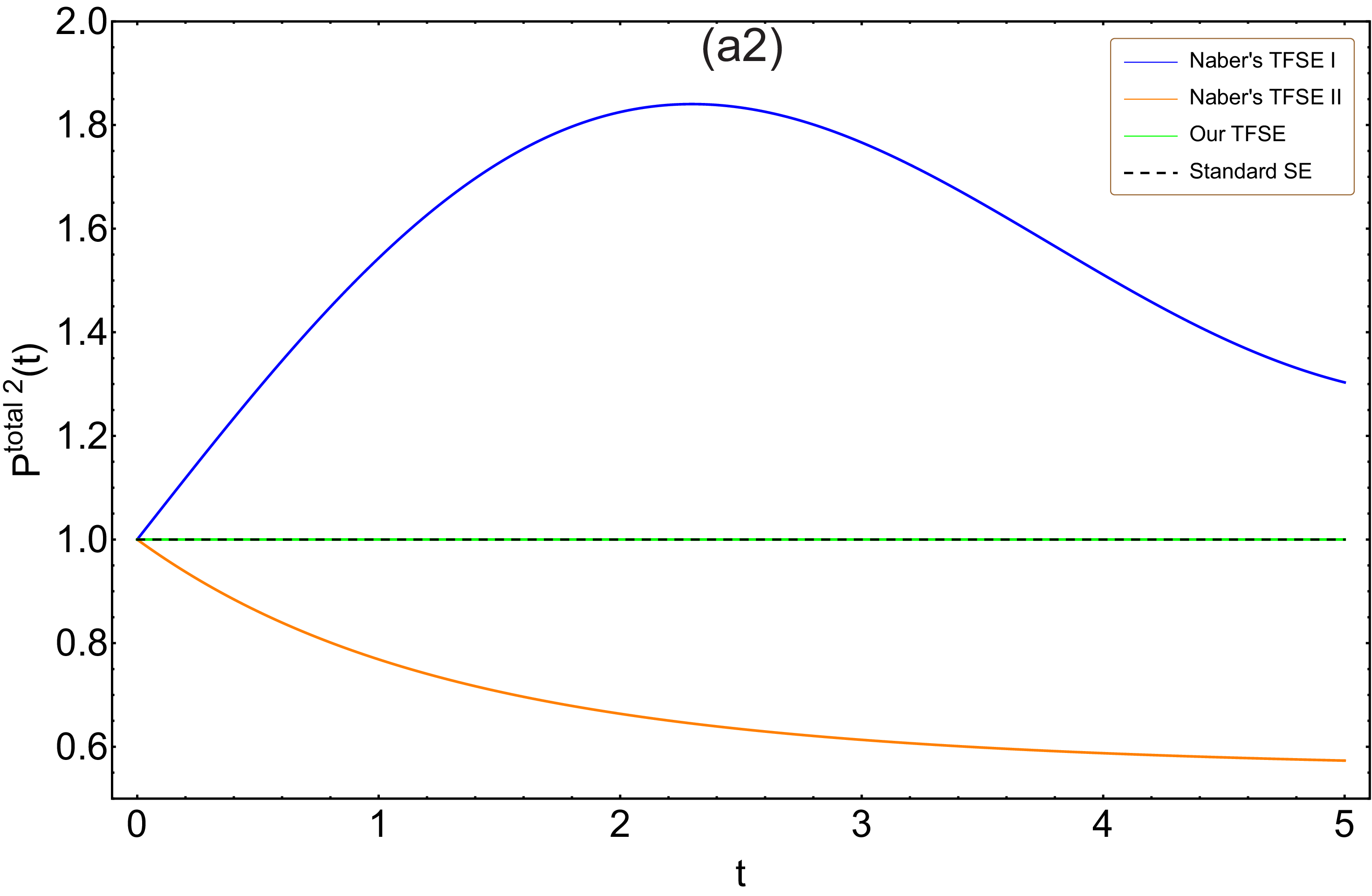}
    \includegraphics[width=0.32\linewidth]{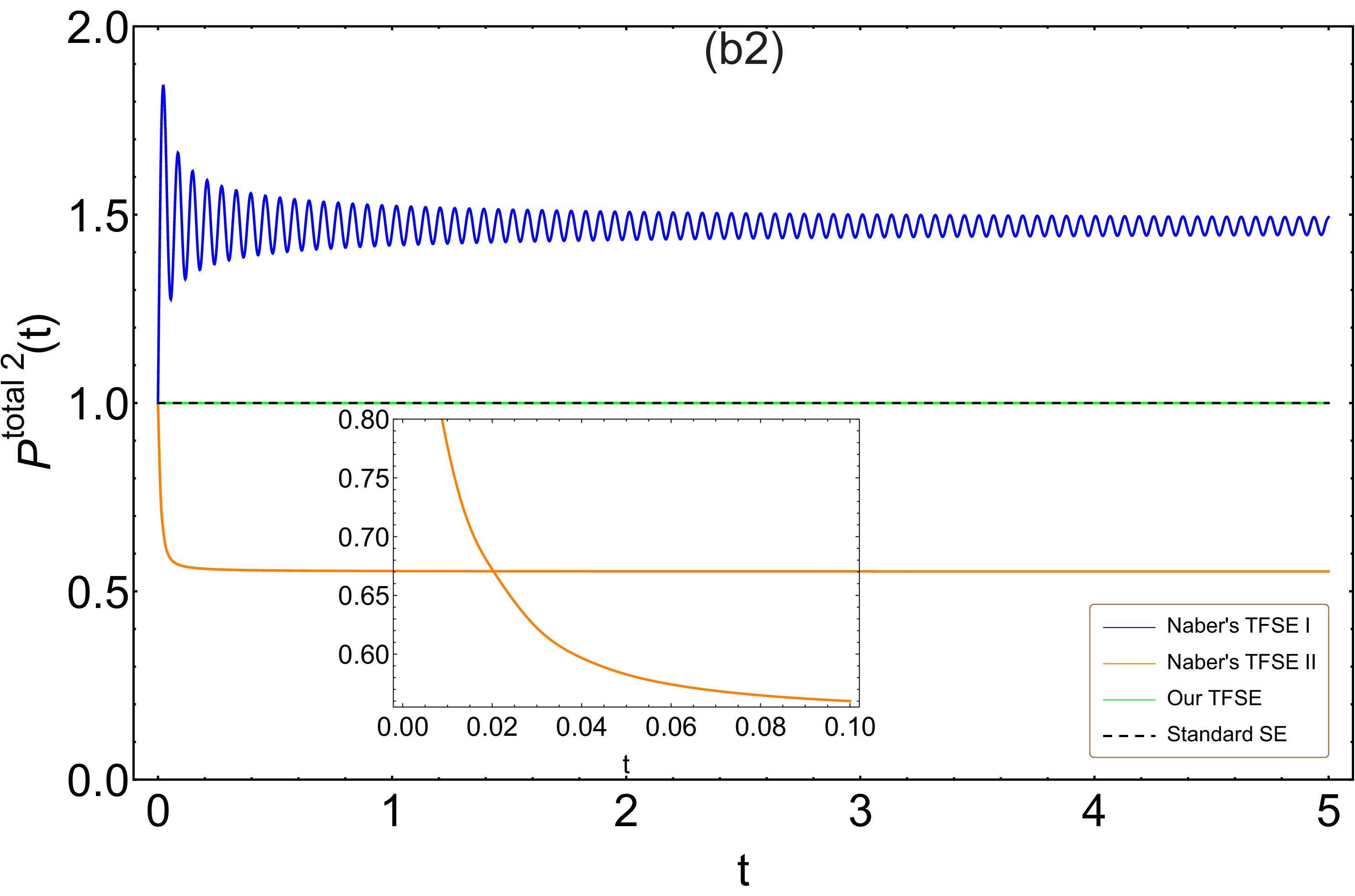}
    \includegraphics[width=0.32\linewidth]{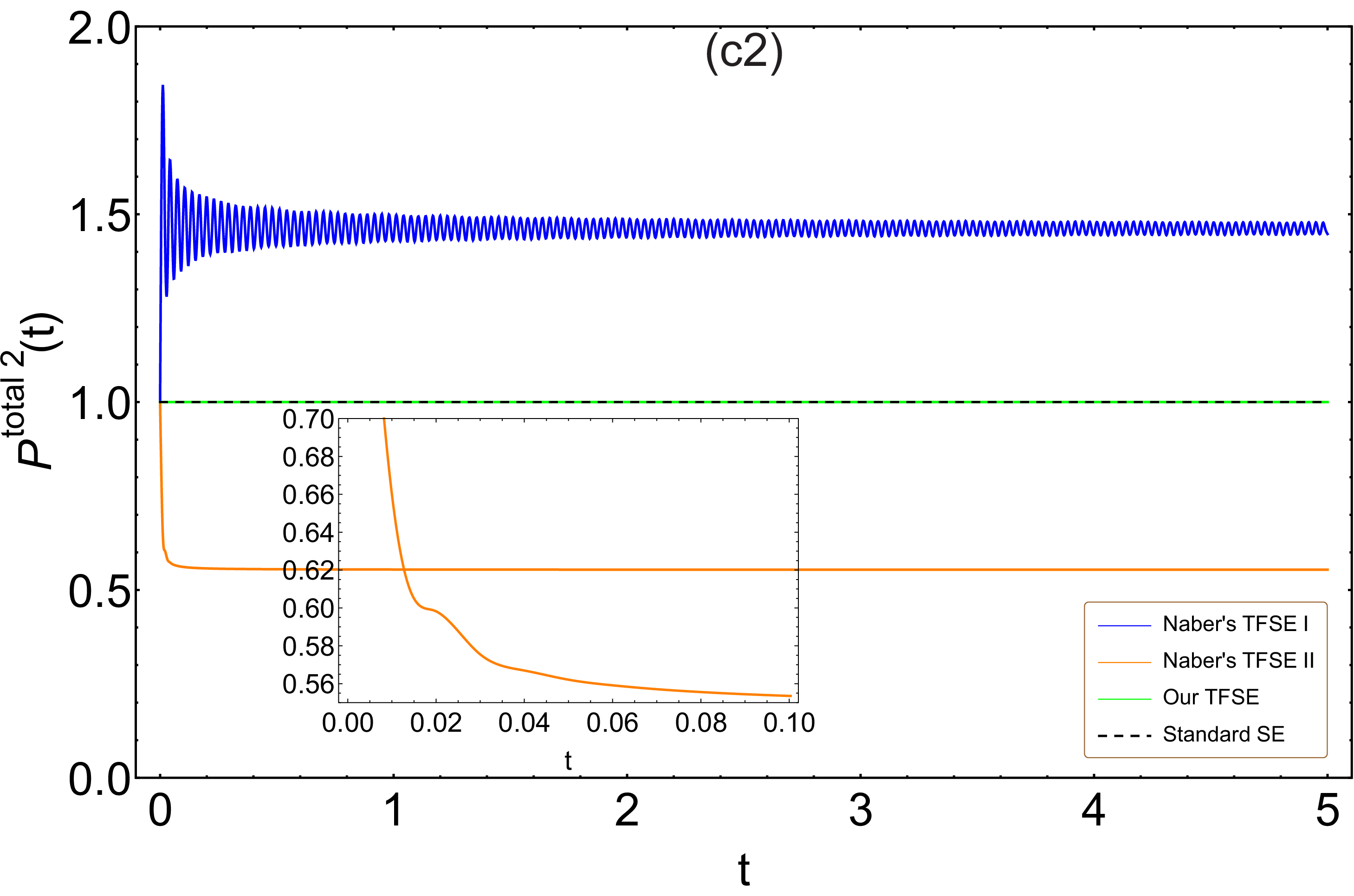}\\
\caption{The time evolution of $P_{\gamma=1,2,4}^{total\,1}(t)$ and $P_{\gamma=1,2,4}^{total\,2}(t)$ is plotted for $n=0, 100, 200$, where we have set $\beta=0.5$, $\lambda=0.5$, and $C_0=0.5$. The insets show the evolution of $P_2^{total\,1}(t)$ and $P_2^{total\,2}(t)$ in a very short time scale.}
\label{Fig3}
\end{figure}

Fig. \ref{Fig3} shows how the change in $n$ affects the time evolution of $P_{\gamma=1,2,4}^{total\,1}(t)$ and $P_{\gamma=1,2,4}^{total\,2}(t)$ for $\beta=0.5$, $\lambda=0.5$, and $C_0=0.5$. If the small $n$ gets larger, $P_1^{total\,1}(t)$ and $P_1^{total\,2}(t)$ of being greater than or equal to one will oscillate faster with a constant amplitude. But only for very large $n$, the oscillations in the time evolution of the $P_2^{total\,1}(t)$ and $P_2^{total\,2}(t)$ may occur. In addition, $P_4^{total\,1}(t)$ and $P_4^{total\,2}(t)$ are always equal to one independently of $n$.

\begin{figure}[htbp]
\centering
    \subfigure{\label{Fig4(a)}
    \includegraphics[width=0.32\linewidth]{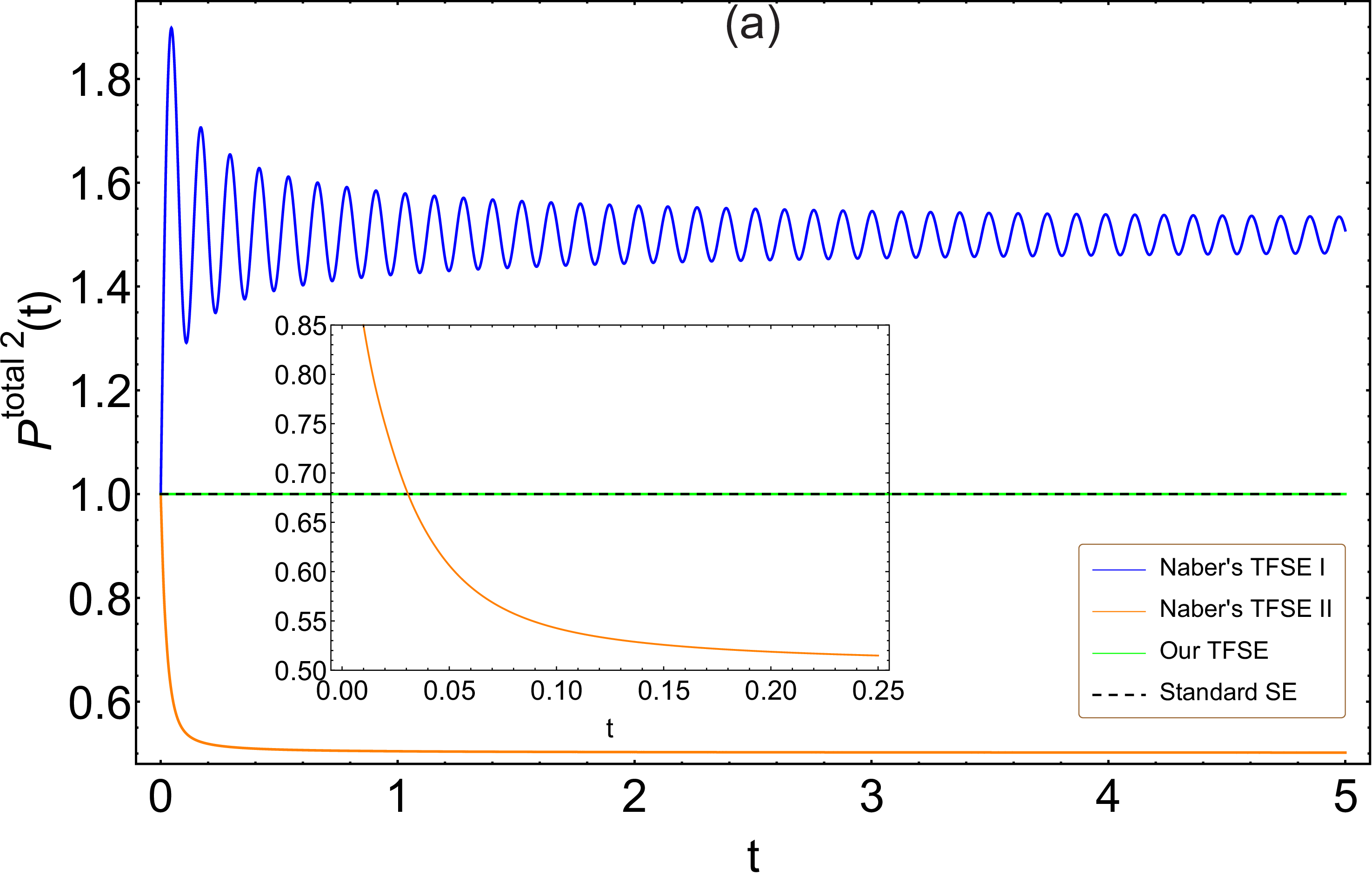}}
    \subfigure{\label{Fig4(b)}
    \includegraphics[width=0.32\linewidth]{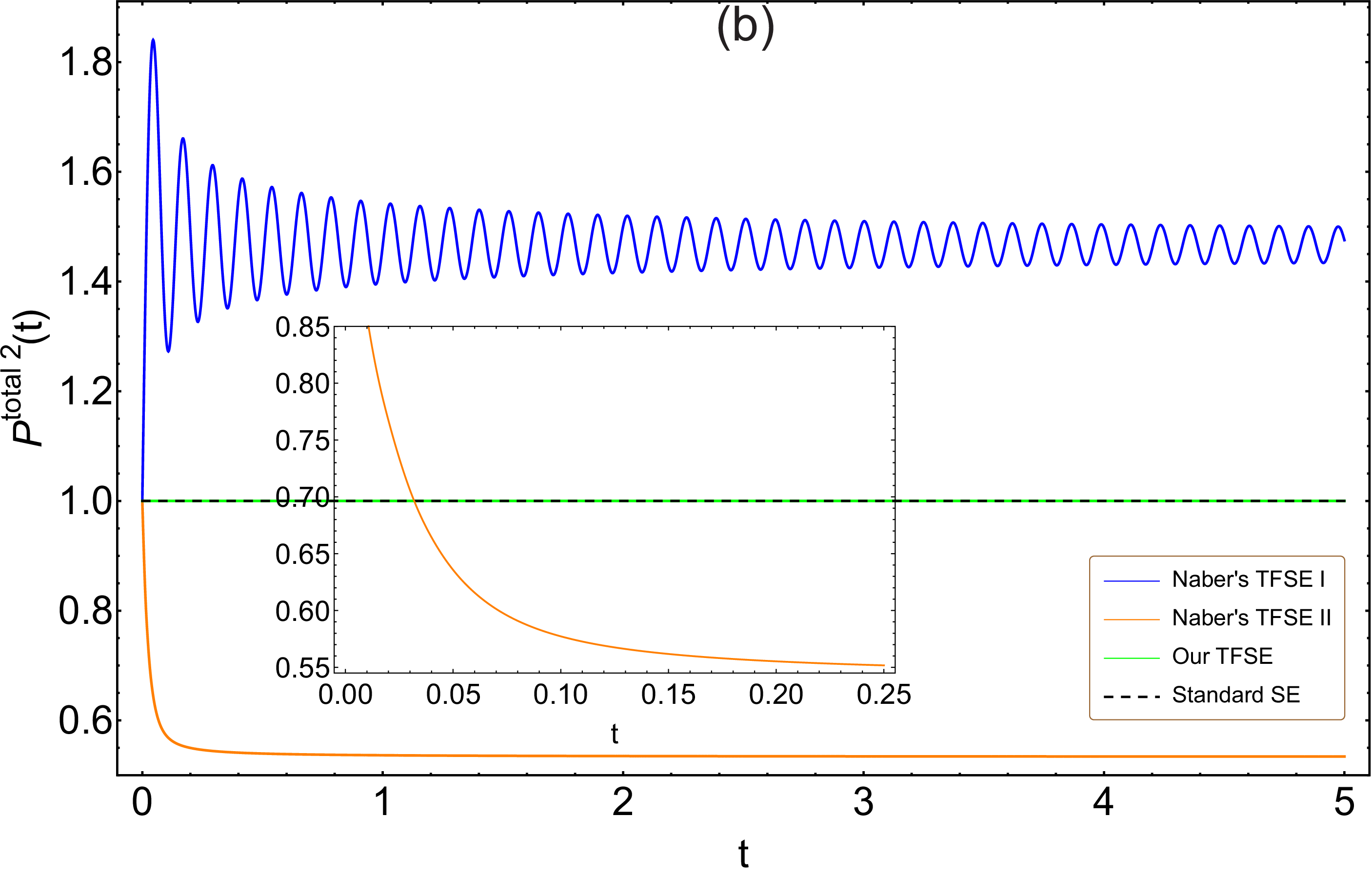}}
    \subfigure{\label{Fig4(c)}
    \includegraphics[width=0.32\linewidth]{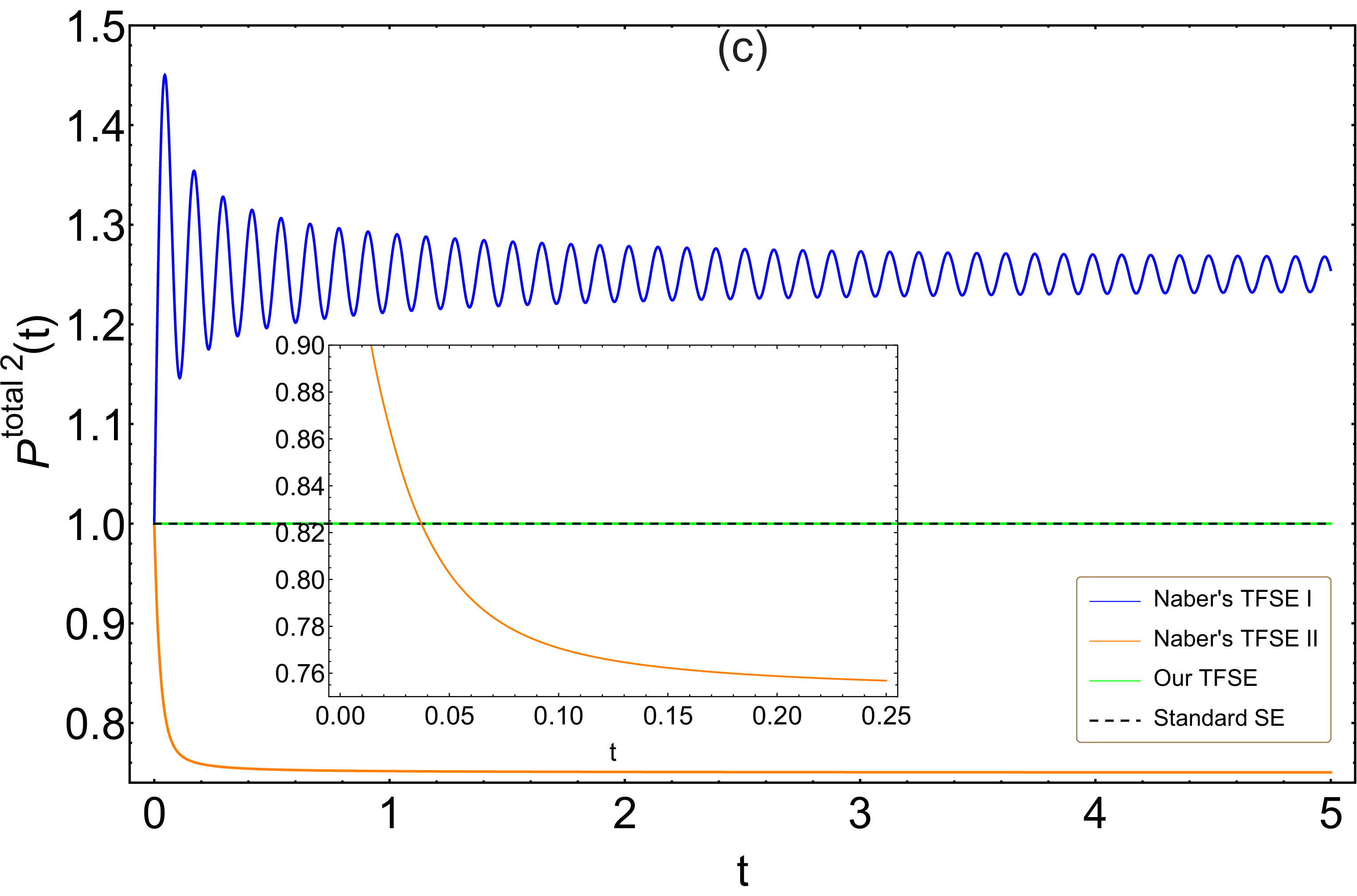}}\\
\caption{The time evolution of $P_{\gamma=1,2,4}^{total\,2}(t)$ is plotted for $C_0=0, 0.5, 1$. In each plot we have chosen $\beta=0.5$, $\lambda=0.5$, and $n=50$. The insets show how $P_2^{total\,2}(t)$ change in a very short time scale.}
\label{Fig4}
\end{figure}

Fig. \ref{Fig4} makes a comparison of the time behavior among $P_{\gamma=1,2,4}^{total\,2}(t)$ for $C_0=0.1, 0.5, 1$ with $\beta=0.5$, $\lambda=0.5$, and $n=50$. We see that $P_1^{total\,2}(t)$ exhibits the non-Markovian oscillations regardless of how $C_0$ changes as time goes on, where an increase in $C_0$ may decrease the oscillating amplitude of $P_1^{total\,2}(t)$ of being always greater than one, but has no effect on its oscillating frequency. $P_2^{total\,2}(t)$ that is less than one for any $t$ moves closer to one as $C_0$ increases, and it as a function of $t$ may be monotonic by viewing from the insets in Figs. \ref{Fig4(a)} to \ref{Fig4(c)}. Significantly, $P_4^{total\,2}(t)$ equals one without dependence on $C_0$ at all $t$.

\begin{figure}[htbp]
\centering
    \subfigure{\label{Fig5(a1)}
    \includegraphics[width=0.32\linewidth]{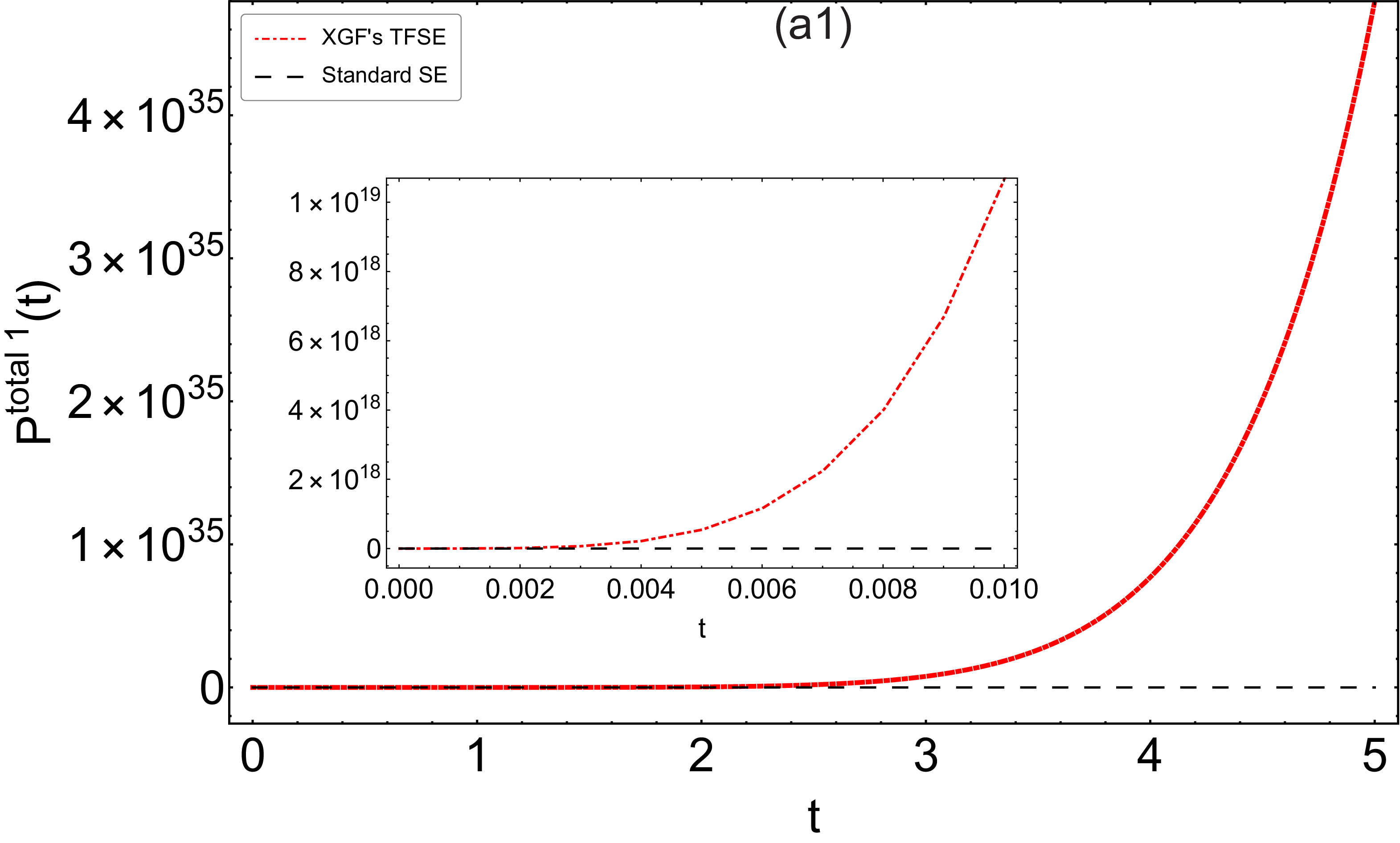}}
    \subfigure{\label{Fig5(b1)}
    \includegraphics[width=0.32\linewidth]{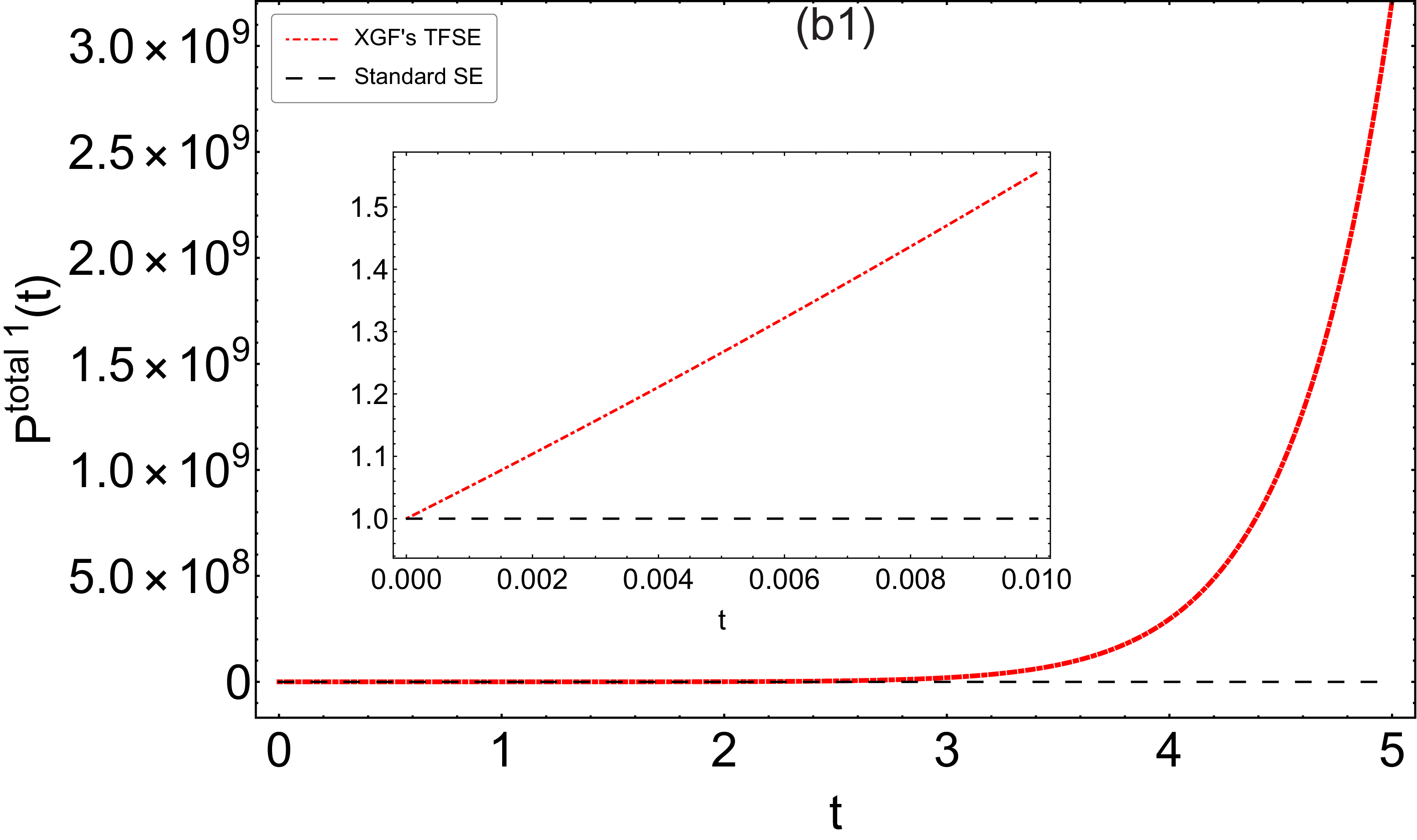}}
    \subfigure{\label{Fig5(c1)}
    \includegraphics[width=0.32\linewidth]{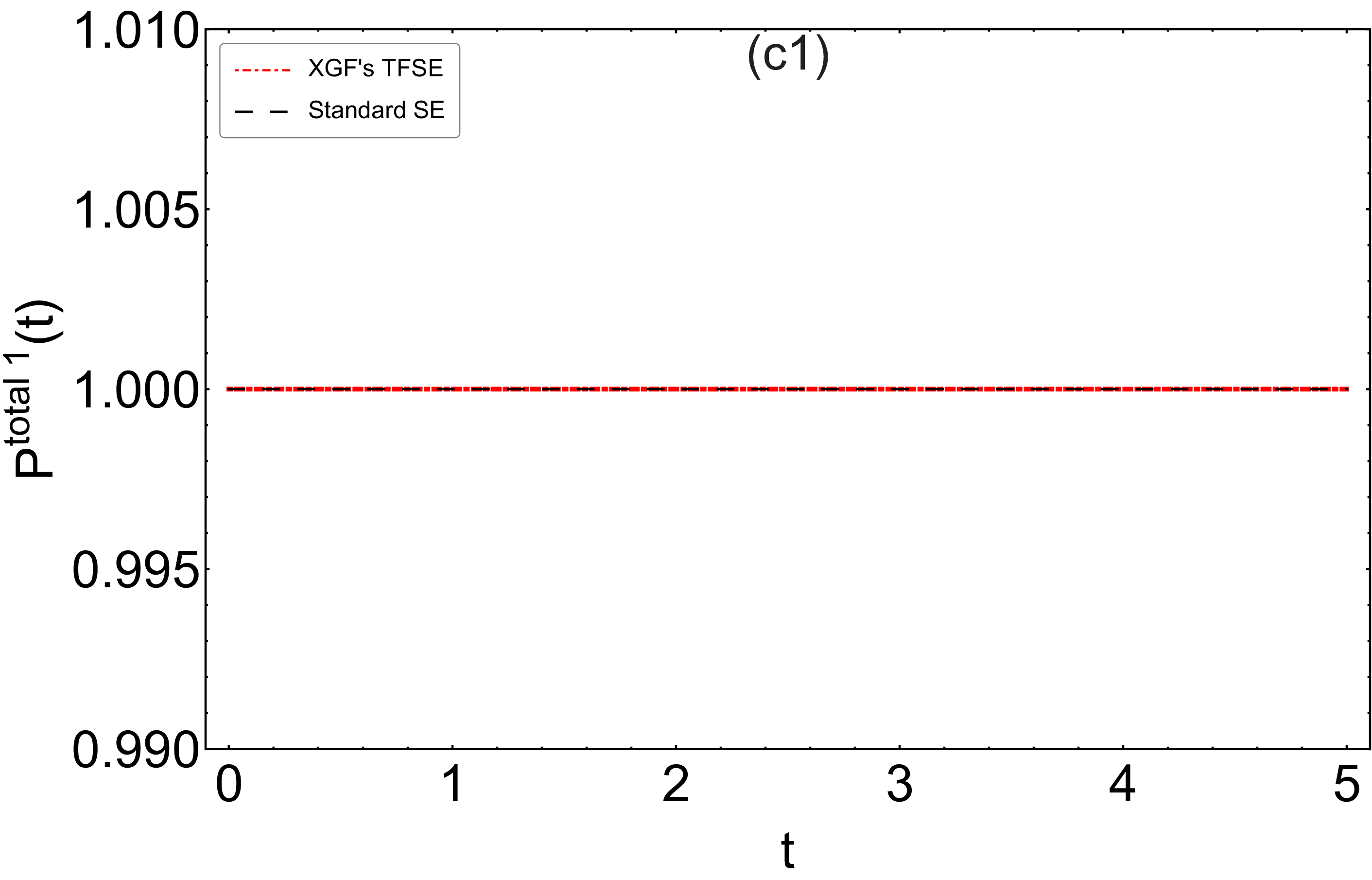}}\\
    \subfigure{\label{Fig5(a2)}
    \includegraphics[width=0.32\linewidth]{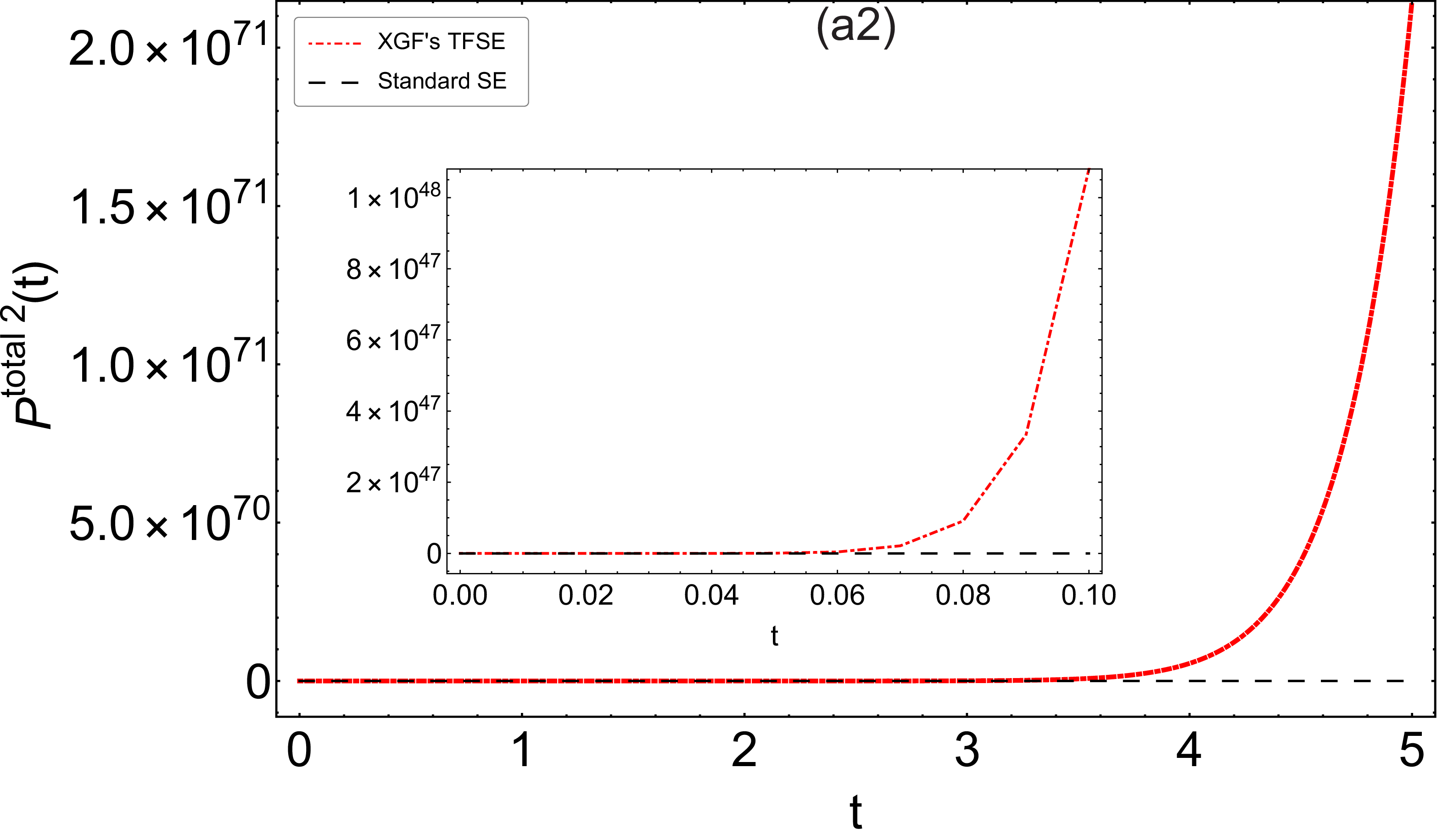}}
    \subfigure{\label{Fig5(b2)}
    \includegraphics[width=0.32\linewidth]{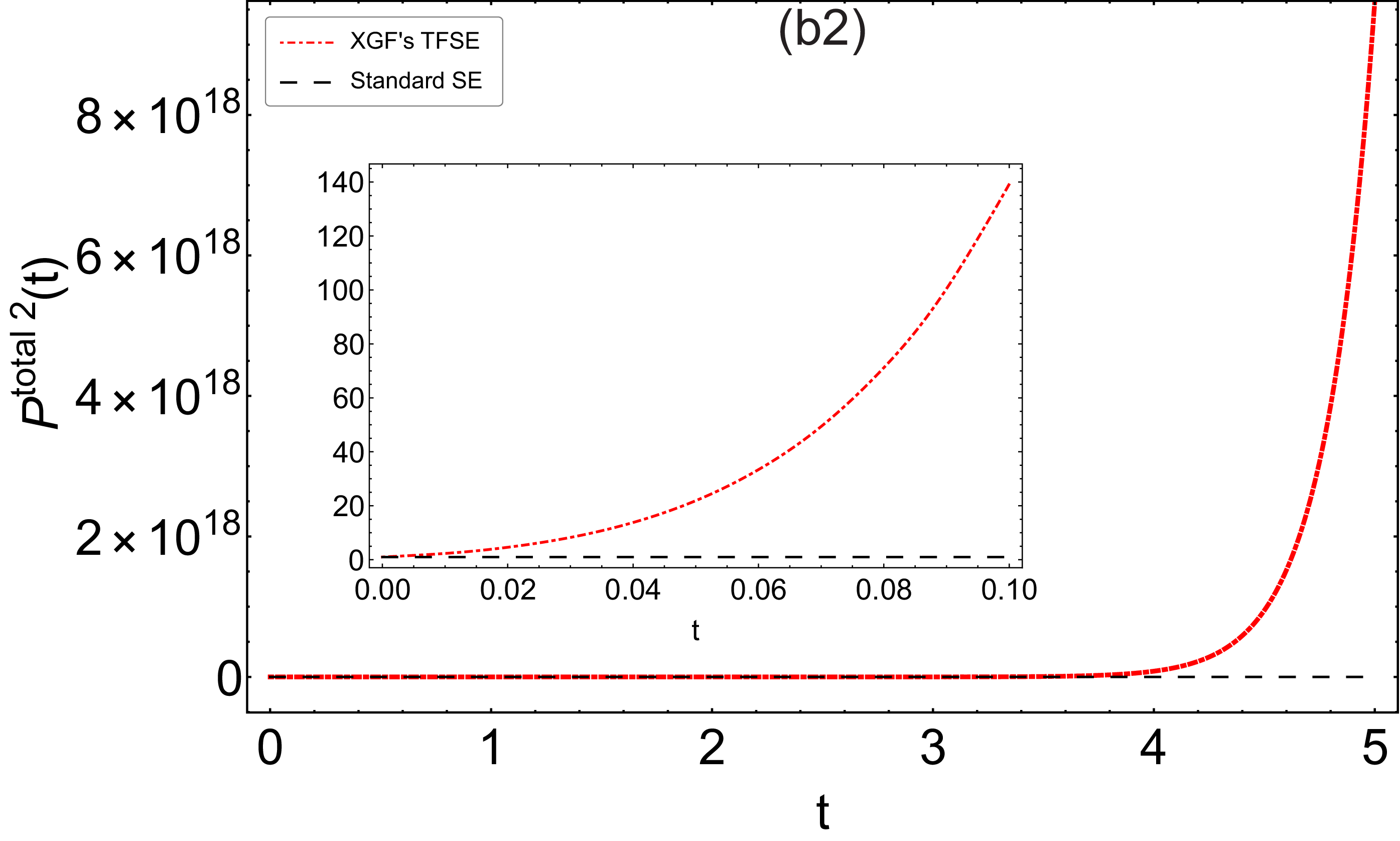}}
    \subfigure{\label{Fig5(c2)}
    \includegraphics[width=0.32\linewidth]{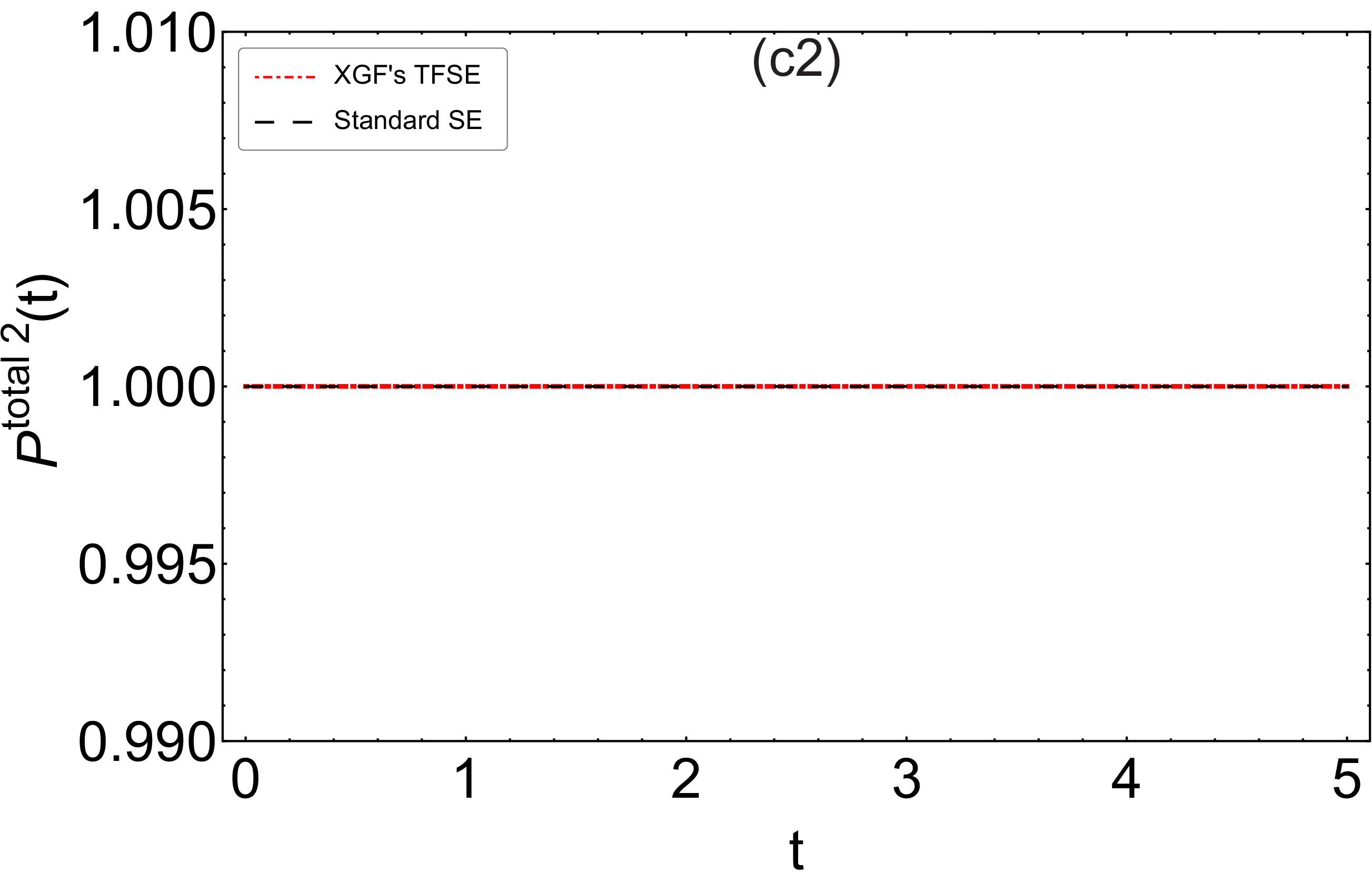}}\\
\caption{The time evolution of $P_{\gamma=3}^{total\,1}(t)$ and $P_{\gamma=3}^{total\,2}(t)$ is plotted for $\beta=0.1, 0.5, 1$. Other input parameters are $\lambda=0.5$, $n=50$, and $C_0=0.5$. The insets show a very-short-time-scale evolution of $P_3^{total\,1}(t)$ and $P_3^{total\,2}(t)$.}
\label{Fig5}
\end{figure}

\begin{figure}[htbp]
\centering
    \includegraphics[width=0.32\linewidth]{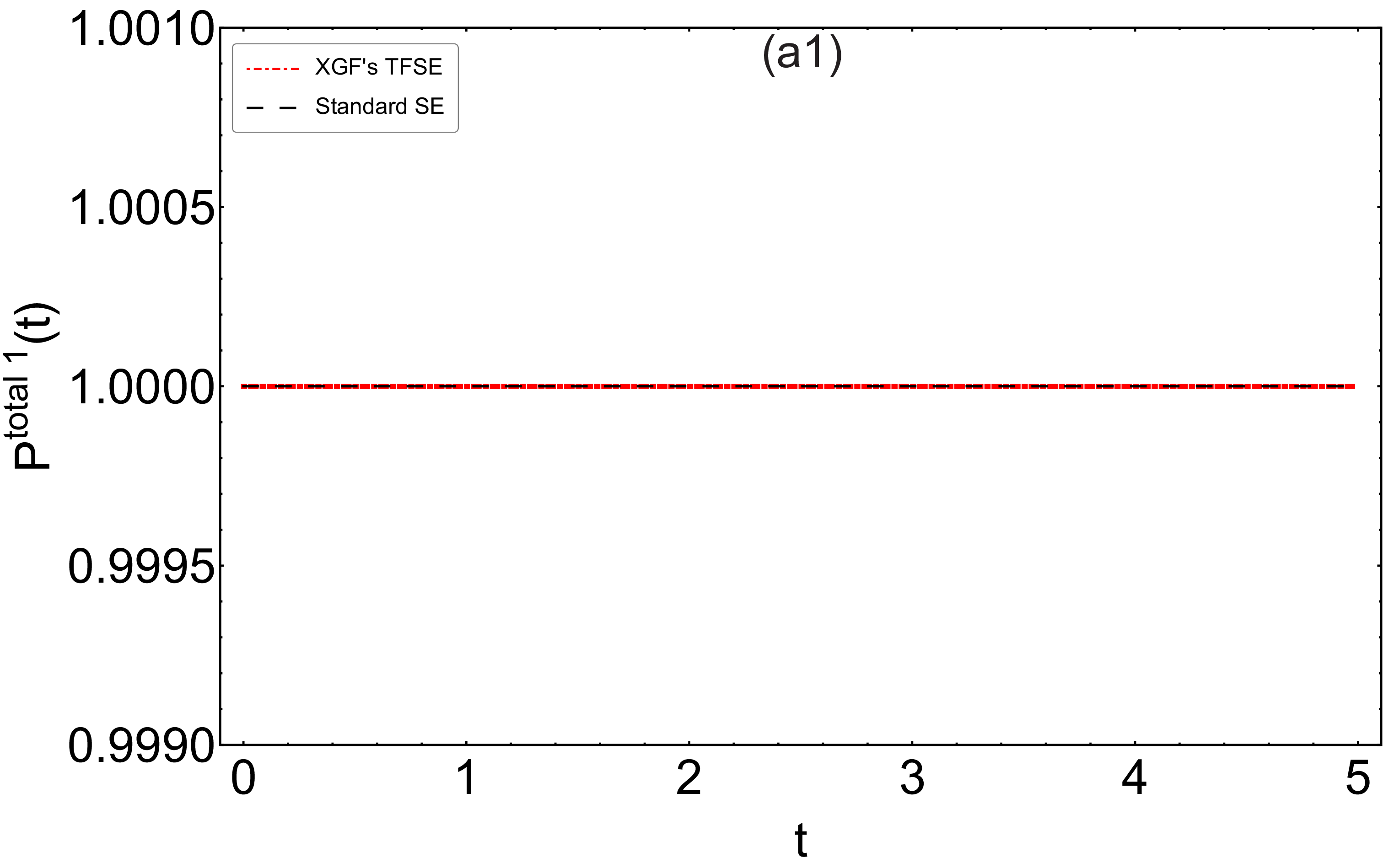}
    \includegraphics[width=0.32\linewidth]{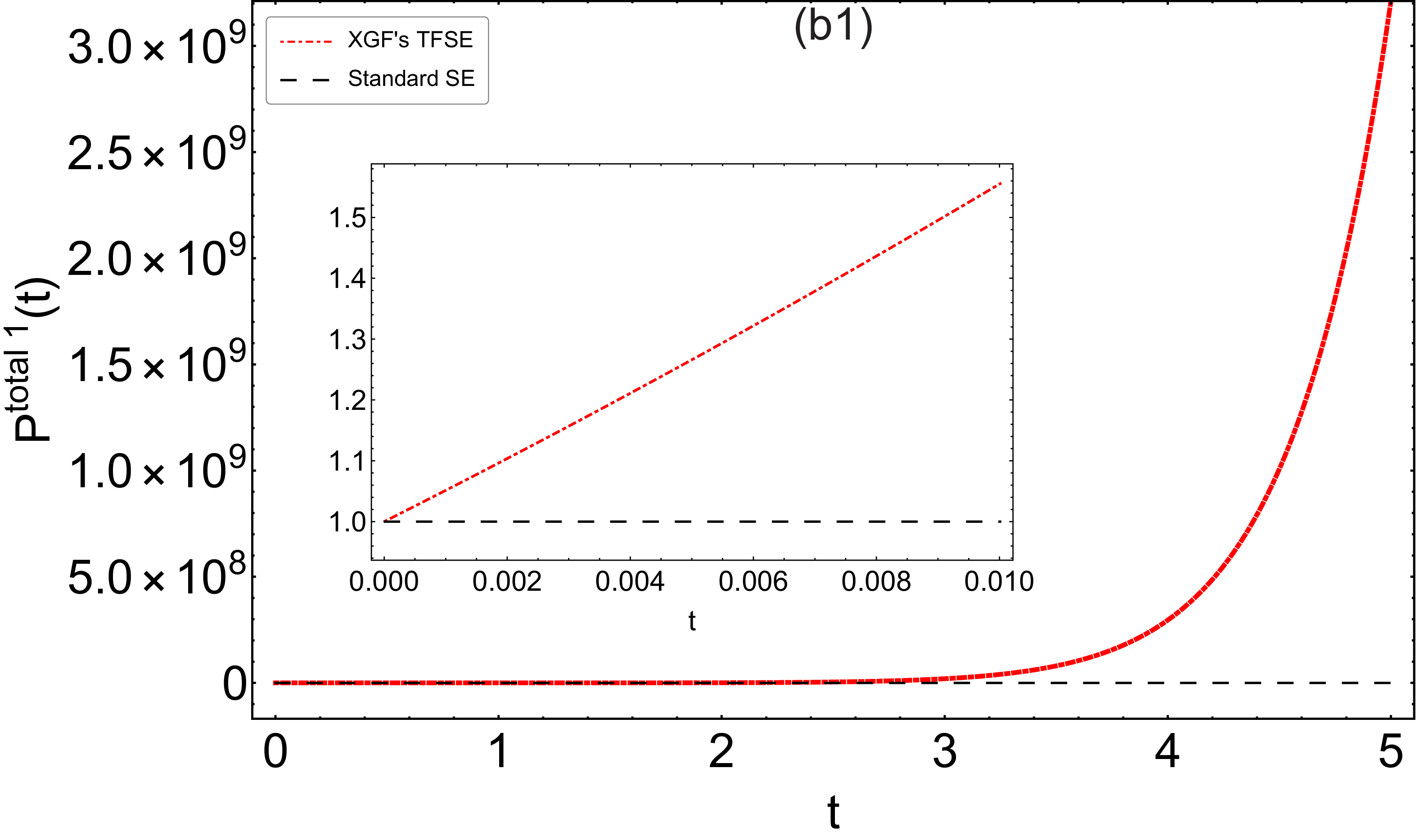}
    \includegraphics[width=0.32\linewidth]{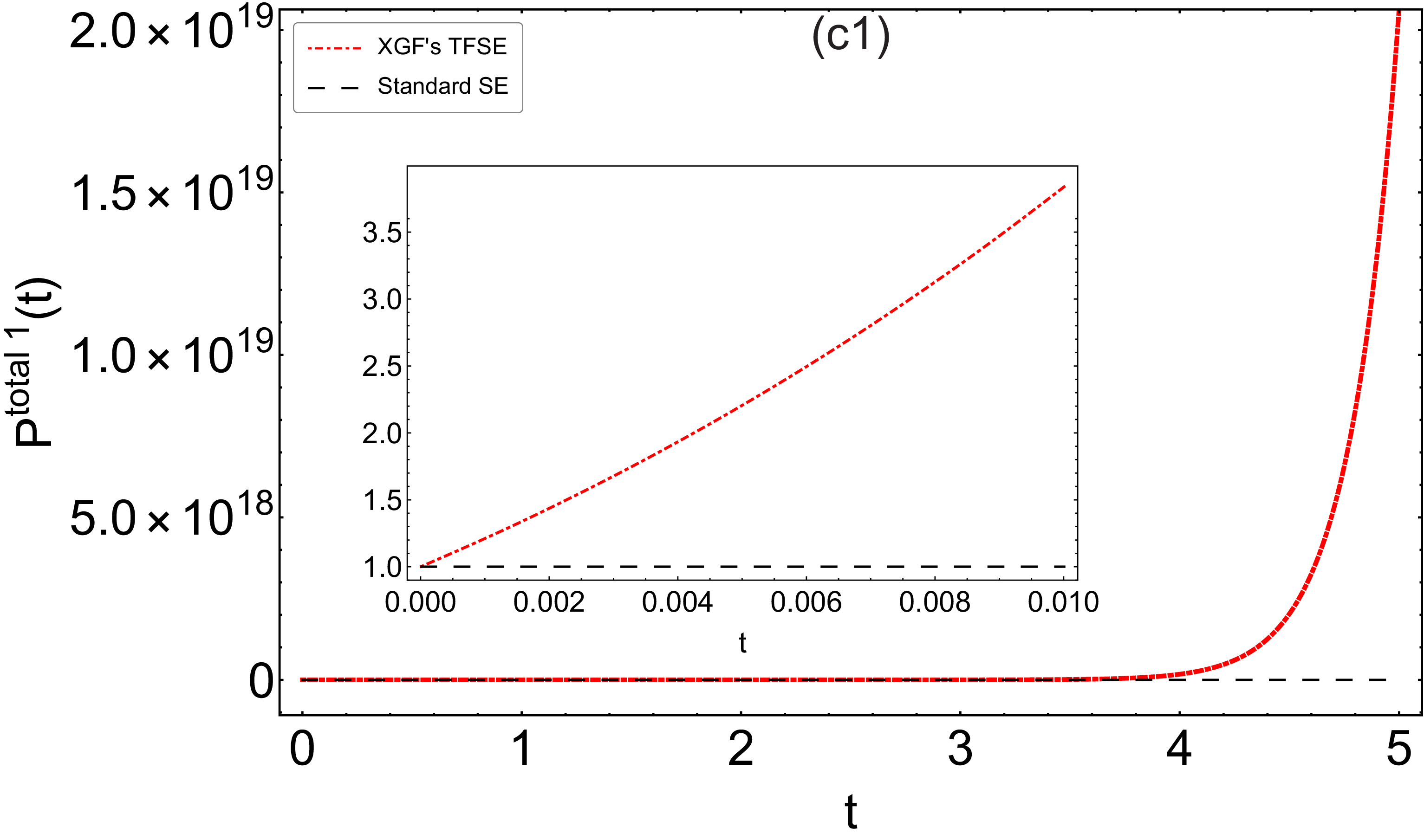}\\
    \includegraphics[width=0.32\linewidth]{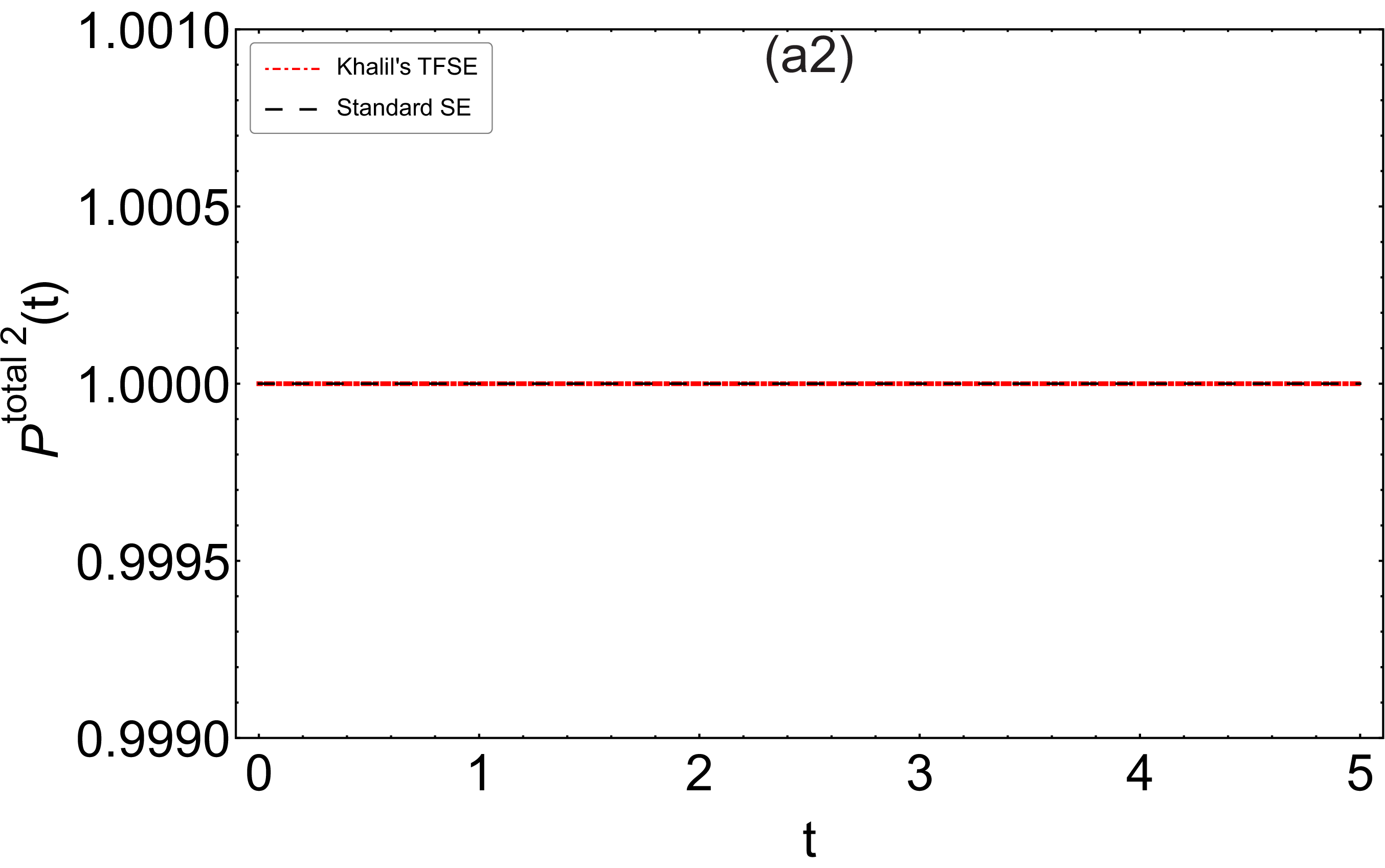}
    \includegraphics[width=0.32\linewidth]{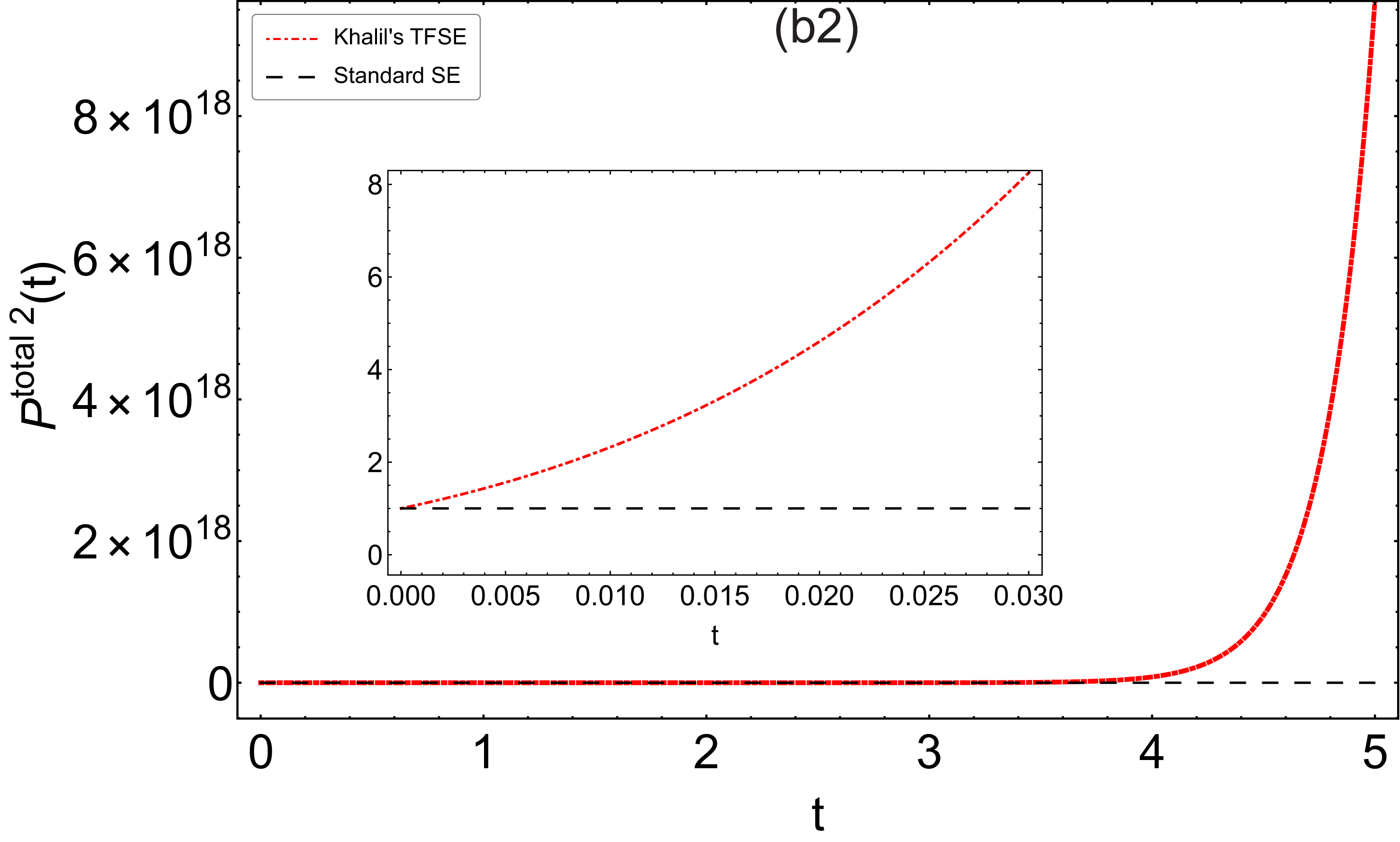}
    \includegraphics[width=0.32\linewidth]{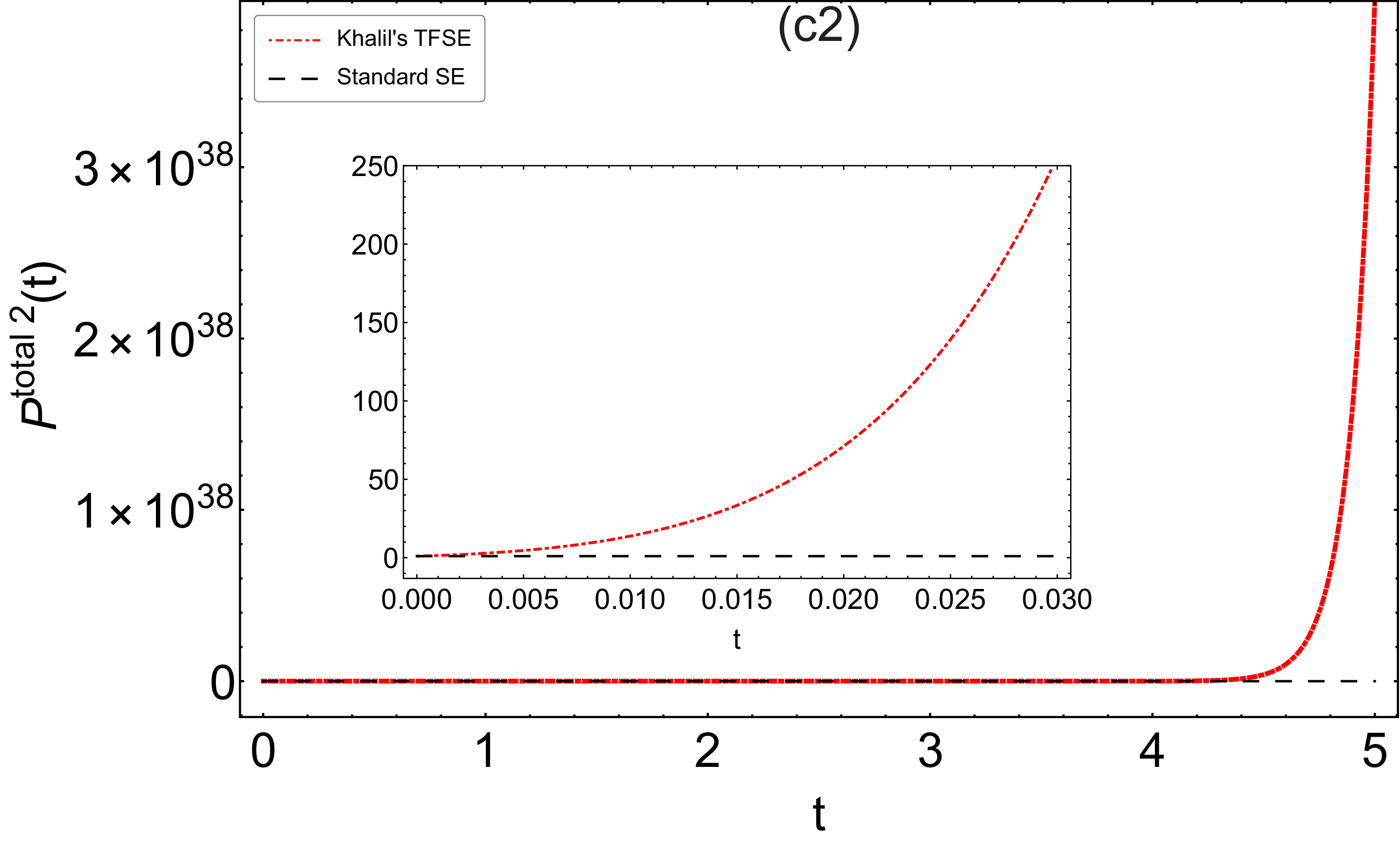}\\
\caption{The time evolution of $P_{\gamma=3}^{total\,1}(t)$ and $P_{\gamma=3}^{total\,2}(t)$ is plotted for $\lambda=0, 0.5, 1$. Other input parameters are $\beta=0.5$, $n=50$, and $C_0=0.5$. The insets show the evolution of $P_3^{total\,1}(t)$ and $P_3^{total\,2}(t)$ in a very short time scale.}
\label{Fig6}
\end{figure}

\begin{figure}[htbp]
\centering
    \includegraphics[width=0.32\linewidth]{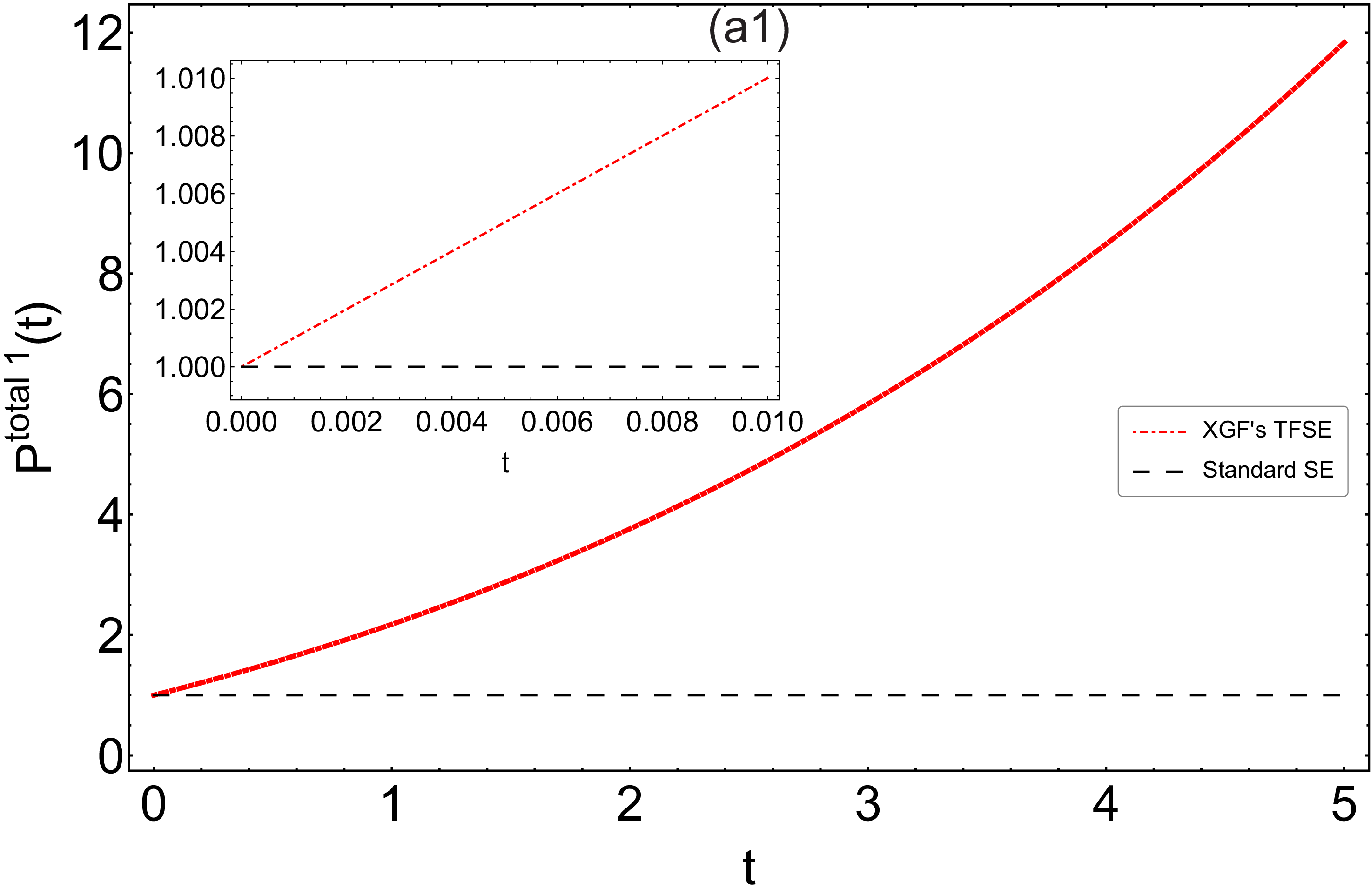}
    \includegraphics[width=0.32\linewidth]{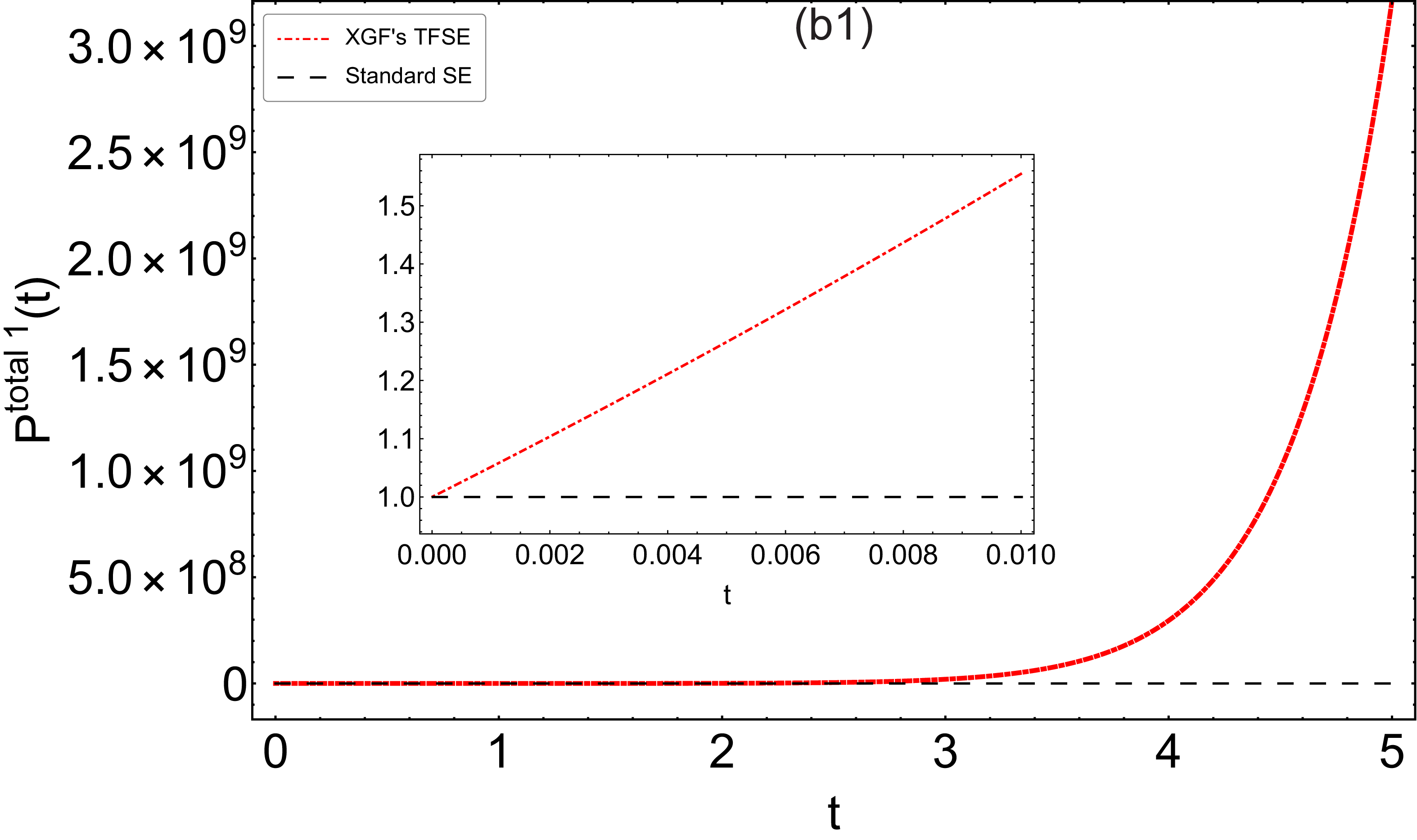}
    \includegraphics[width=0.32\linewidth]{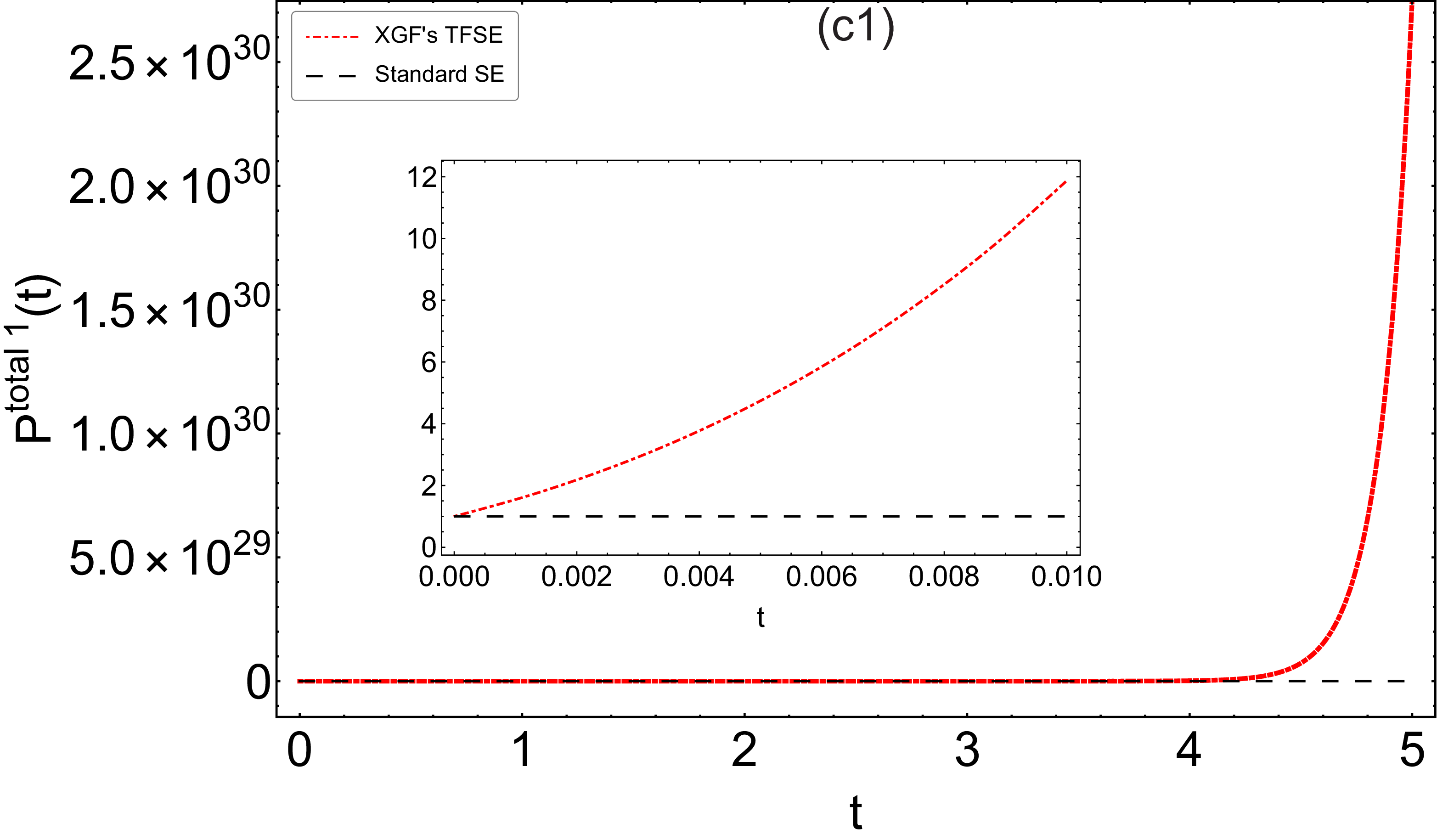}\\
    \includegraphics[width=0.32\linewidth]{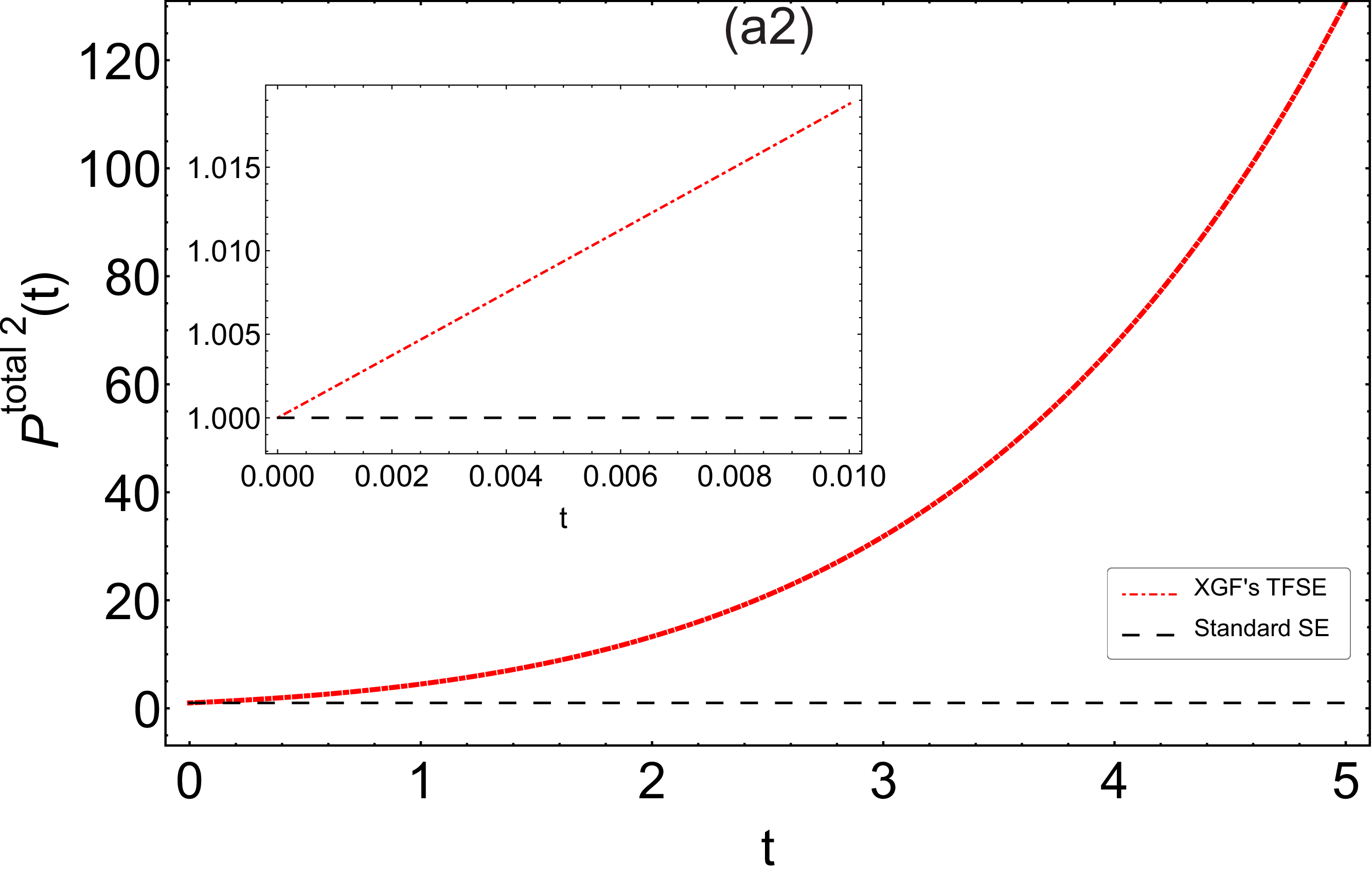}
    \includegraphics[width=0.32\linewidth]{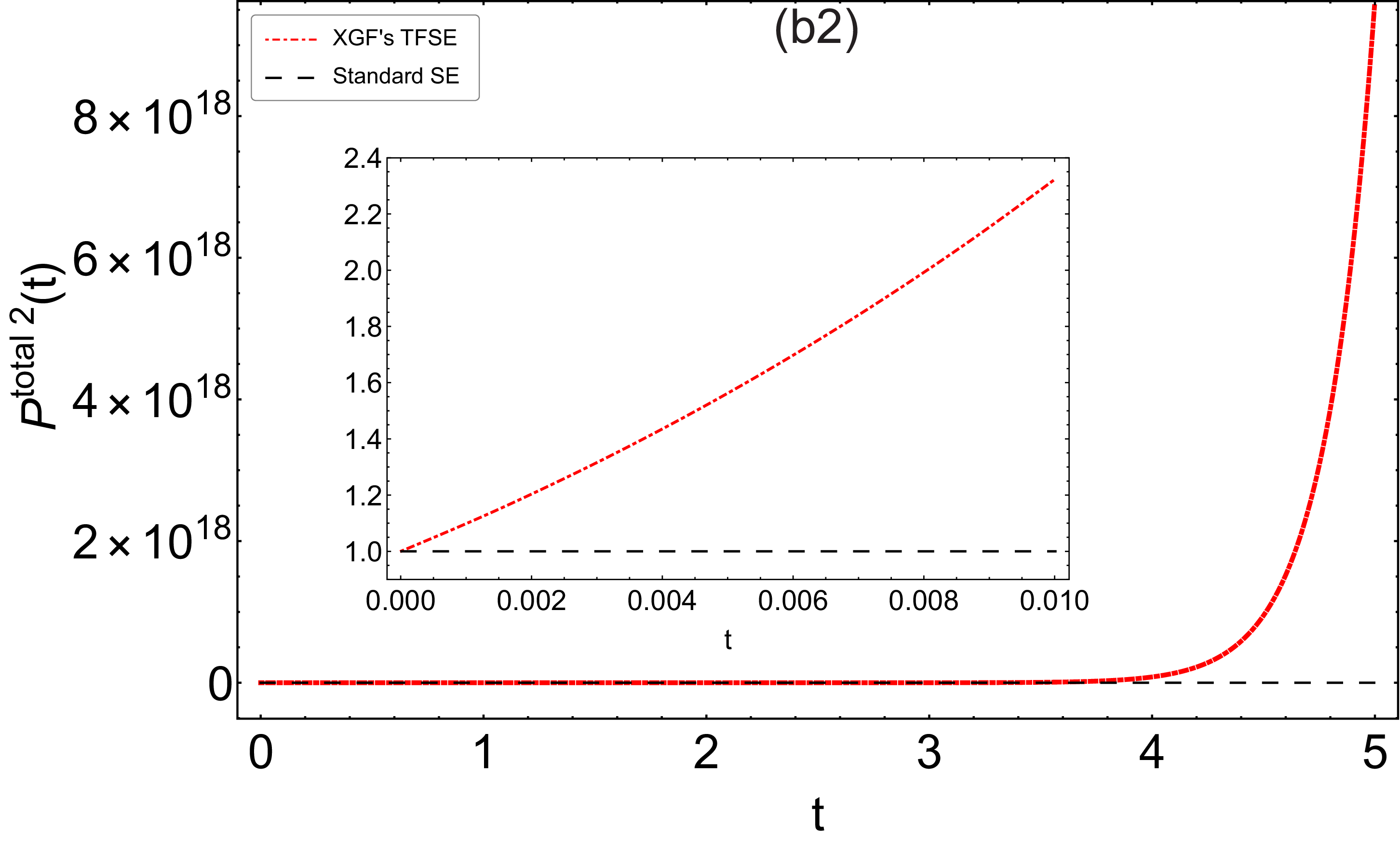}
    \includegraphics[width=0.32\linewidth]{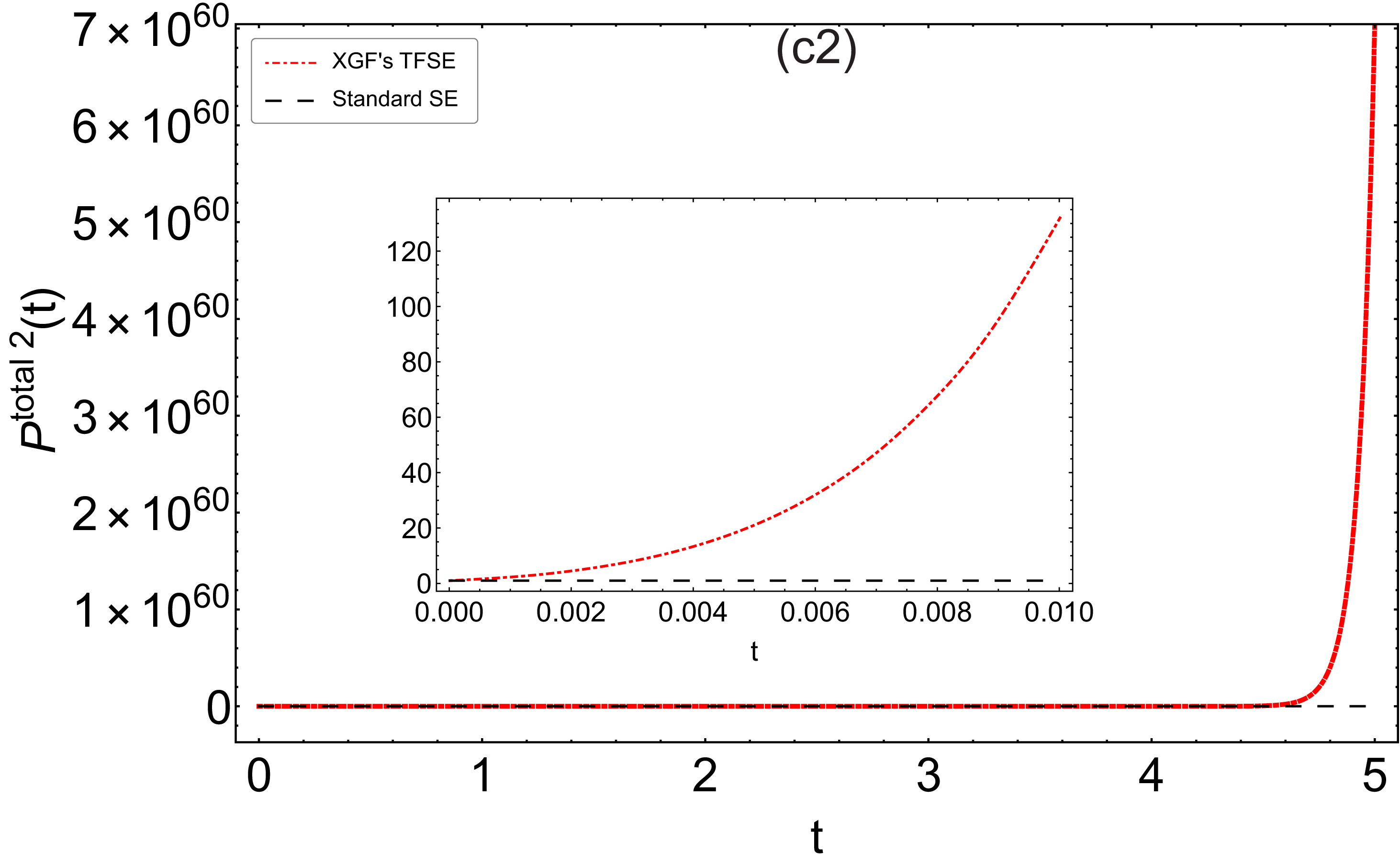}\\
\caption{The time evolution of $P_{\gamma=3}^{total\,1}(t)$ and $P_{\gamma=3}^{total\,2}(t)$ is plotted for $n=0, 50, 500$. Other input parameters are $\beta=0.5$, $\lambda=0.5$, and $C_0=0.5$. The insets show how $P_3^{total\,1}(t)$ and $P_3^{total\,2}(t)$ change in a very short time scale.}
\label{Fig7}
\end{figure}

\begin{figure}[htbp]
\centering
    \includegraphics[width=0.32\linewidth]{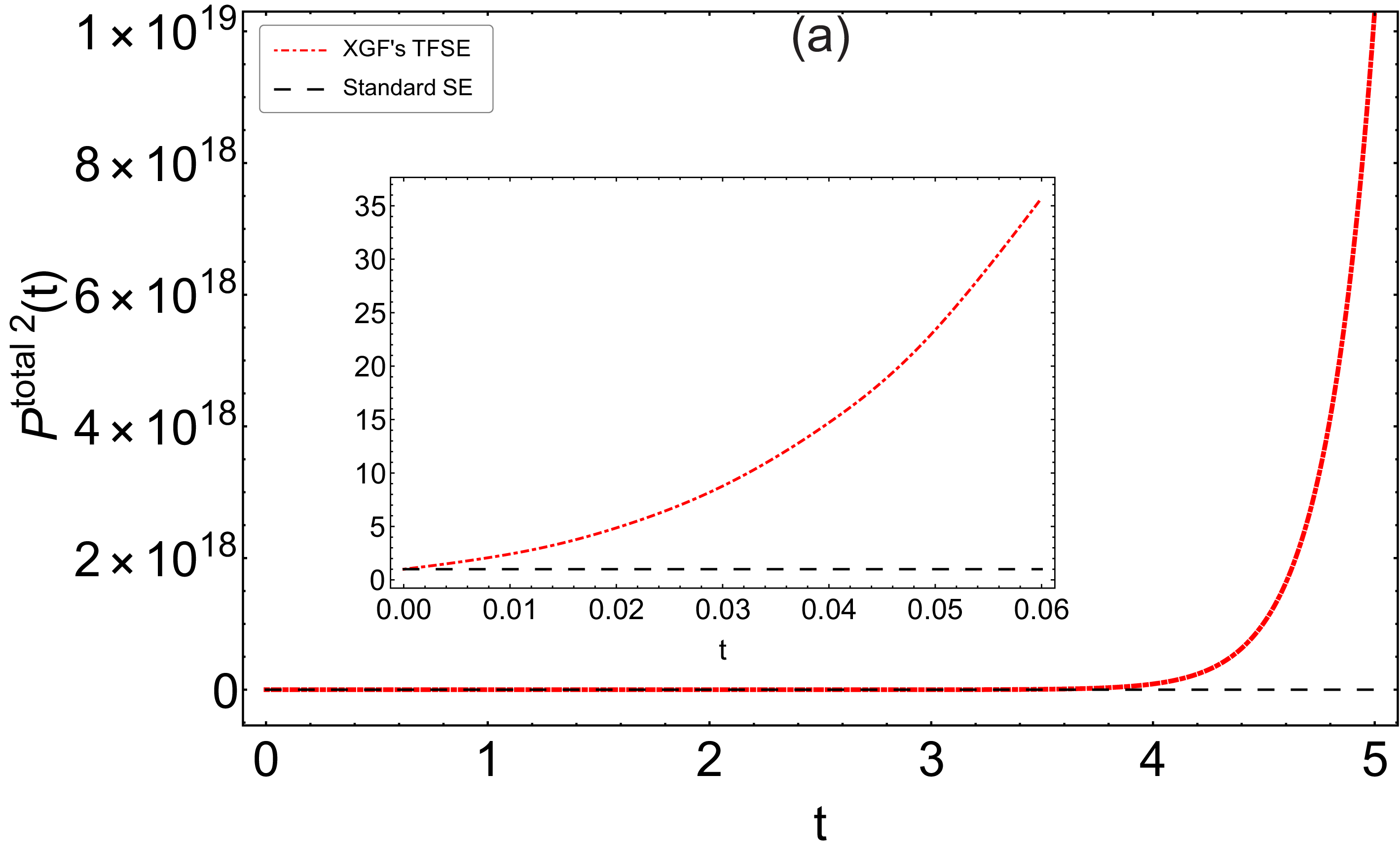}
    \includegraphics[width=0.32\linewidth]{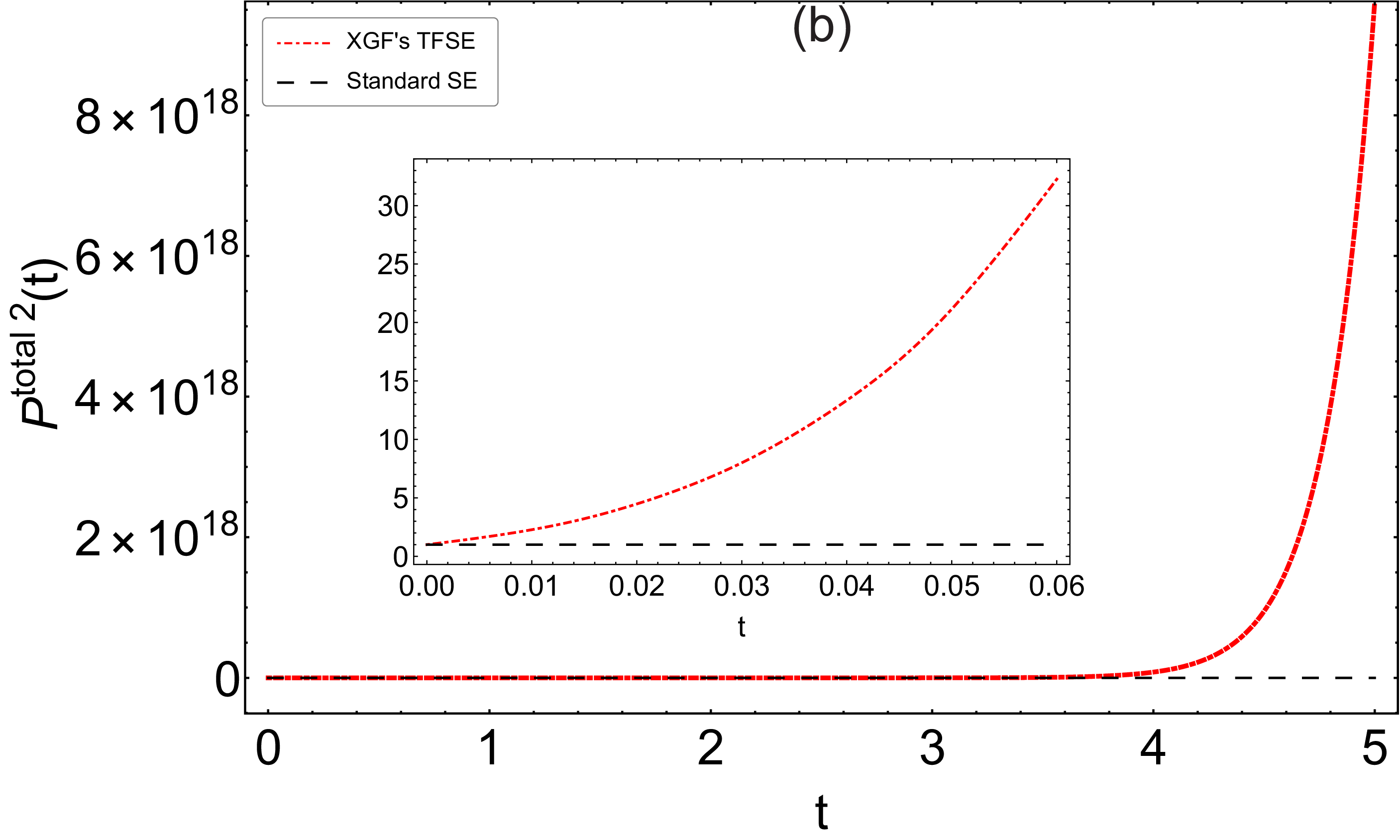}
    \includegraphics[width=0.32\linewidth]{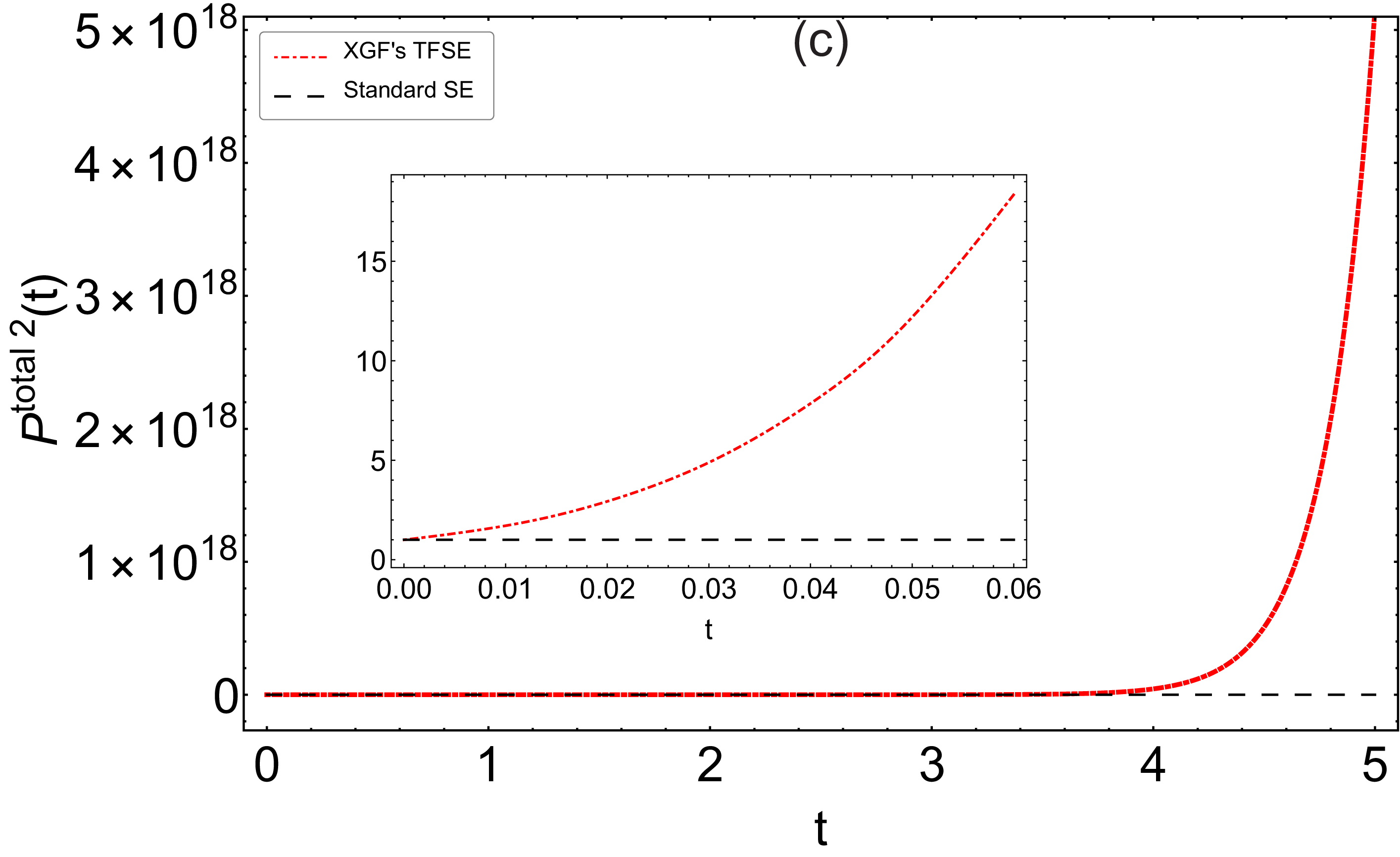}\\
\caption{The time evolution of $P_{\gamma=3}^{total\,2}(t)$ is plotted for $C_0=0, 0.5, 1$. Other input parameters are $\beta=0.5$, $\lambda=0.5$, and $n=50$. The insets show a very-short-time-scale evolution of $P_3^{total\,2}(t)$.}
\label{Fig8}
\end{figure}

We now focus on the time behaviour of $P_{\gamma=3}^{total\,1}(t)$ and $P_{\gamma=3}^{total\,2}(t)$ described by XGF's TFSE. In Fig. \ref{Fig5} we note that for $\beta\in(0,1)$, $P_3^{total\,1}(t)$ and $P_3^{total\,2}(t)$ as the functions of $t$ increase monotonically from one, where the larger $\beta$ is, the smaller $P_3^{total\,1}(t)$ and $P_3^{total\,2}(t)$ are at the same $t$. Only when $\beta=1$, $P_3^{total\,1}(t)$ and $P_3^{total\,2}(t)$ are conserved for any $t$. Fig. \ref{Fig6} displays for $\lambda=0$, $P_3^{total\,1}(t)$ and $P_3^{total\,2}(t)$ are always conserved independently of $\beta$, $n$, and $C_0$. But for $\lambda\in(0,1]$, $P_3^{total\,1}(t)$ and $P_3^{total\,2}(t)$ are conspicuous monotonic increase over time from a minimal value one and increase as $\lambda$ increases for the same $t$. It is shown in Fig. \ref{Fig7} with $n\in{N}$ that $P_3^{total\,1}(t)$ and $P_3^{total\,2}(t)$ as the functions of $t$ are monotonically increasing from their minimal value one, and grow with increasing $n$. In Fig. \ref{Fig8}, $P_3^{total\,1}(t)$ and $P_3^{total\,2}(t)$ serve as the monotone increasing functions of $t$ from one in the case of $C_0\in[0,1]$. Moreover, the smaller $C_0$ induces the larger $P_3^{total\,1}(t)$ and $P_3^{total\,2}(t)$ for the same $t$.

\subsection{Exact dynamics of the probabilities for the TFSQOS in the excited state $\left|{1}\right\rangle$ and the TFTQOS in the excited states $\left|{11}\right\rangle$}
\label{subsec:42}
Let us concentrate on $P^{\left|{1}\right\rangle}_{\gamma=1,2,3,4}(t)$ and $P^{\left|{11}\right\rangle}_{\gamma=1,2,3,4}(t)$, which are the probabilities of finding the TFSQOS in the excited state $\left|{1}\right\rangle$ and the TFTQOS in the excited state $\left|{11}\right\rangle$, given by
\begin{equation}
\label{e36}
\begin{aligned}
P^{\left|{1}\right\rangle}_{\gamma=1,2,3,4}(t)=\frac{\rho_{S{\gamma}}^{1(22)}}{\rho_{S{\gamma}}^{1(11)} + \rho_{S{\gamma}}^{1(22)}},\,\,\,\,\,\,\,\,\,\,
P^{\left|{11}\right\rangle}_{\gamma=1,2,3,4}(t)=\frac{\rho_{S{\gamma}}^{2(44)}}{\rho_{S{\gamma}}^{2(11)} + \rho_{S{\gamma}}^{2(22)} + \rho_{S{\gamma}}^{2(33)} + \rho _{S{\gamma}}^{2(44)}}.
\end{aligned}
\end{equation}

\begin{figure}[htbp]
\centering
    \includegraphics[width=0.32\linewidth]{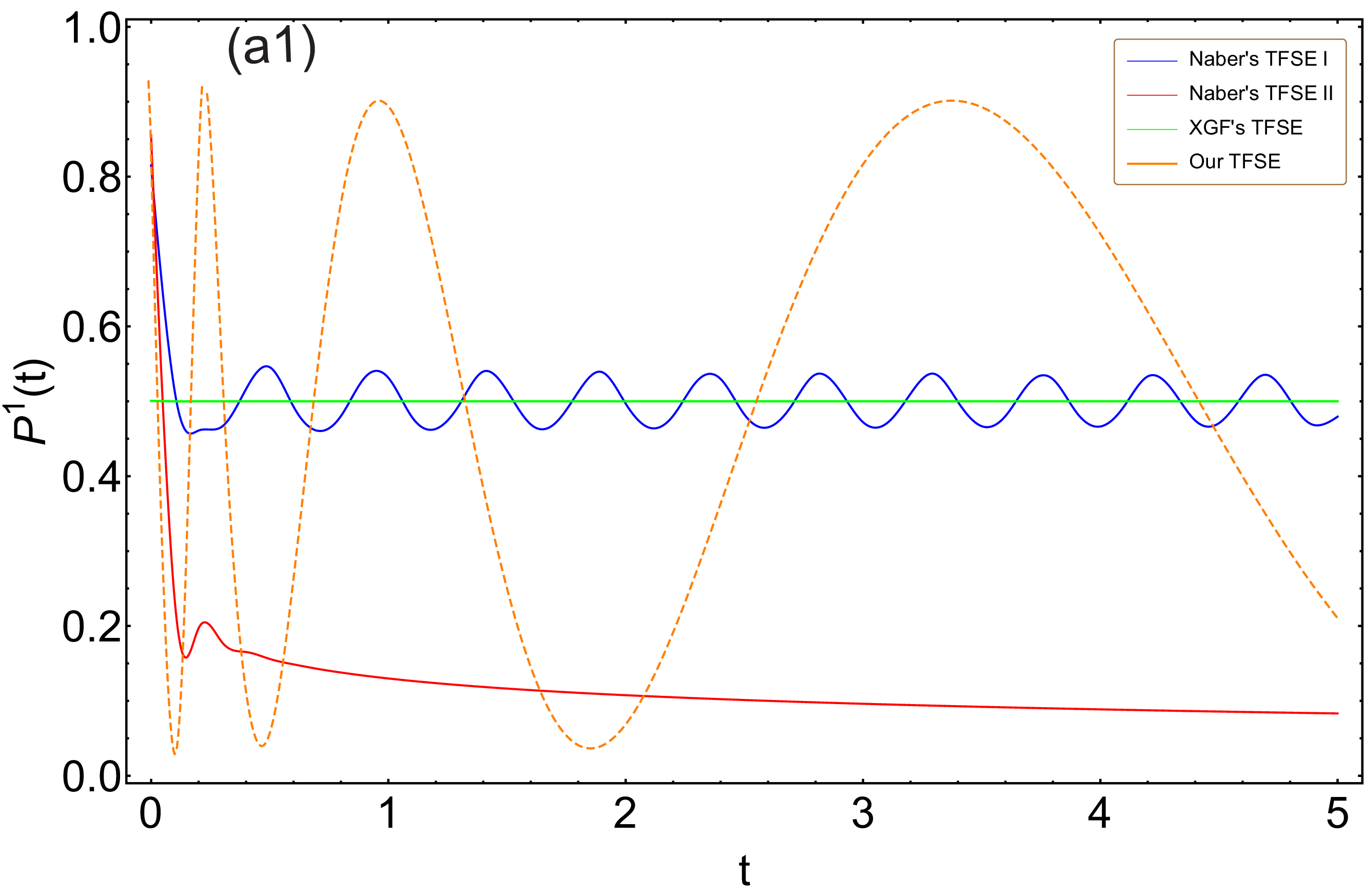}
    \includegraphics[width=0.32\linewidth]{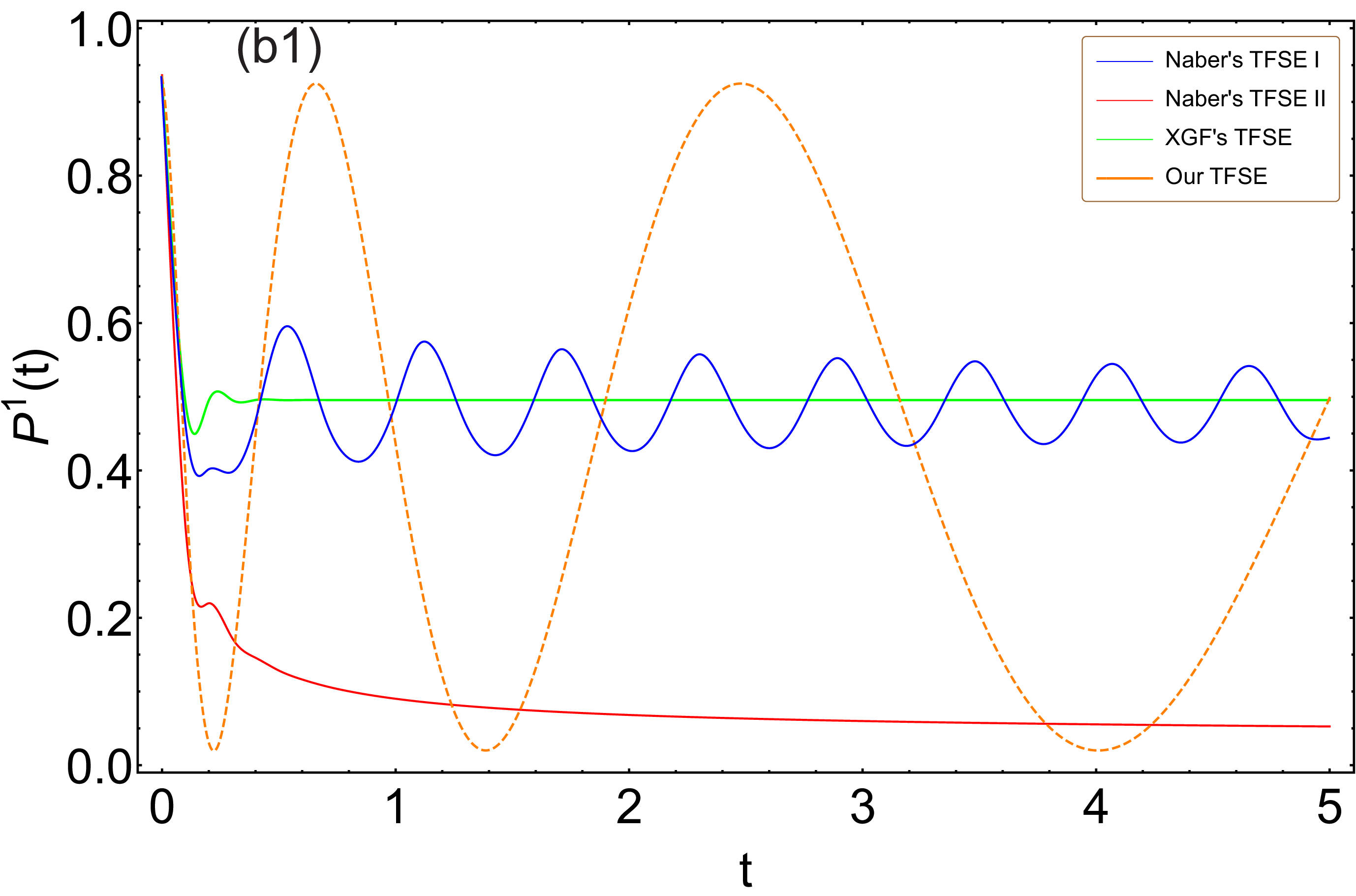}
    \includegraphics[width=0.32\linewidth]{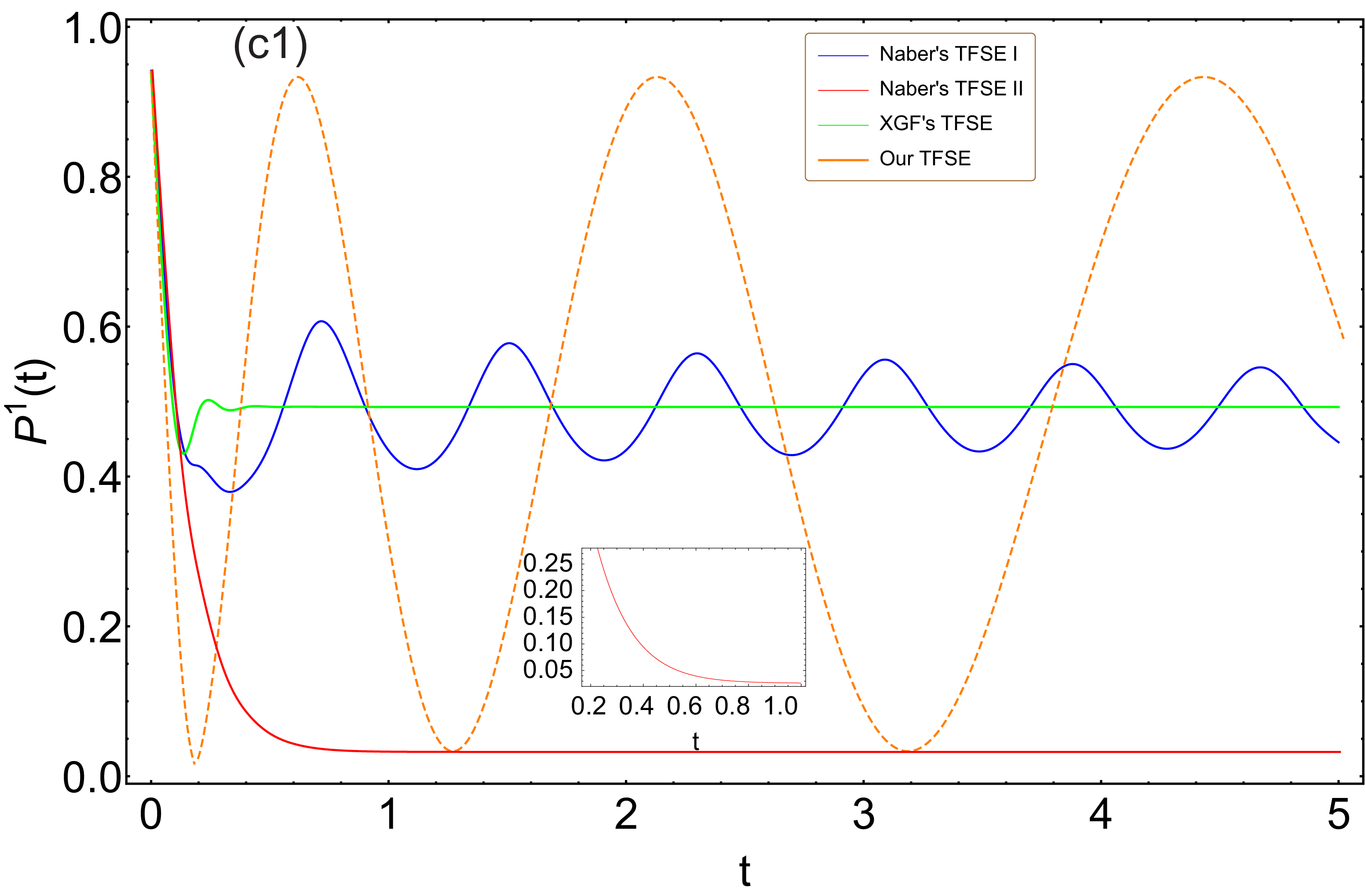}\\
    \includegraphics[width=0.32\linewidth]{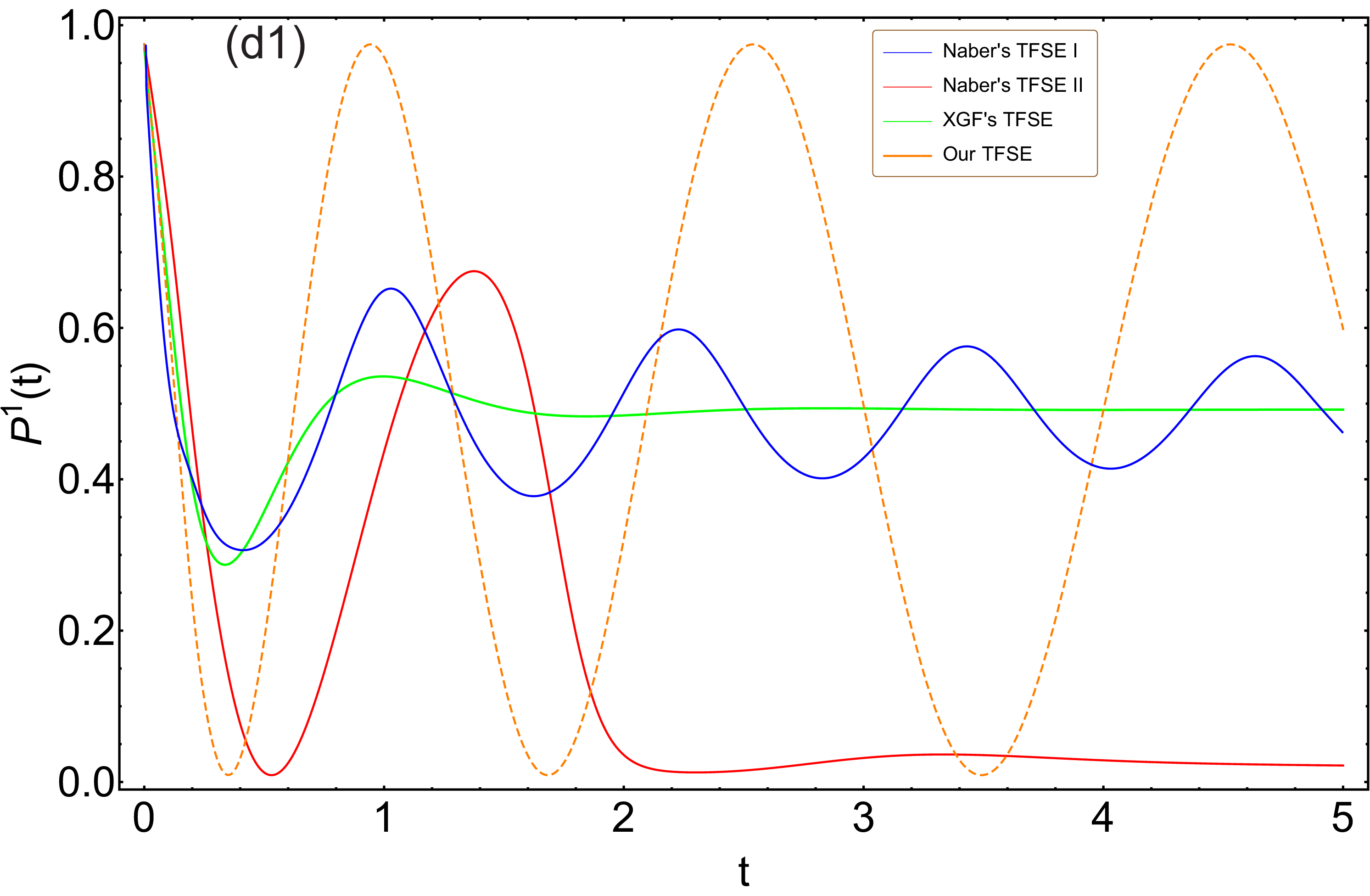}
    \includegraphics[width=0.32\linewidth]{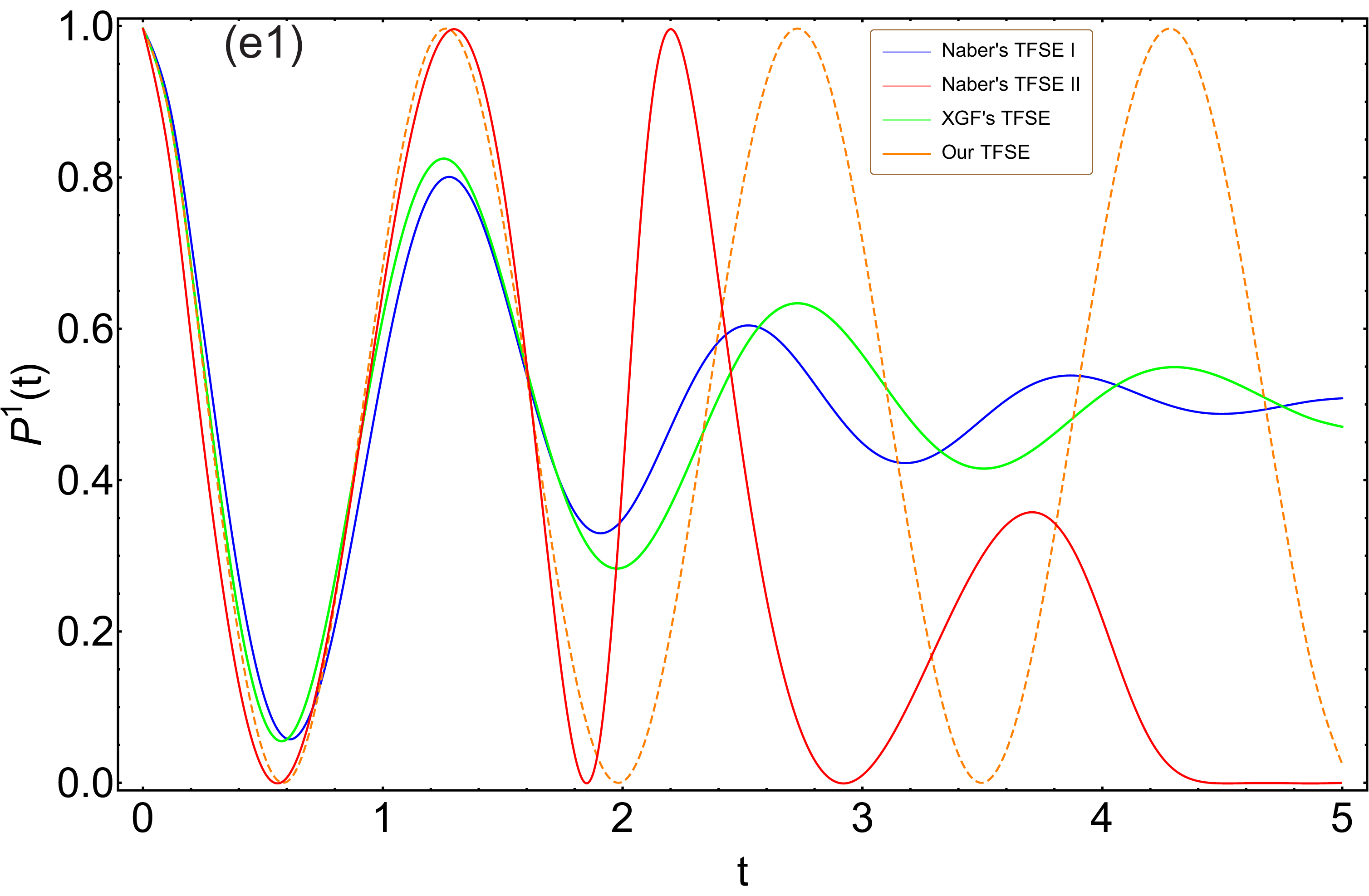}
    \includegraphics[width=0.32\linewidth]{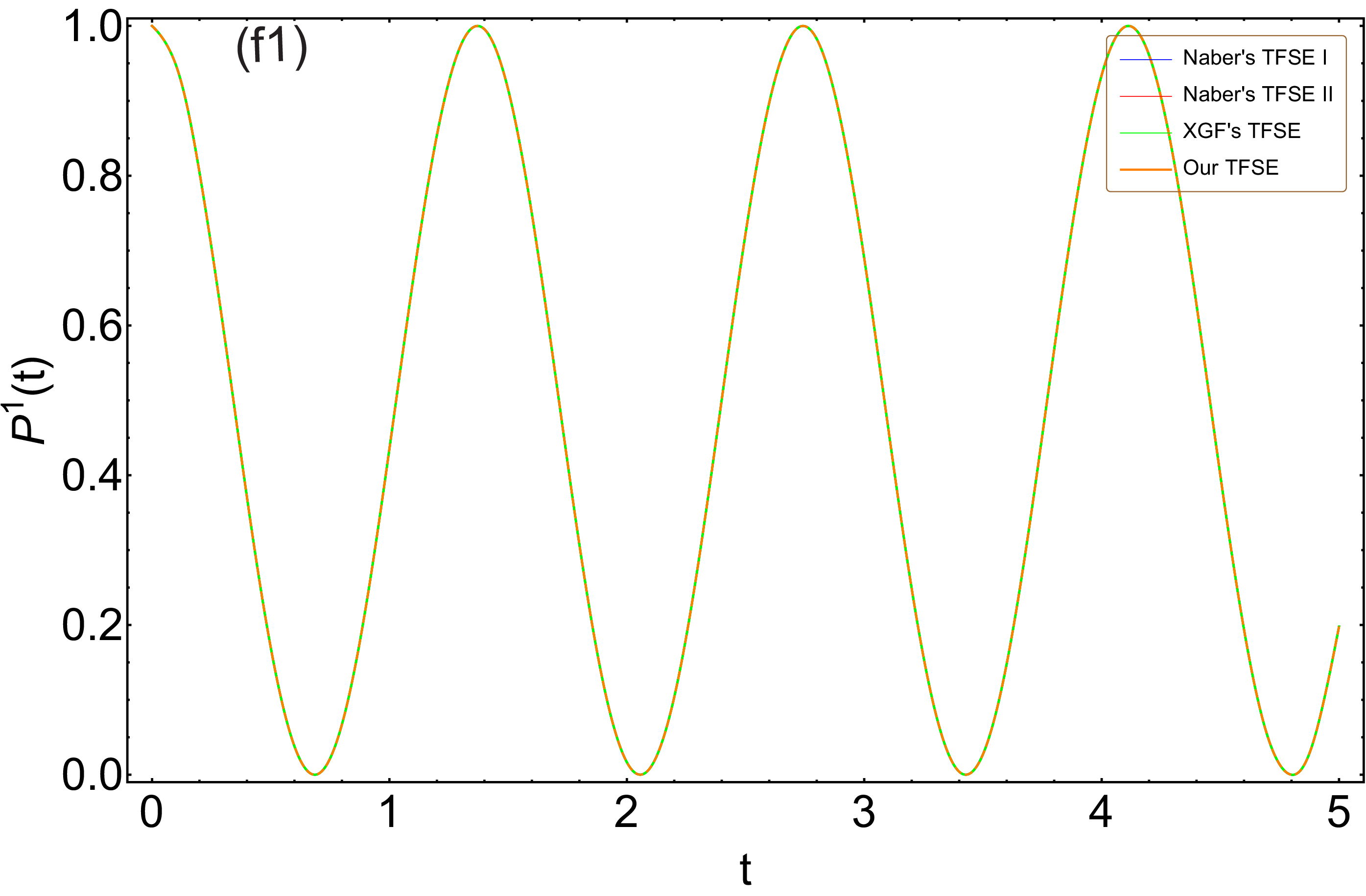}\\
    \includegraphics[width=0.32\linewidth]{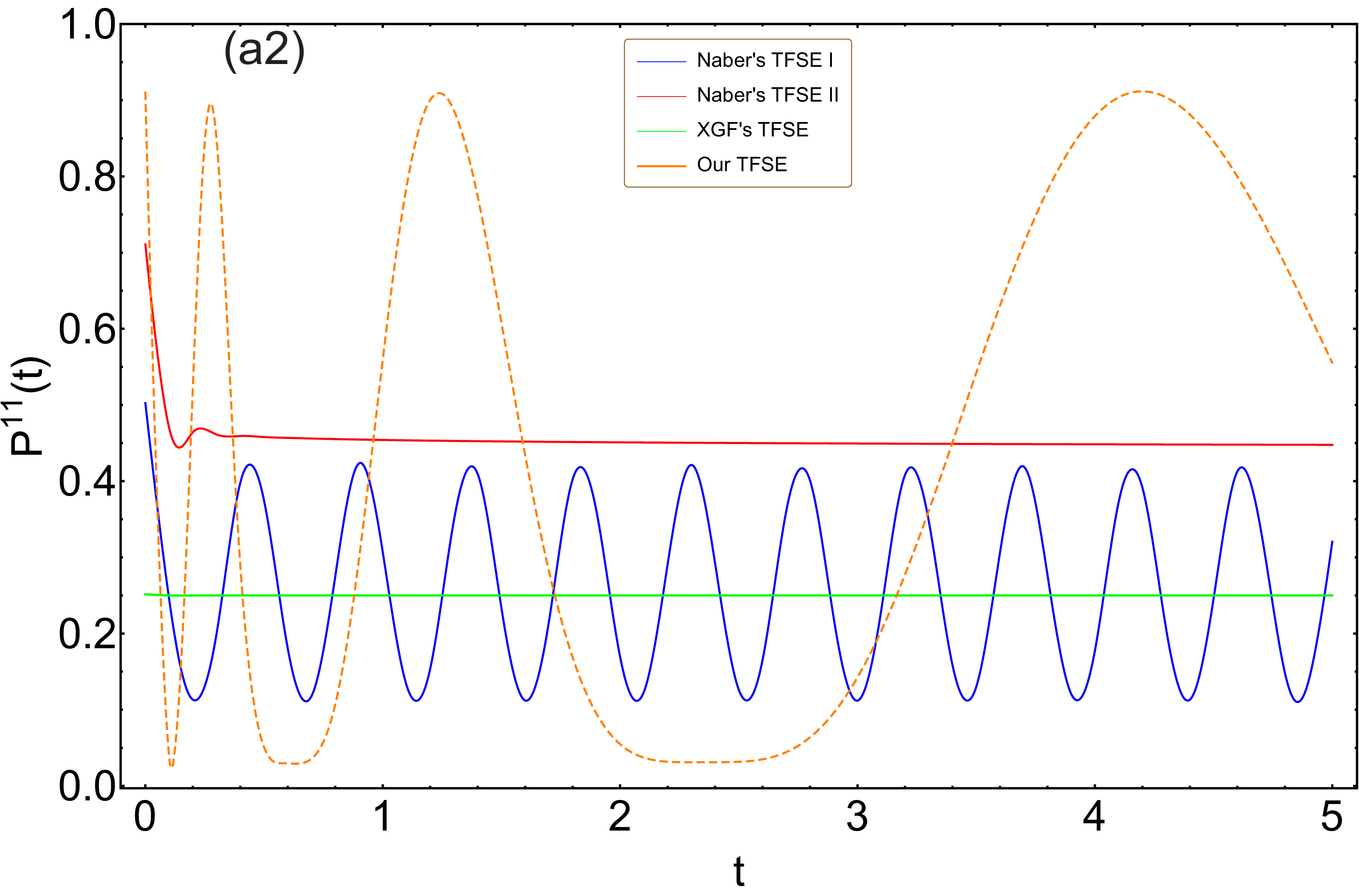}
    \includegraphics[width=0.32\linewidth]{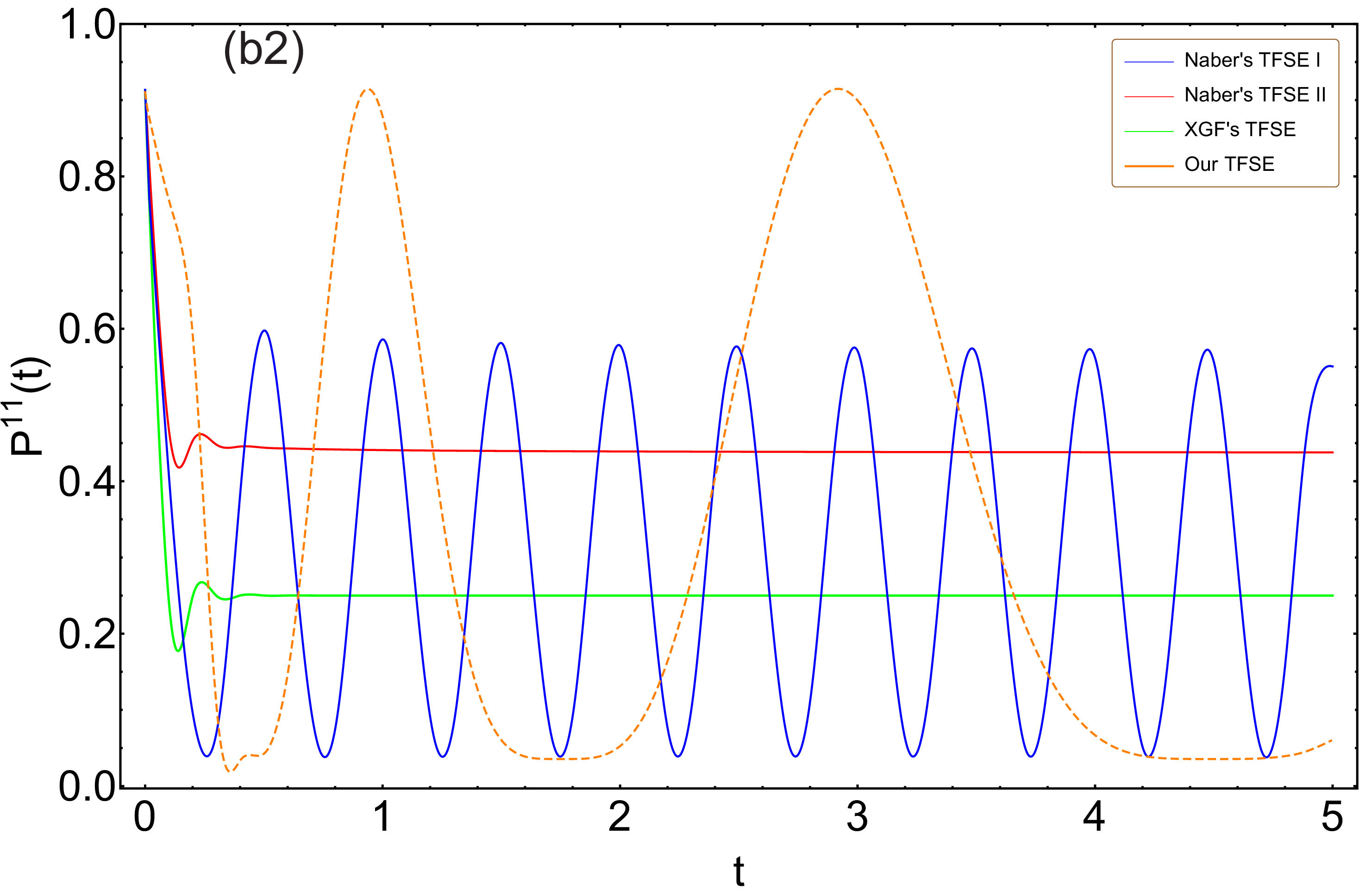}
    \includegraphics[width=0.32\linewidth]{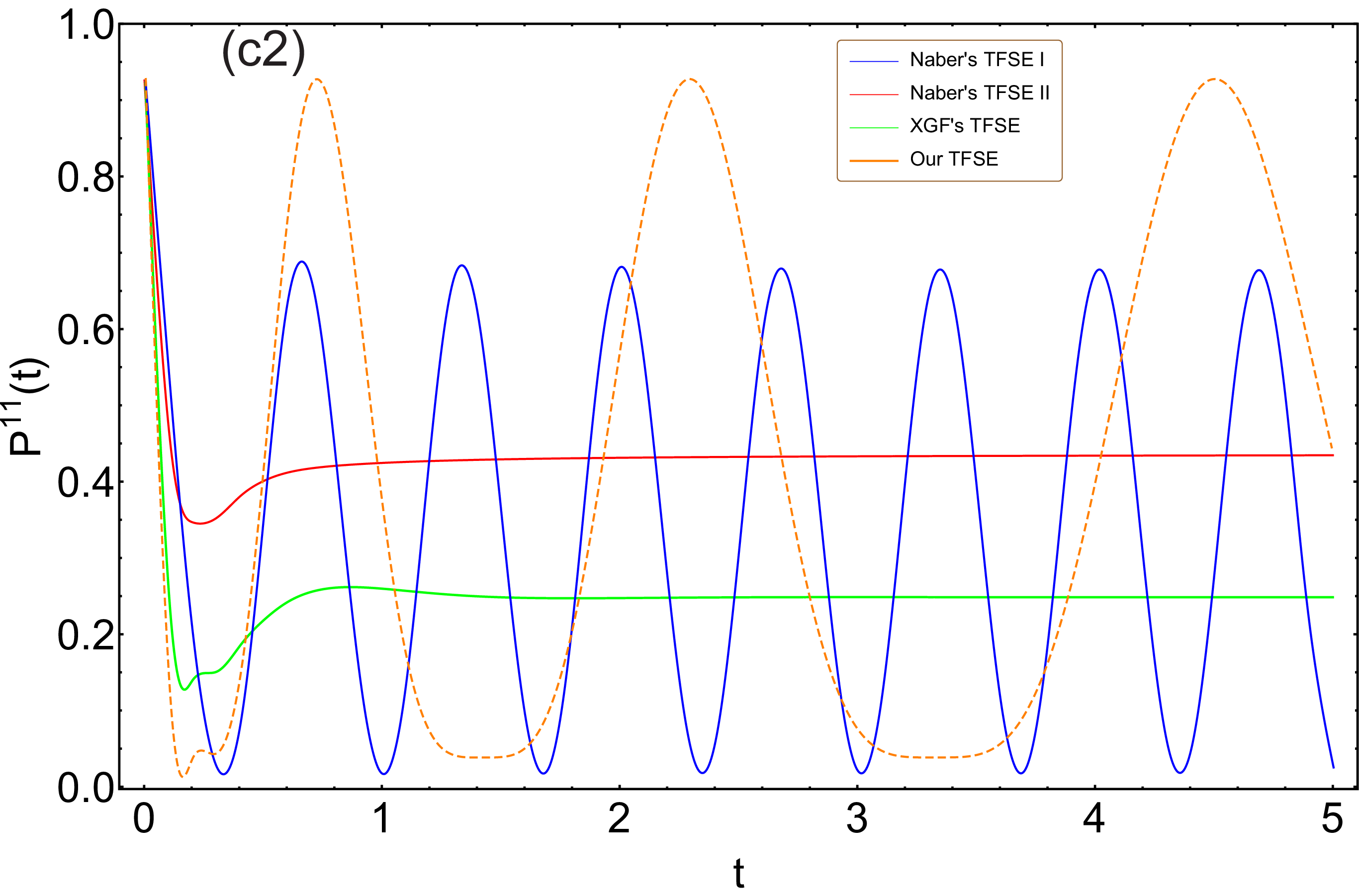}\\
    \includegraphics[width=0.32\linewidth]{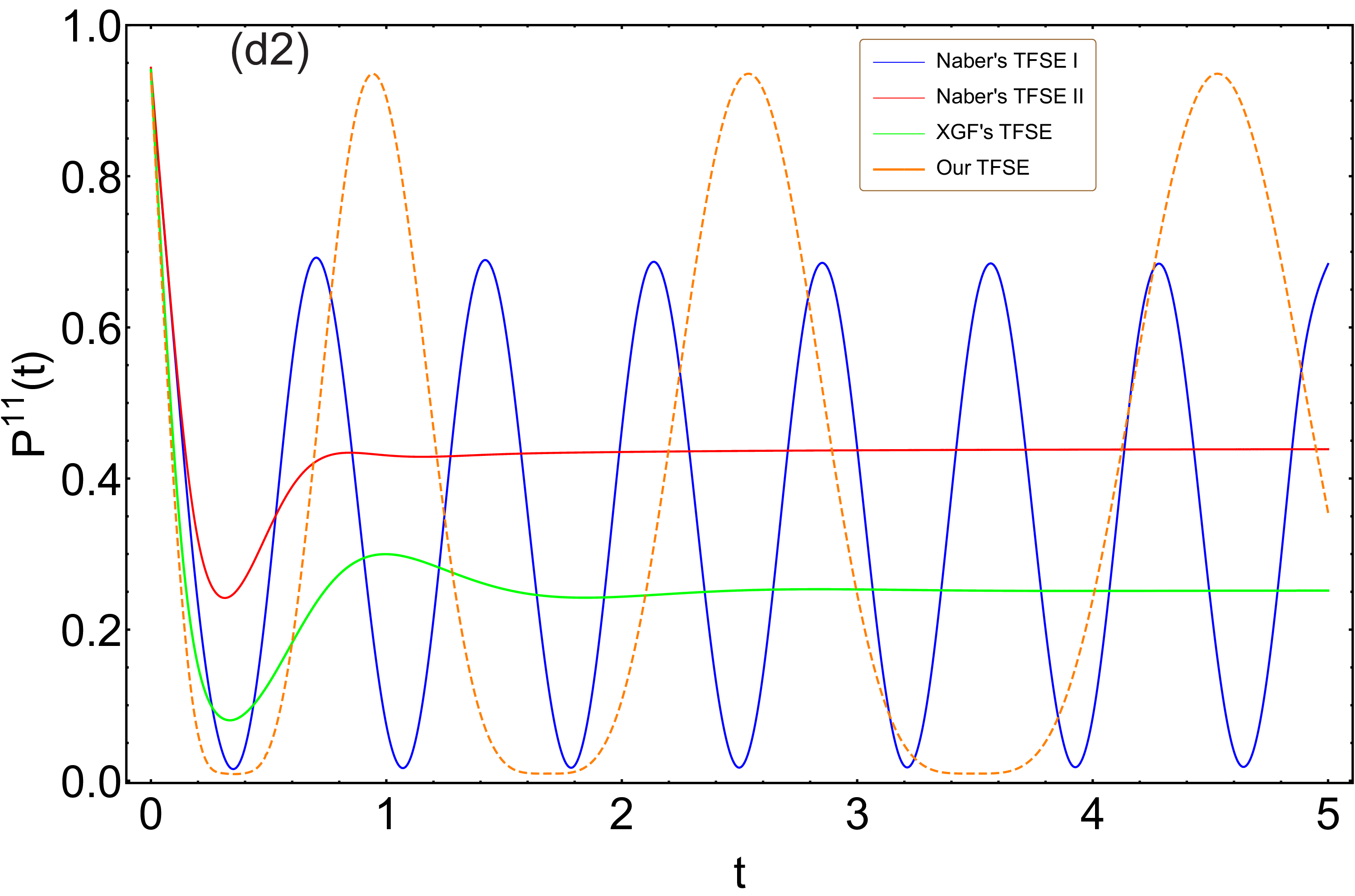}
    \includegraphics[width=0.32\linewidth]{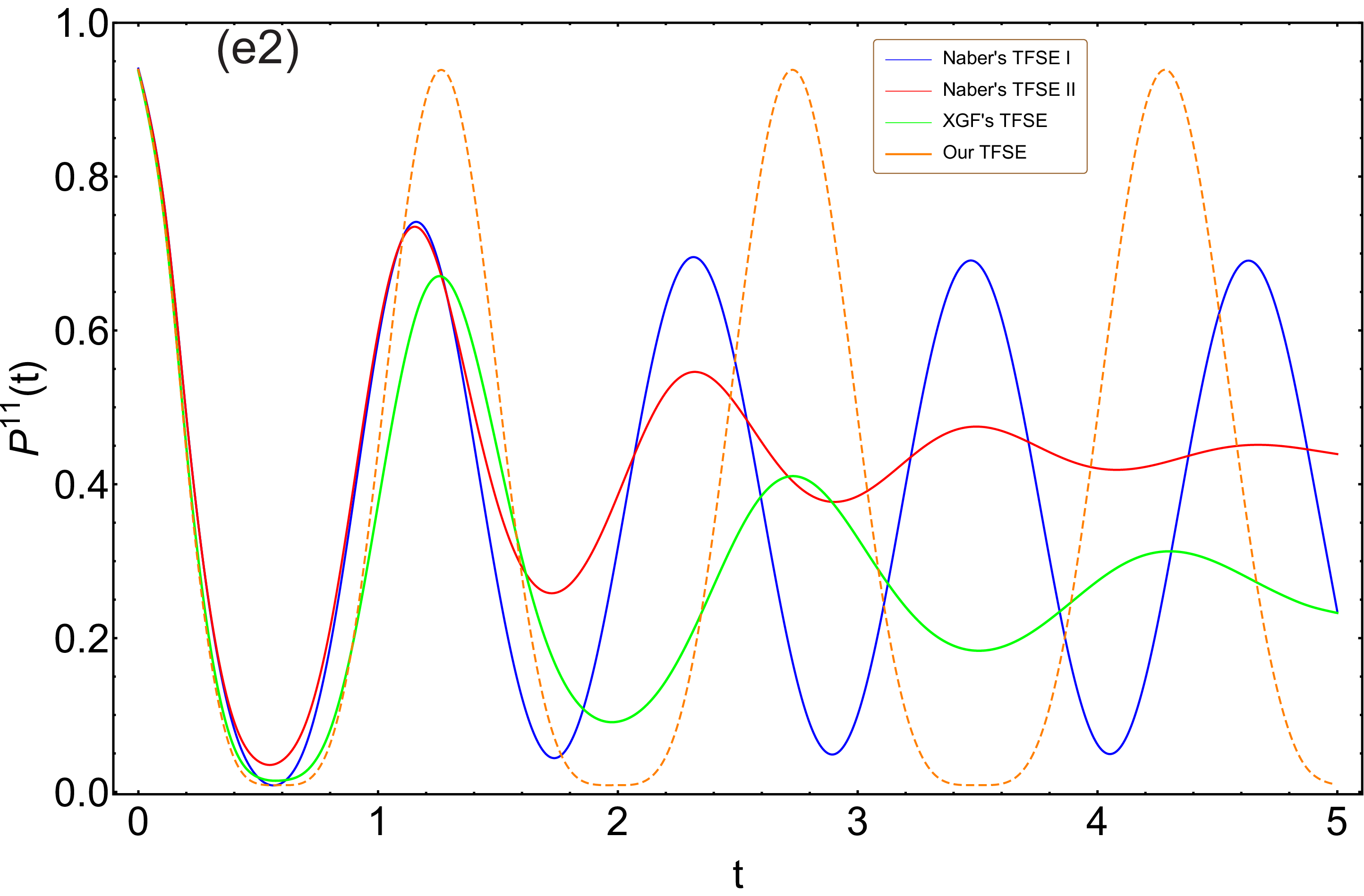}
    \includegraphics[width=0.32\linewidth]{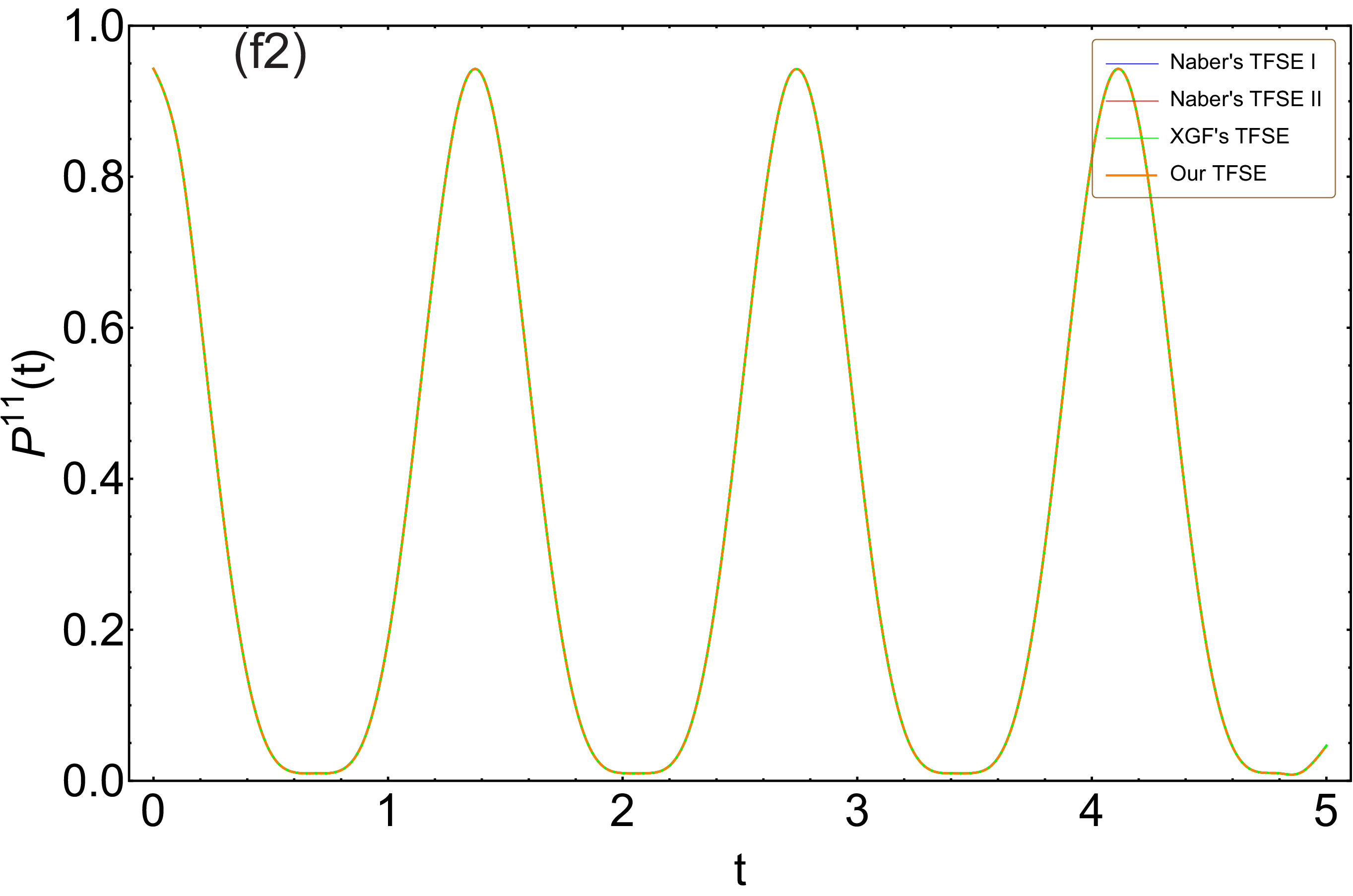}\\
\caption{The time evolution of $P^{\left|{1}\right\rangle}_{\gamma=1,2,3,4}(t)$ and $P^{\left|{11}\right\rangle}_{\gamma=1,2,3,4}(t)$ is plotted for $\beta=0.1, 0.3, 0.5, 0.7, 0.9, 1$. The other parameters are set as $\lambda=0.5$, $n=20$, and $C_0=0.5$. The inset shows how $P^{\left|{1}\right\rangle}_2(t)$ changes in a very short time scale.}
\label{Fig9}
\end{figure}

Fig. \ref{Fig9} shows the time behavior of $P^{\left|{1}\right\rangle}_{\gamma=1,2,3,4}(t)$ and $P^{\left|{11}\right\rangle}_{\gamma=1,2,3,4}(t)$ for $\beta=0.1, 0.3, 0.5, 0.7, 0.9, 1$ by fixing $\lambda=0.5$, $n=20$, and $C_0=0.5$. We note for the smaller $\beta$, the smaller $P^{\left|{1}\right\rangle}_1(t)$ and $P^{\left|{11}\right\rangle}_1(t)$ oscillate more rapidly, which suggests that Naber's TFSE I can describe the non-Markovian dynamics for all $\beta$. Interestingly, when small $\beta$ is larger than a certain critical value, the oscillations in the time evolution of $P^{\left|{1}\right\rangle}_2(t)$ and $P^{\left|{11}\right\rangle}_2(t)$ weaken and even disappear. While large $\beta$ is greater than this certain critical value, their oscillations appear and enhance. This indicates that Naber's TFSE II is incapable of describing the non-Markovian dynamics at intermediate $\beta$. Clearly, the oscillatory behavior of $P^{\left|{1}\right\rangle}_3(t)$ and $P^{\left|{11}\right\rangle}_3(t)$ will be observed only when small $\beta$ is larger than a certain critical value. In this case, XGF's TFSE fails to describe the non-Markovian dynamics at small $\beta$. It is noteworthy that the oscillations in the time evolution of $P^{\left|{1}\right\rangle}_4(t)$ and $P^{\left|{11}\right\rangle}_4(t)$ occur for all $\beta$, implying our TFSE are capable of describing the non-Markovian dynamics at all $\beta$. Therefore, we argue that Naber's TFSE I and our TFSE can better describe the non-Markovian dynamics than Naber's TFSE I and XGF's TFSE since the former applies to all $\beta$. Additionally, when $\beta$ becomes larger until it reaches one, the time behavior of all probabilities gets closer until they are in complete agreement.

\begin{figure}[htbp]
\centering
    \subfigure{\label{Fig10(a1)}
    \includegraphics[width=0.32\linewidth]{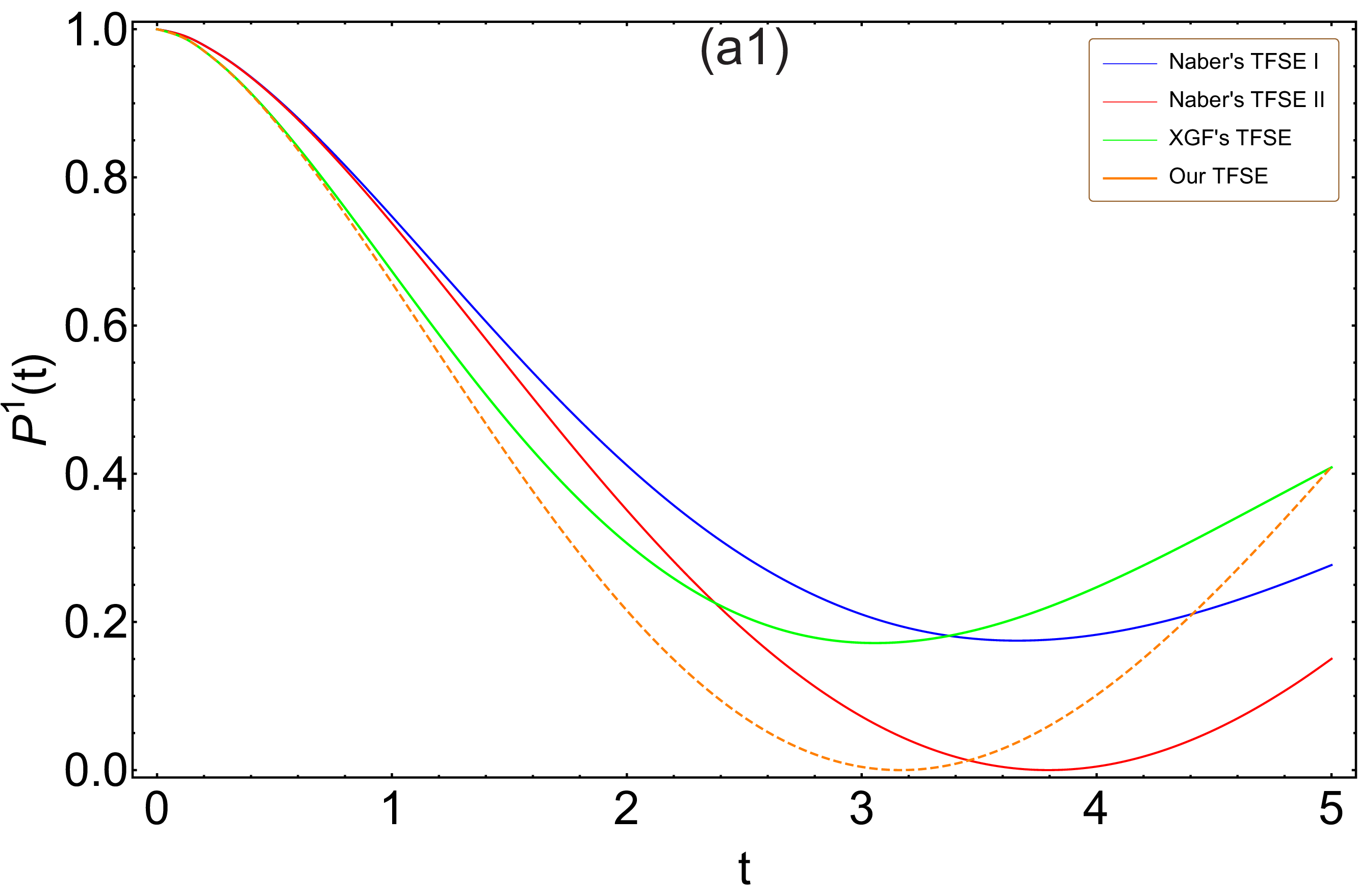}}
    \subfigure{\label{Fig10(b1)}
    \includegraphics[width=0.32\linewidth]{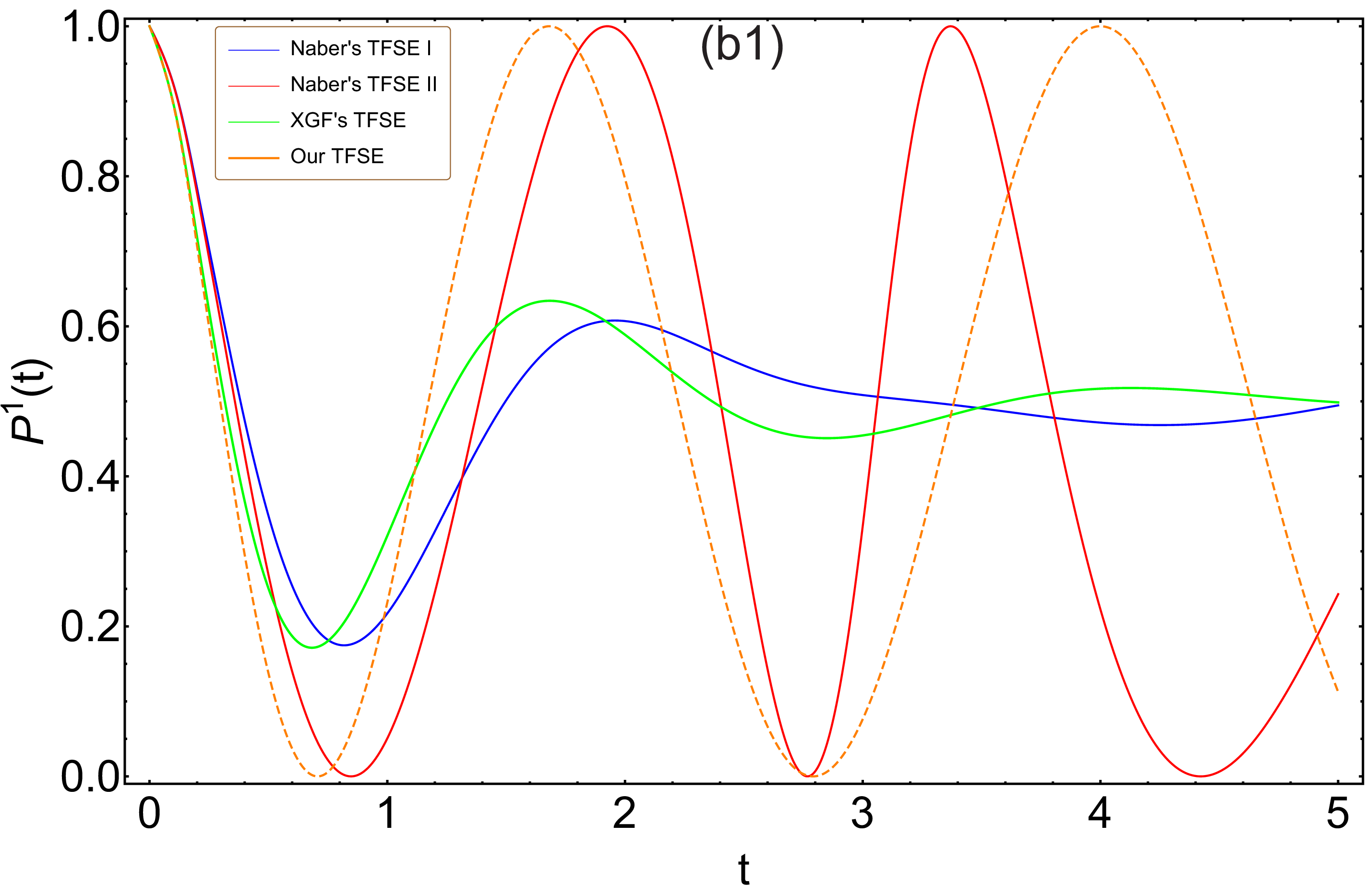}}
    \subfigure{\label{Fig10(c1)}
    \includegraphics[width=0.32\linewidth]{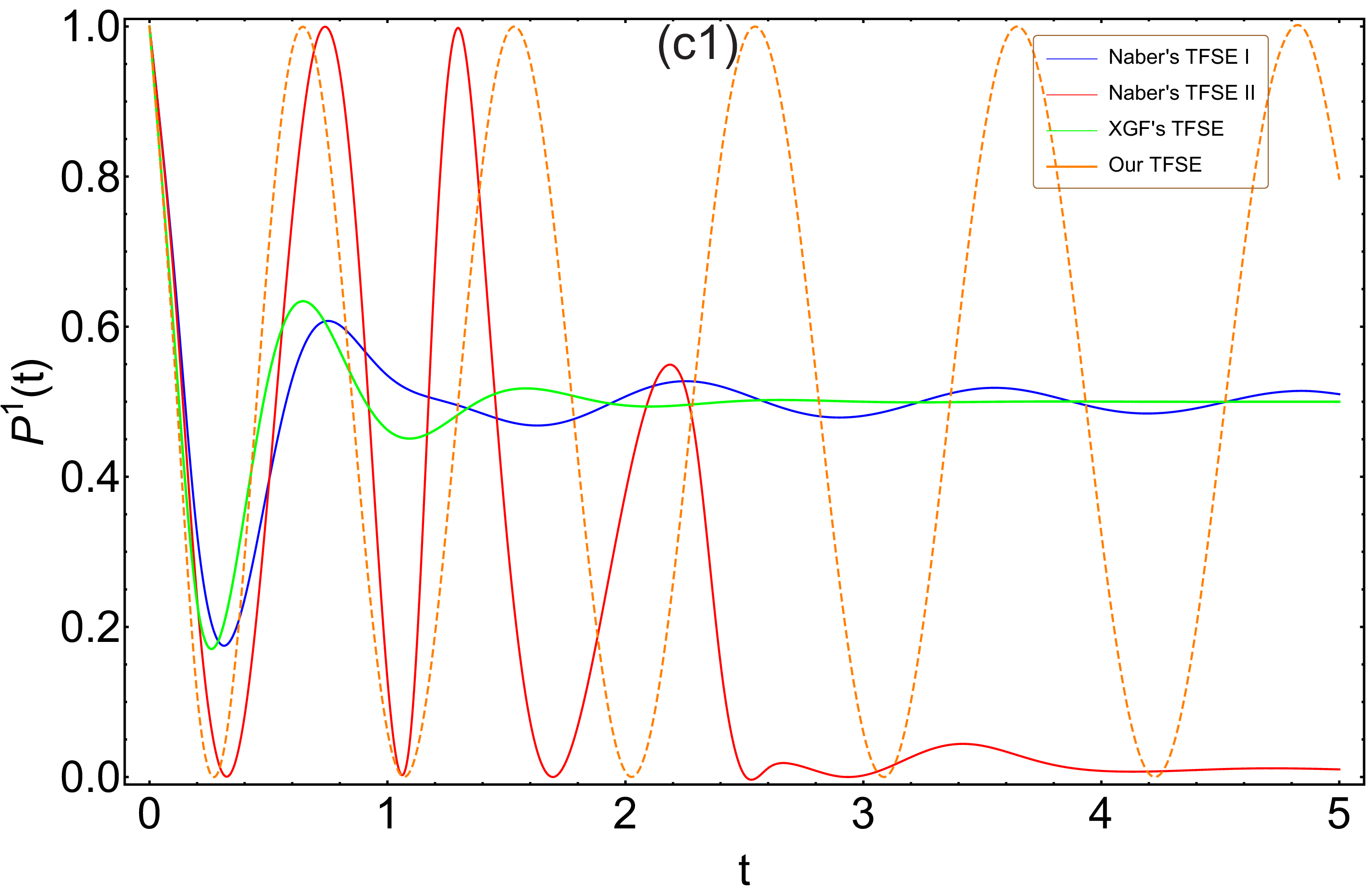}}\\
    \subfigure{\label{Fig10(a2)}
    \includegraphics[width=0.32\linewidth]{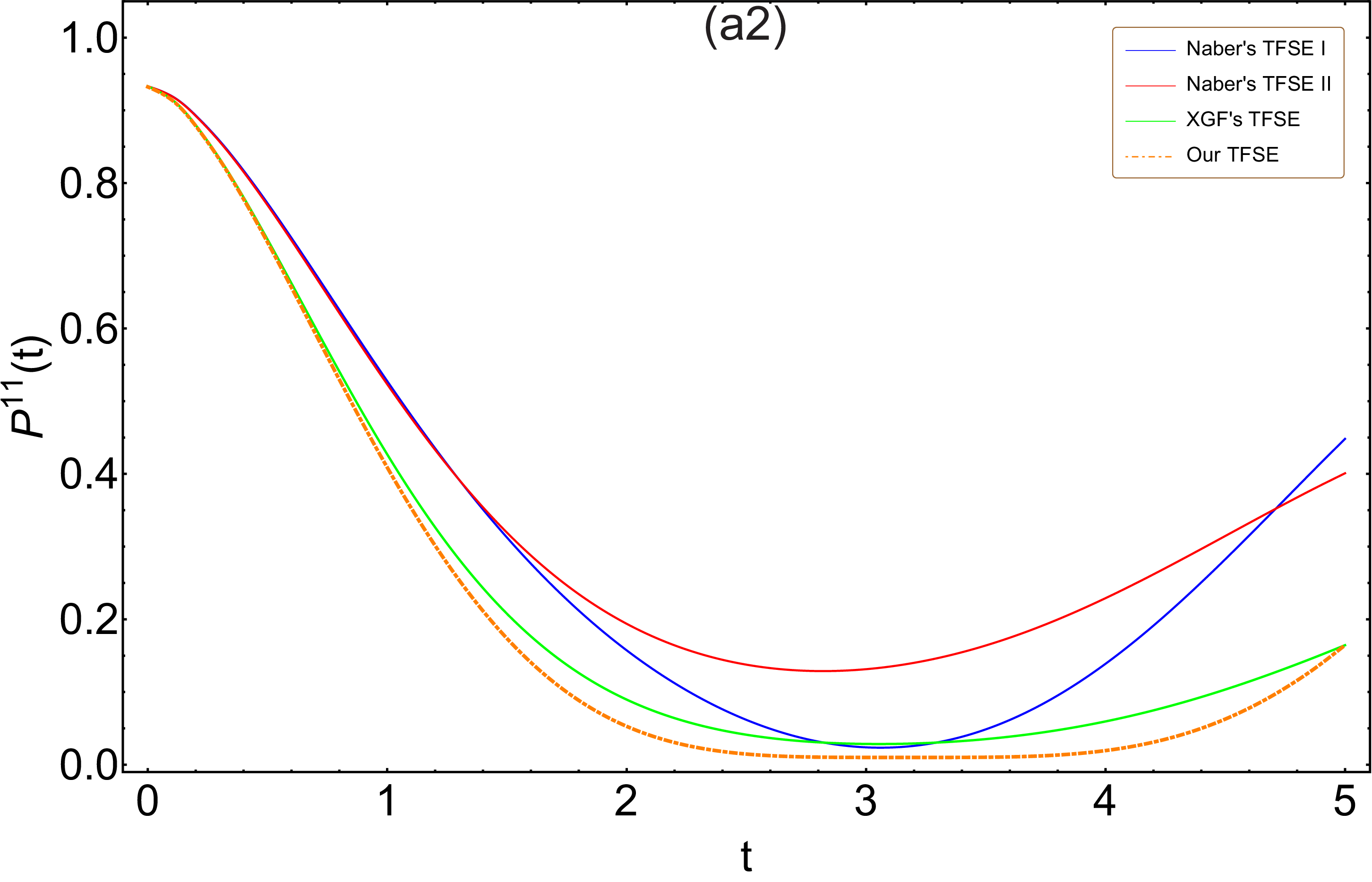}}
    \subfigure{\label{Fig10(b2)}
    \includegraphics[width=0.32\linewidth]{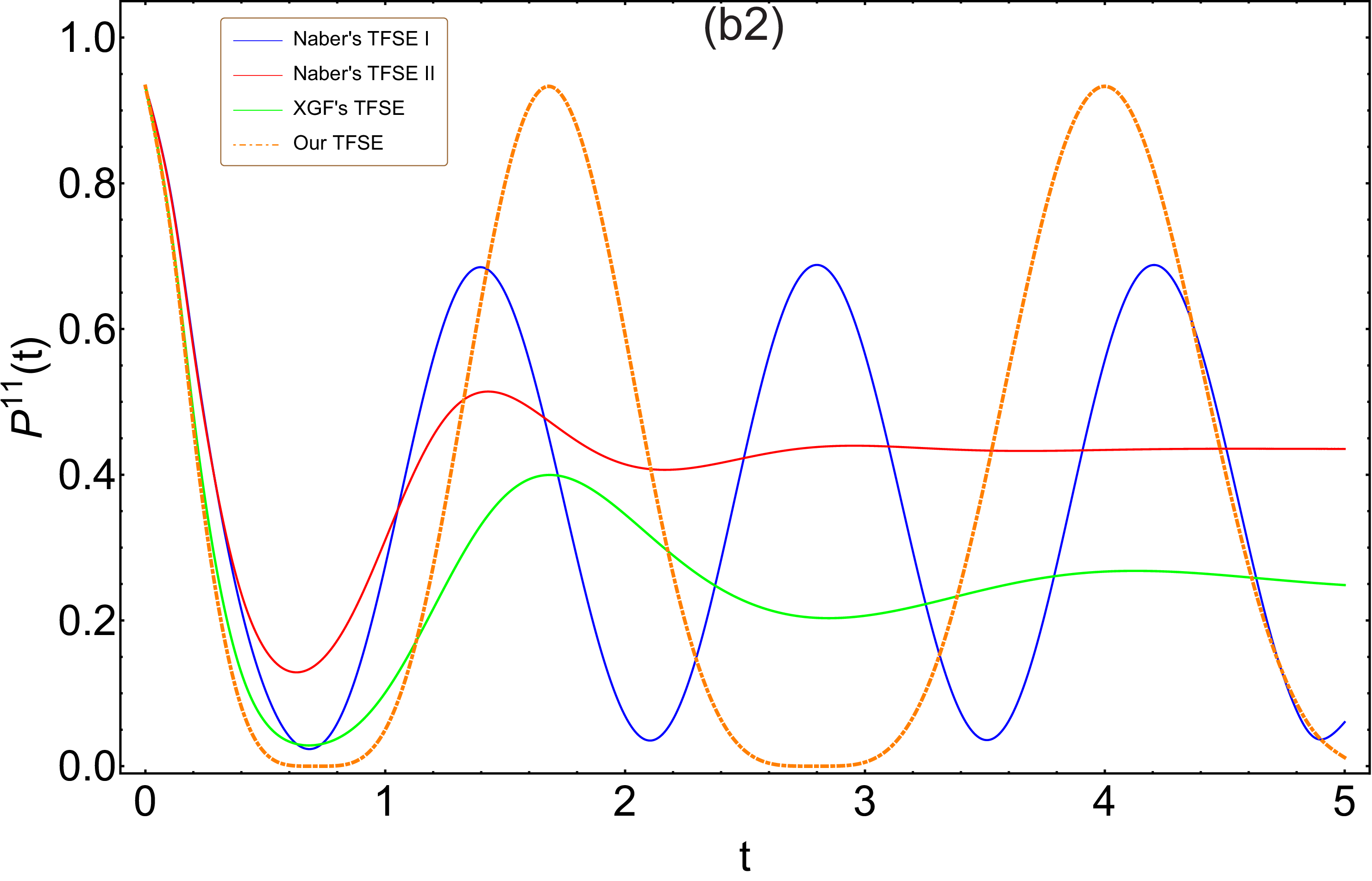}}
    \subfigure{\label{Fig10(c2)}
    \includegraphics[width=0.32\linewidth]{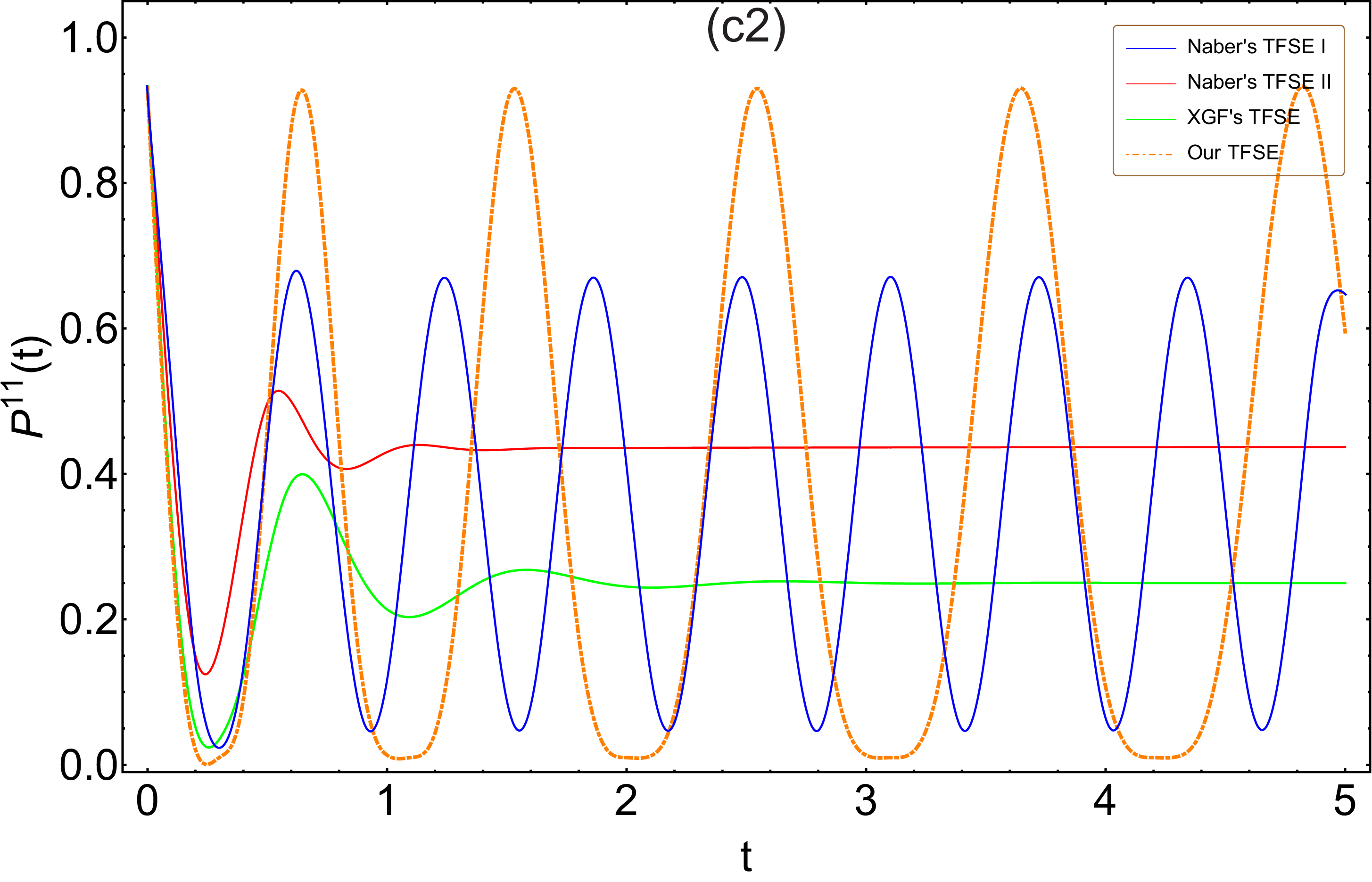}}\\
\caption{The time evolution of $P^{\left|{1}\right\rangle}_{\gamma=1,2,3,4}(t)$ and $P^{\left|{11}\right\rangle}_{\gamma=1,2,3,4}(t)$ is plotted for $n=0, 10, 50$. In each plot we have set $\beta=0.8$, $\lambda=0.5$, and $C_0=0.5$.}
\label{Fig10}
\end{figure}

In Fig. \ref{Fig10} we observe $t$ and $n$ dependence of $P^{\left|{1}\right\rangle}_{\gamma=1,2,3,4}(t)$ and $P^{\left|{11}\right\rangle}_{\gamma=1,2,3,4}(t)$ with $\beta=0.8$, $\lambda=0.5$, and $C_0=0.5$. Obviously, although for $n=0$ the non-Markovian oscillations of all probabilities have taken shape from intermediate to large $t$. Moreover, as very small $n$ increases, the oscillations are enhanced at all $t$. In fact, the increase in $n$ raises their oscillating frequencies but hardly changes their amplitudes. This means that the four TFSEs can give a great performance in describing the non-Markovian features of time-fractional open quantum dynamics for all $n$.

\begin{figure}[htbp]
\centering
    \includegraphics[width=0.32\linewidth]{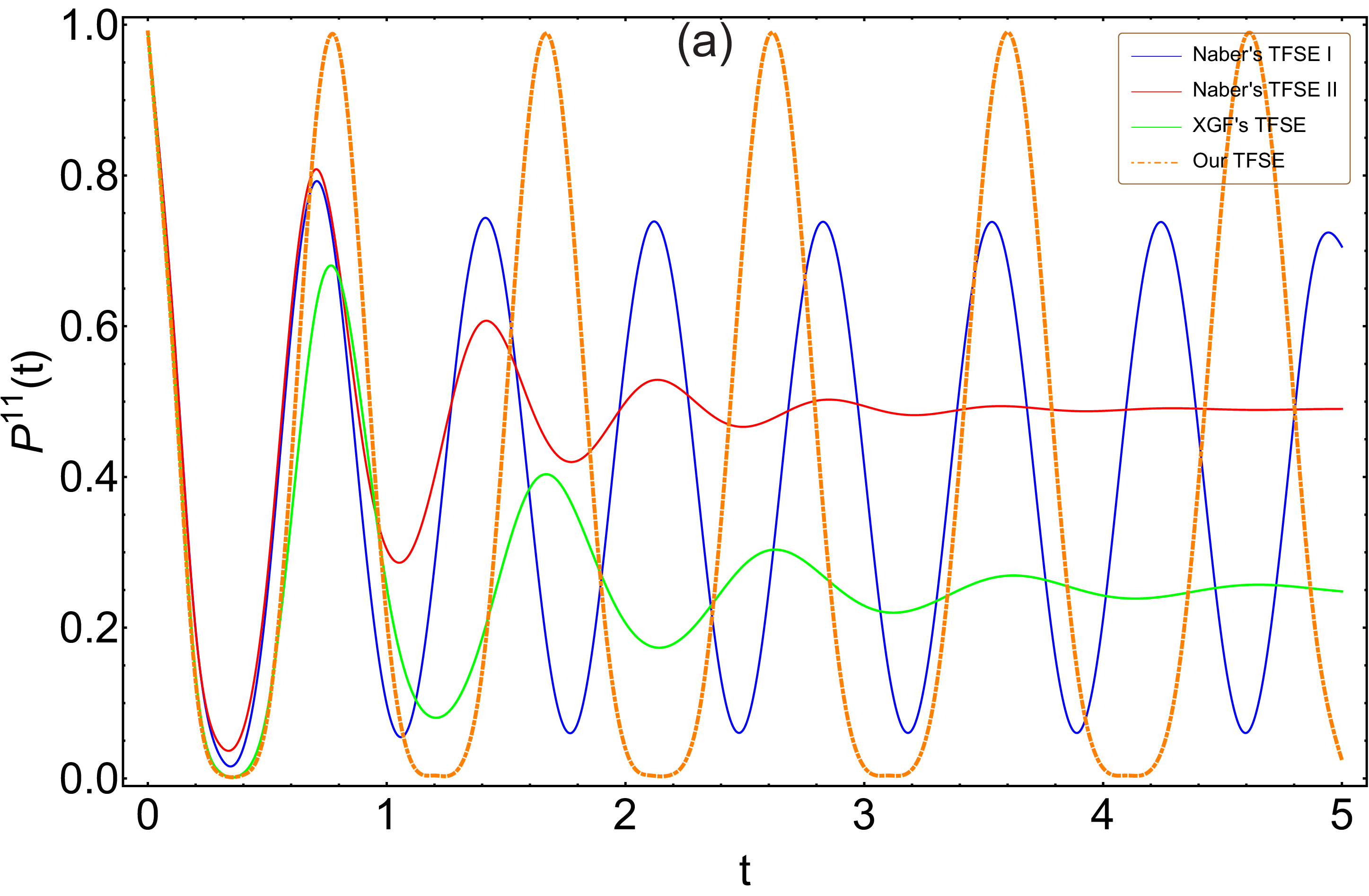}
    \includegraphics[width=0.32\linewidth]{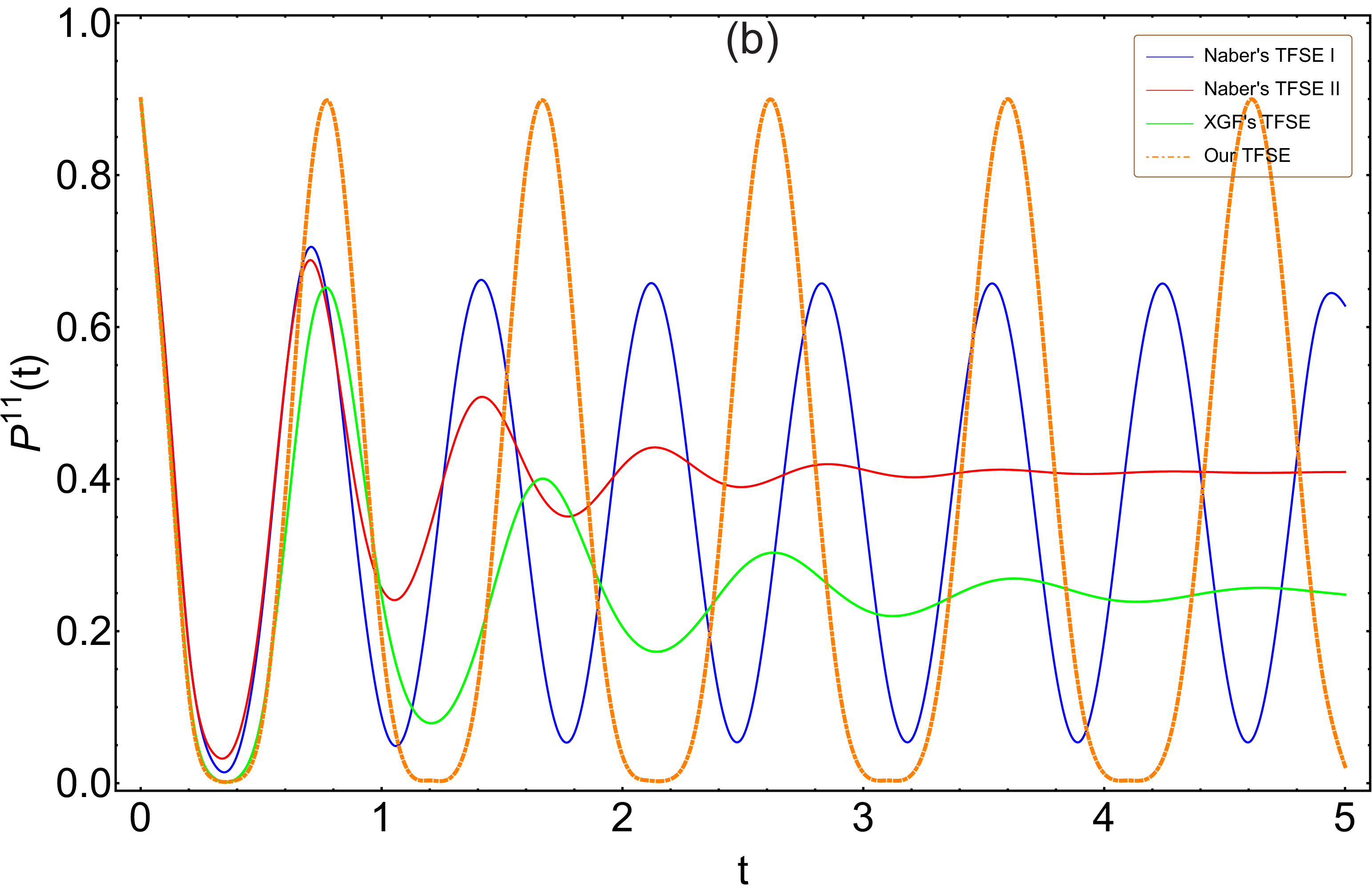}
    \includegraphics[width=0.32\linewidth]{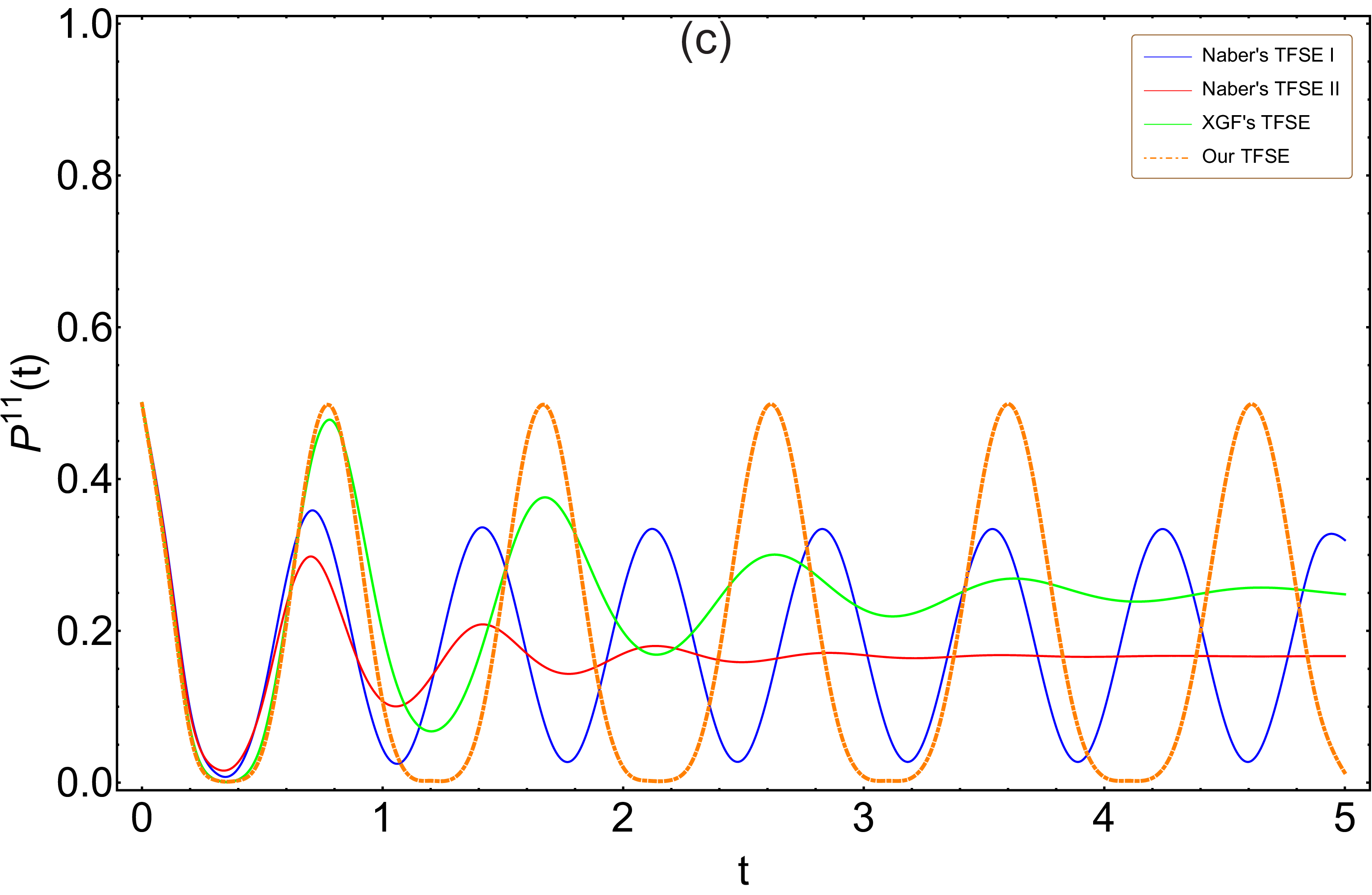}\\
\caption{The time evolution of $P^{\left|{11}\right\rangle}_{\gamma=1,2,3,4}(t)$ is plotted for $C_0=0.2, 0.6, 1$. Other input parameters are $\beta=0.9$, $\lambda=0.5$, $n=50$.}
\label{Fig11}
\end{figure}

Fig. \ref{Fig11} shows how the change in $C_0$ affects the time behavior of $P^{\left|{11}\right\rangle}_{\gamma=1,2,3,4}(t)$ for $\beta=0.9$, $\lambda=0.5$, and $n=50$. We see that the larger $C_0$ results in the smaller oscillating amplitudes for all probabilities but has no influence on their frequencies, which indicates that the four TFSEs are able to describe the non-Markovian features of time-fractional open quantum dynamics for all $C_0$. Besides, for a short-time evolution, all probabilities behave in similar oscillations, while for the long time behavior, $P^{\left|{11}\right\rangle}_2(t)$ and $P^{\left|{11}\right\rangle}_3(t)$ oscillate slowly and $P^{\left|{11}\right\rangle}_2(t)$ may even level off at the end.

\begin{figure}[htbp]
\centering
    \subfigure{\label{Fig12(a1)}
    \includegraphics[width=0.32\linewidth]{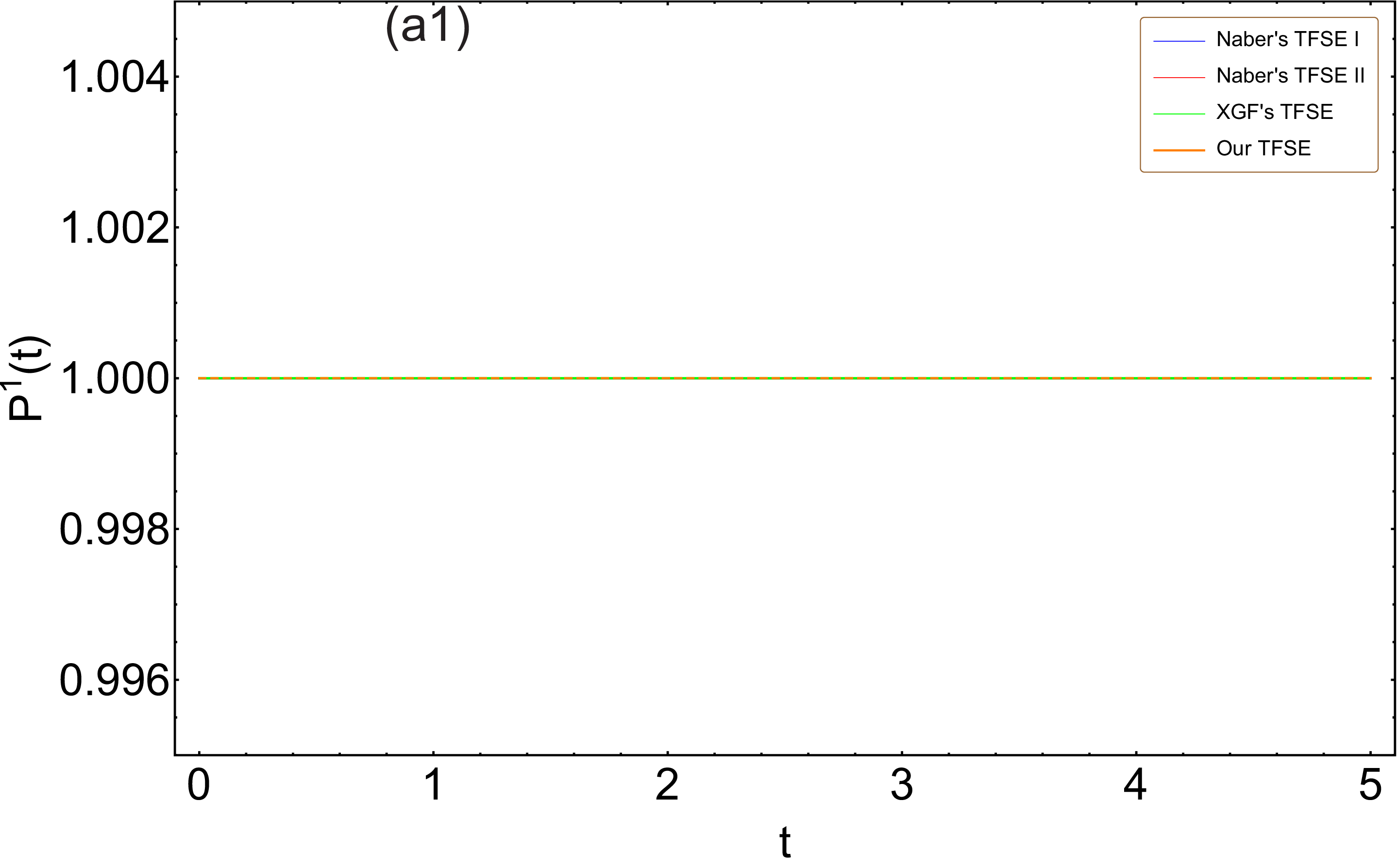}}
    \subfigure{\label{Fig12(b1)}
    \includegraphics[width=0.32\linewidth]{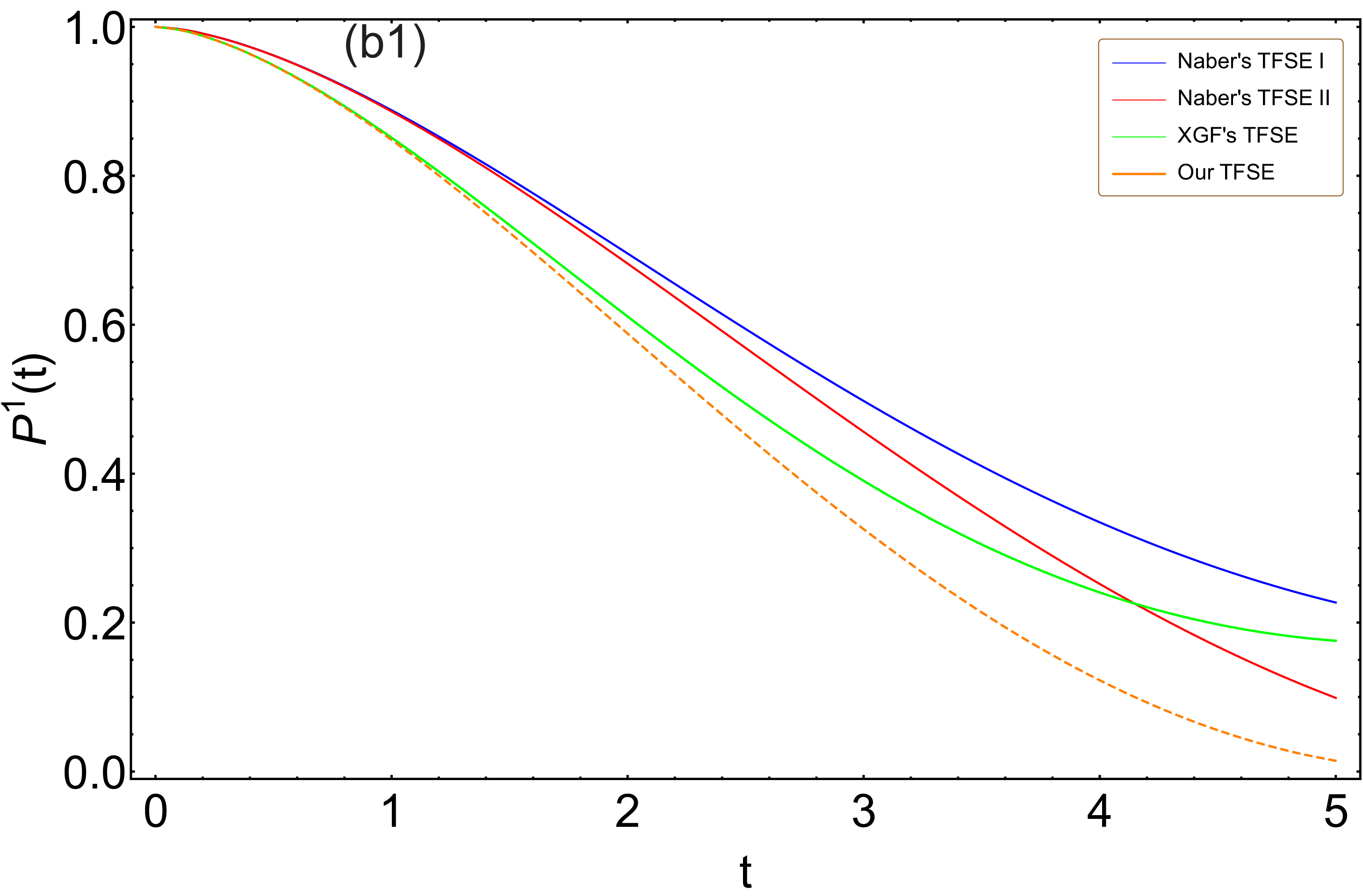}}
    \subfigure{\label{Fig12(c1)}
    \includegraphics[width=0.32\linewidth]{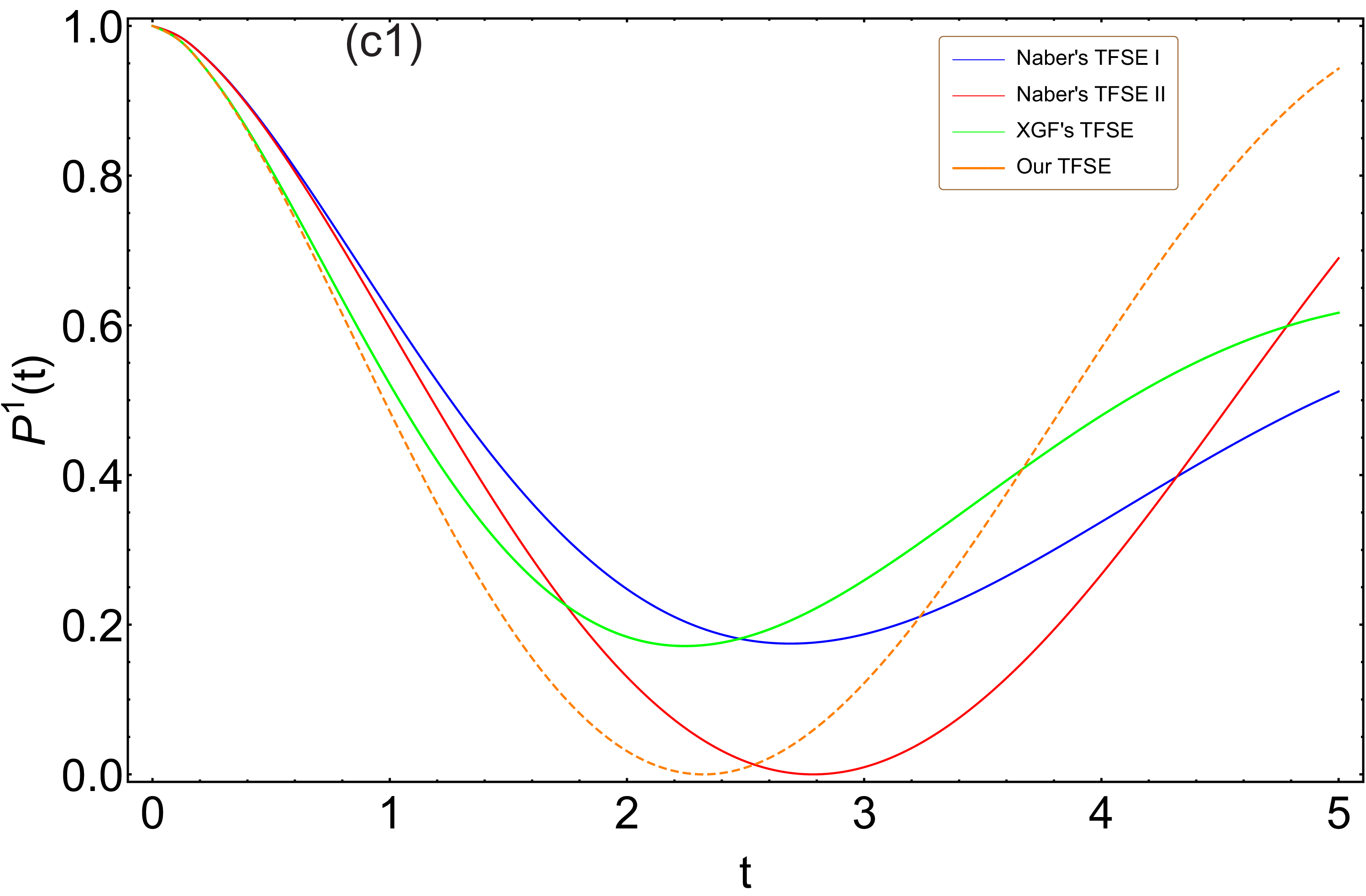}}\\
    \subfigure{\label{Fig12(d1)}
    \includegraphics[width=0.32\linewidth]{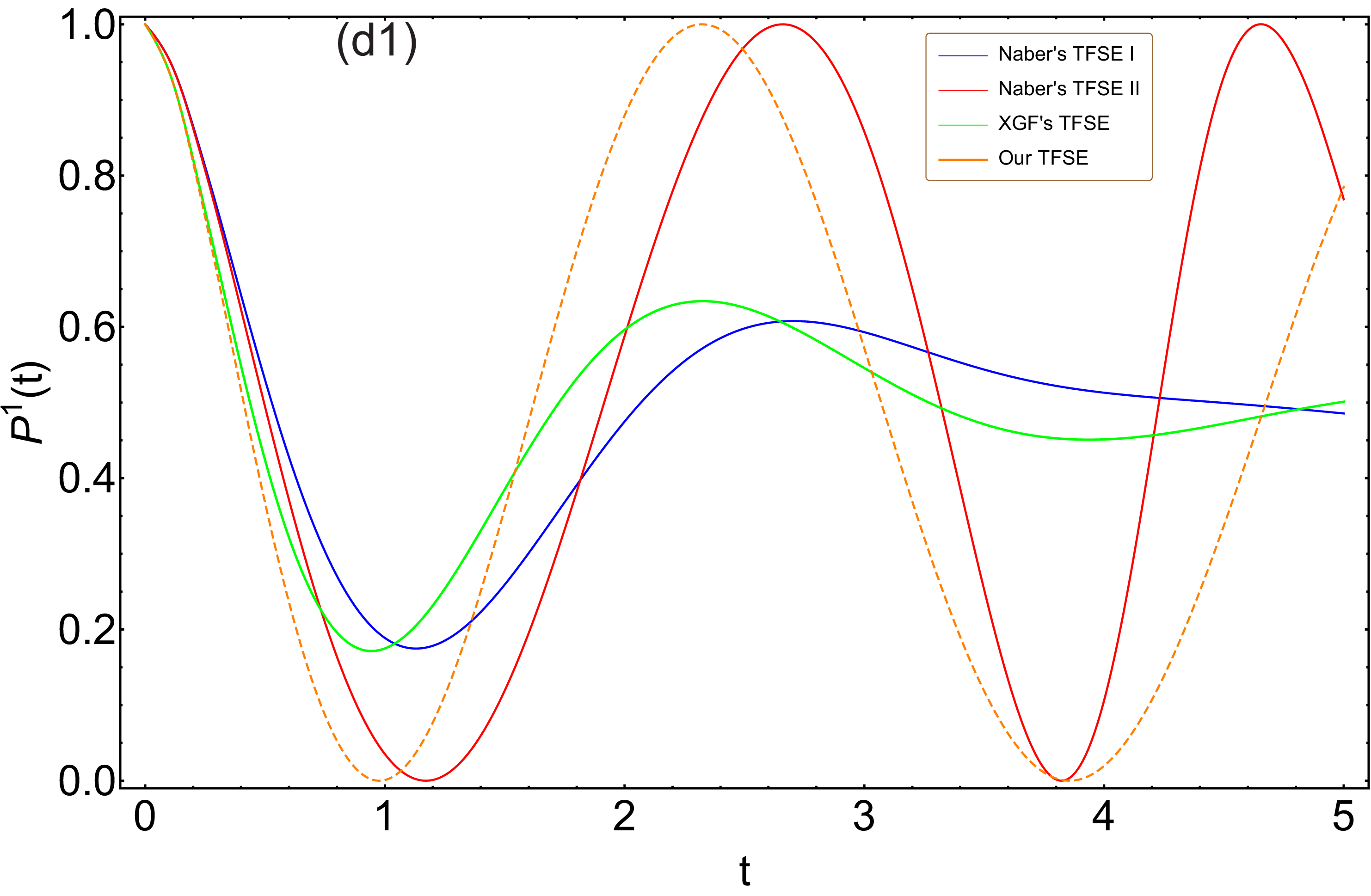}}
    \subfigure{\label{Fig12(e1)}
    \includegraphics[width=0.32\linewidth]{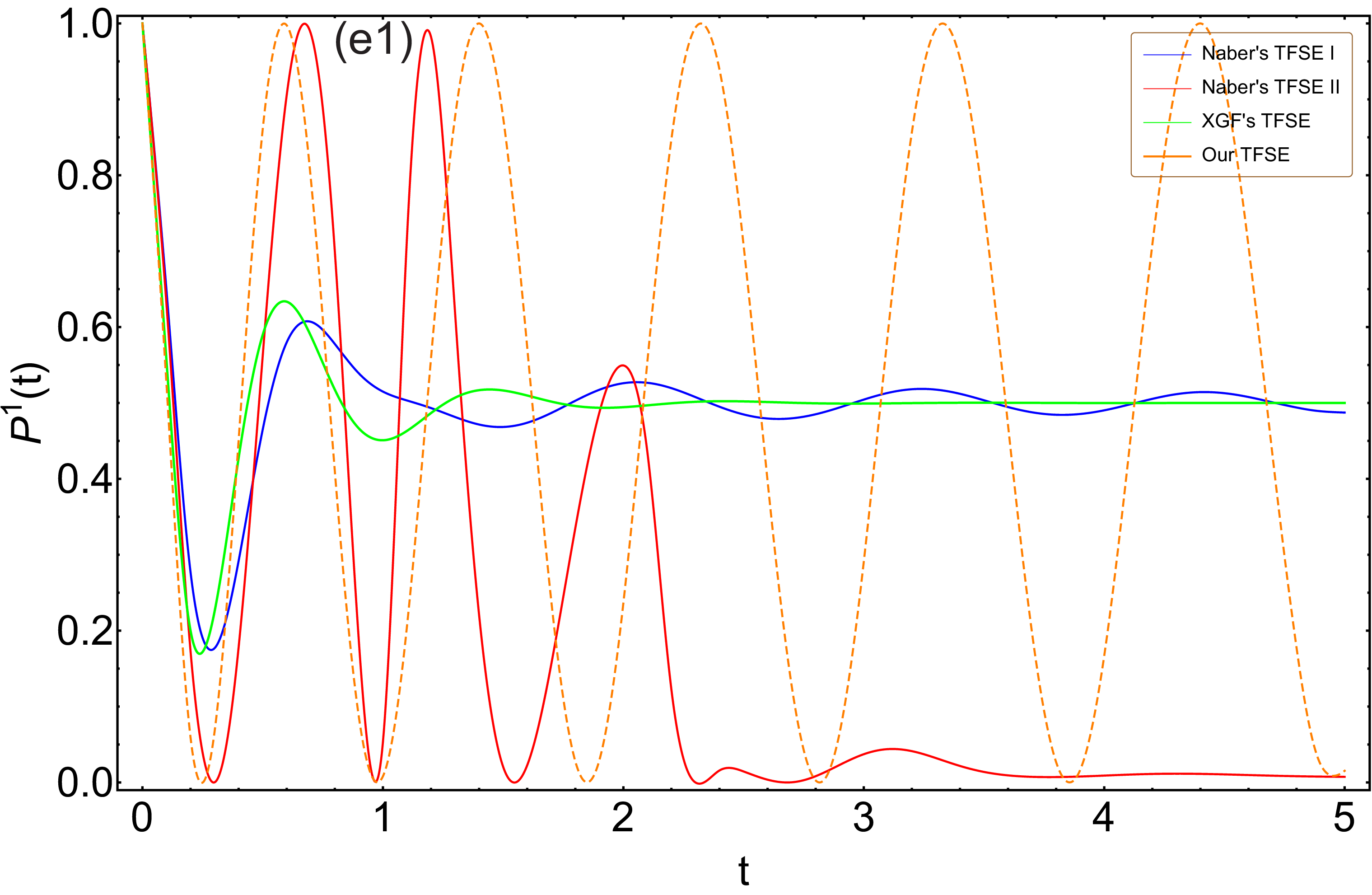}}
    \subfigure{\label{Fig12(f1)}
    \includegraphics[width=0.32\linewidth]{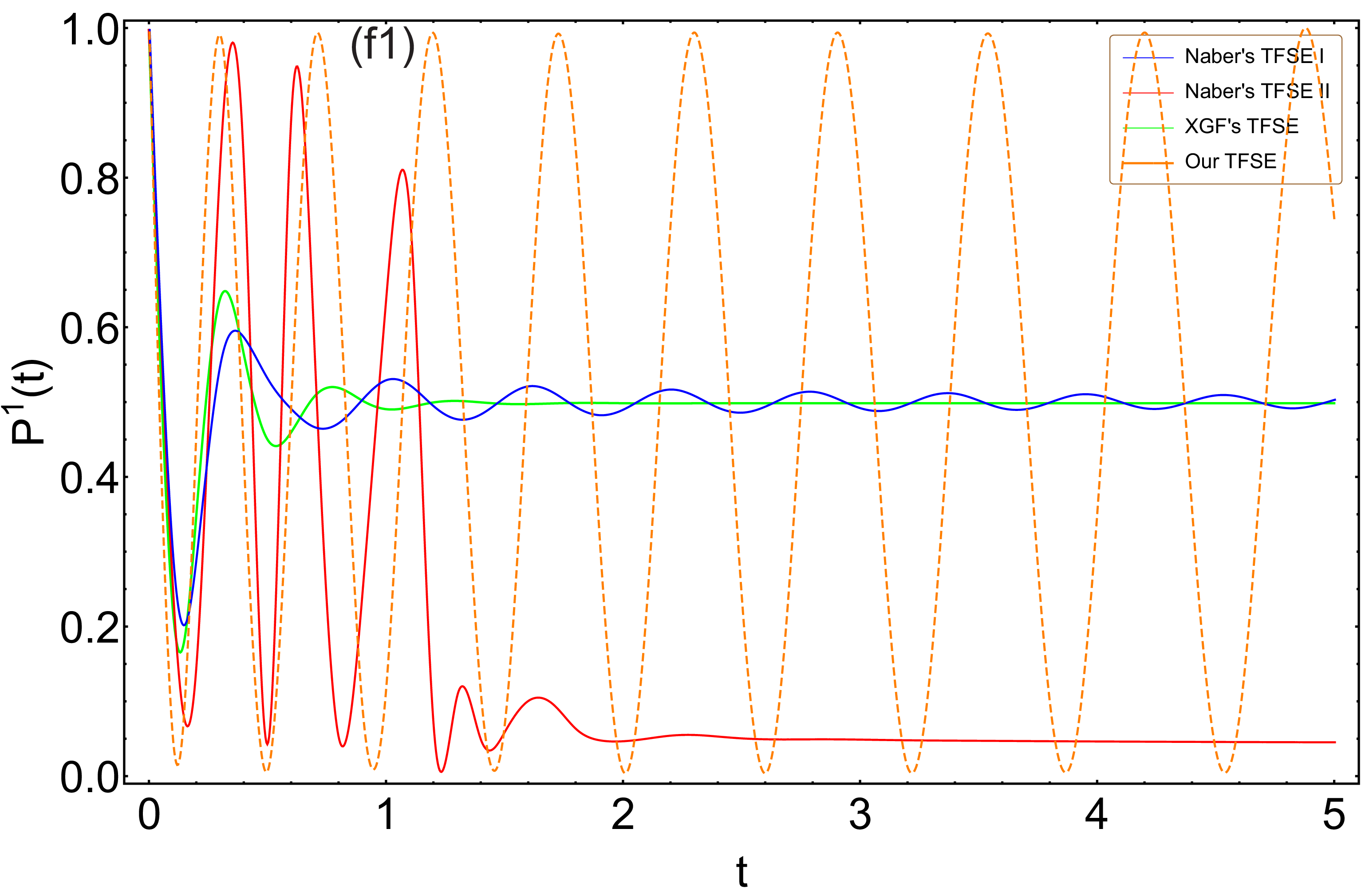}}\\
    \subfigure{\label{Fig12(a2)}
    \includegraphics[width=0.32\linewidth]{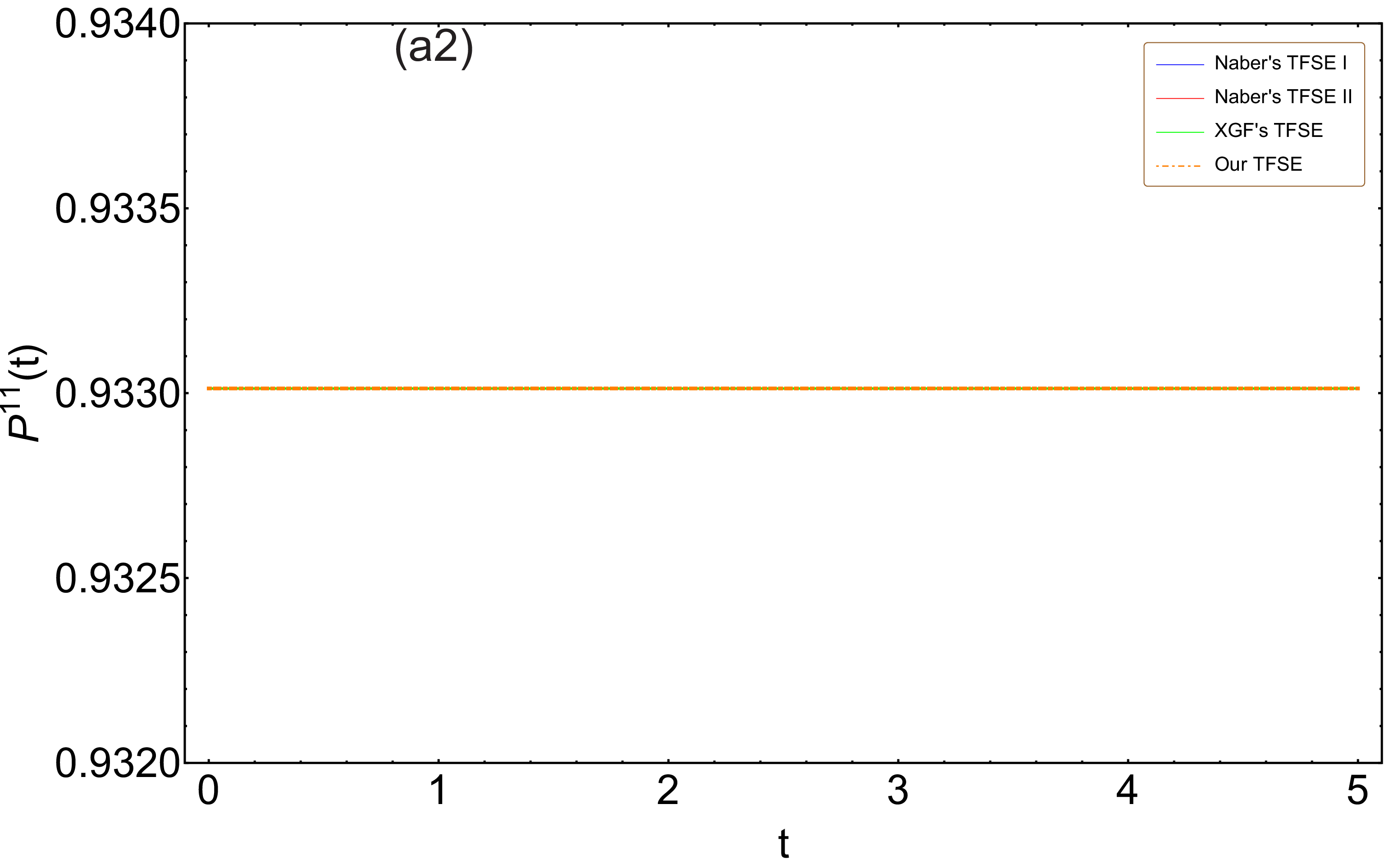}}
    \subfigure{\label{Fig12(b2)}
    \includegraphics[width=0.32\linewidth]{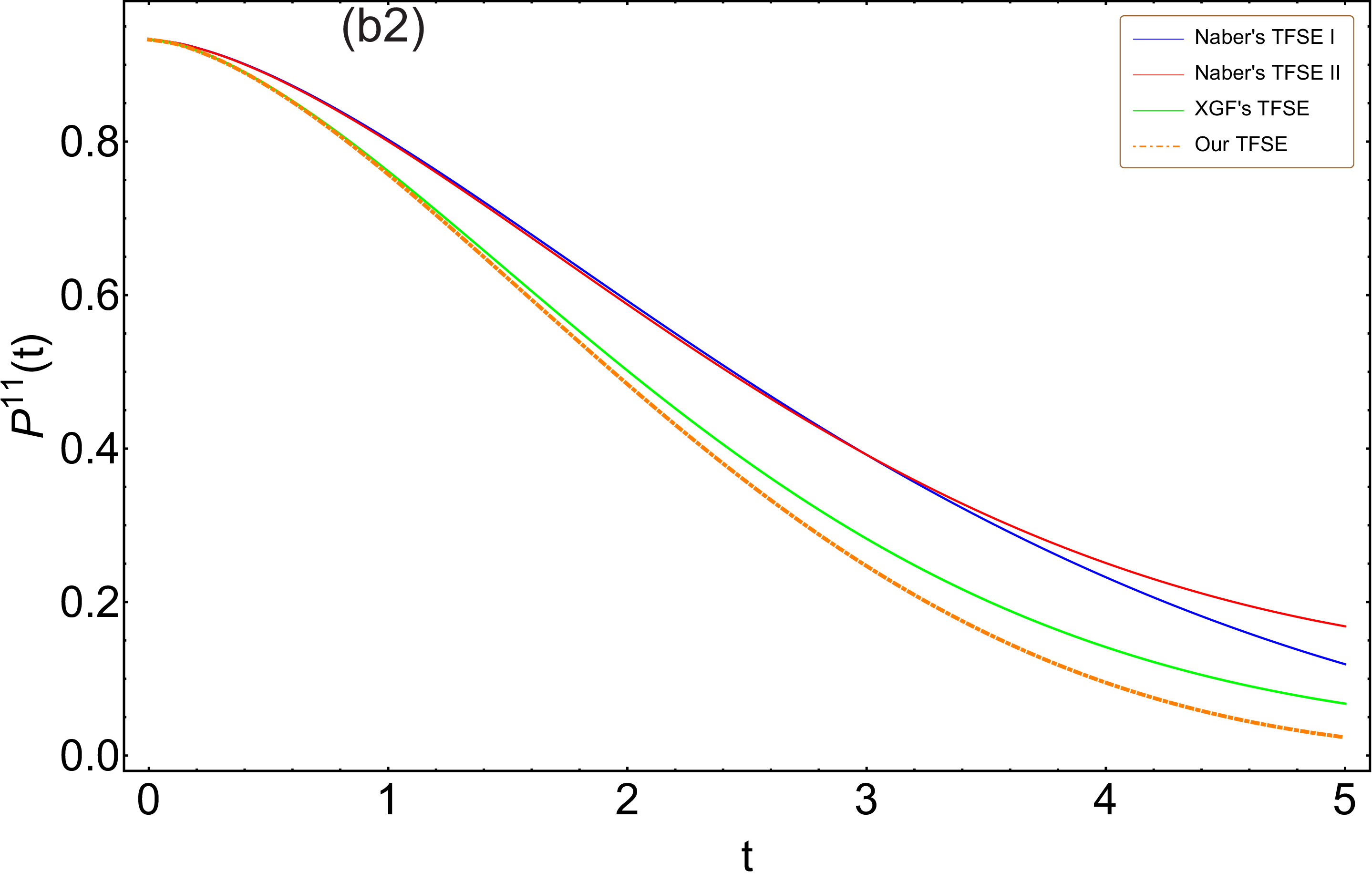}}
    \subfigure{\label{Fig12(c2)}
    \includegraphics[width=0.32\linewidth]{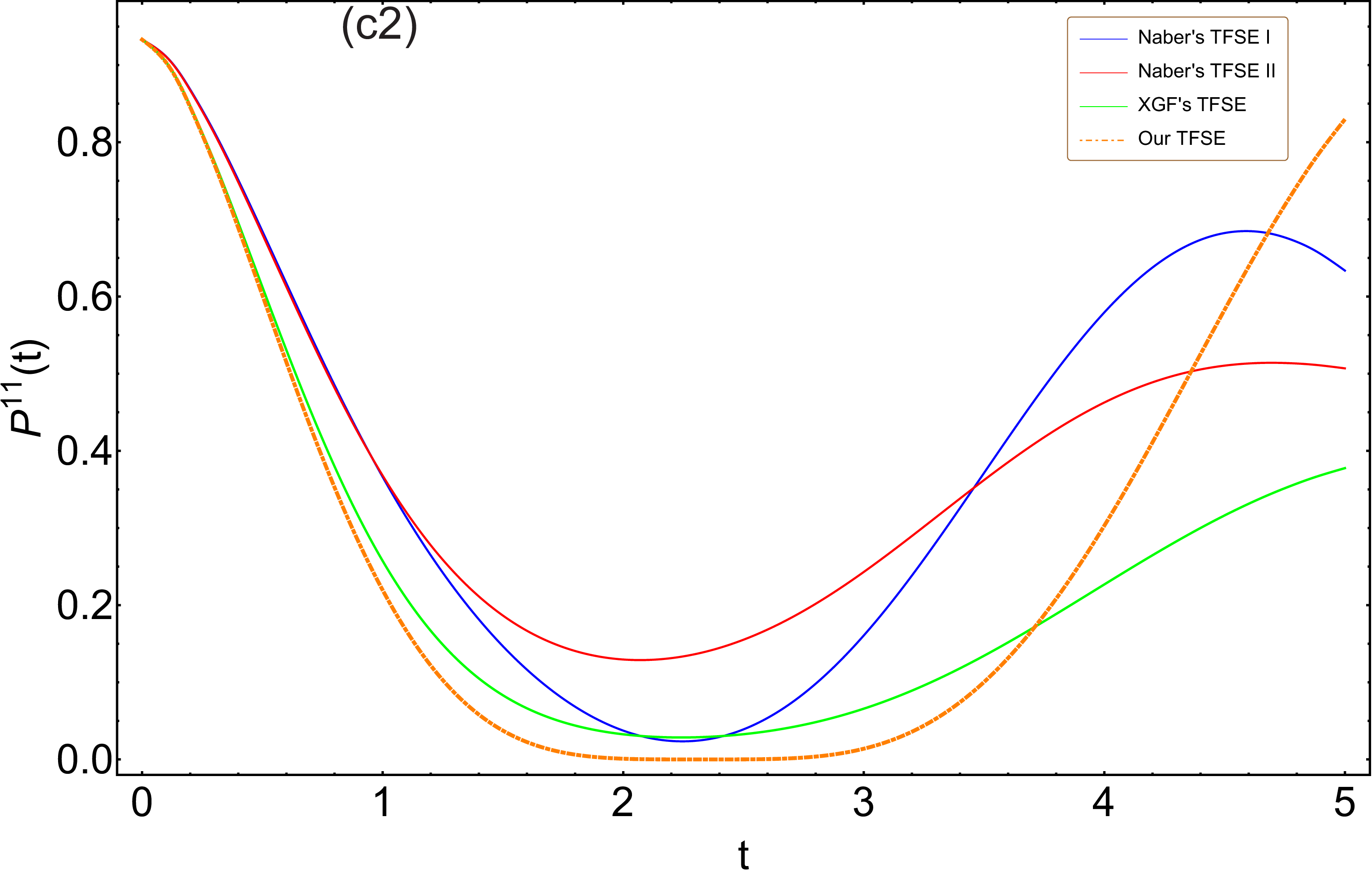}}\\
    \subfigure{\label{Fig12(d2)}
    \includegraphics[width=0.32\linewidth]{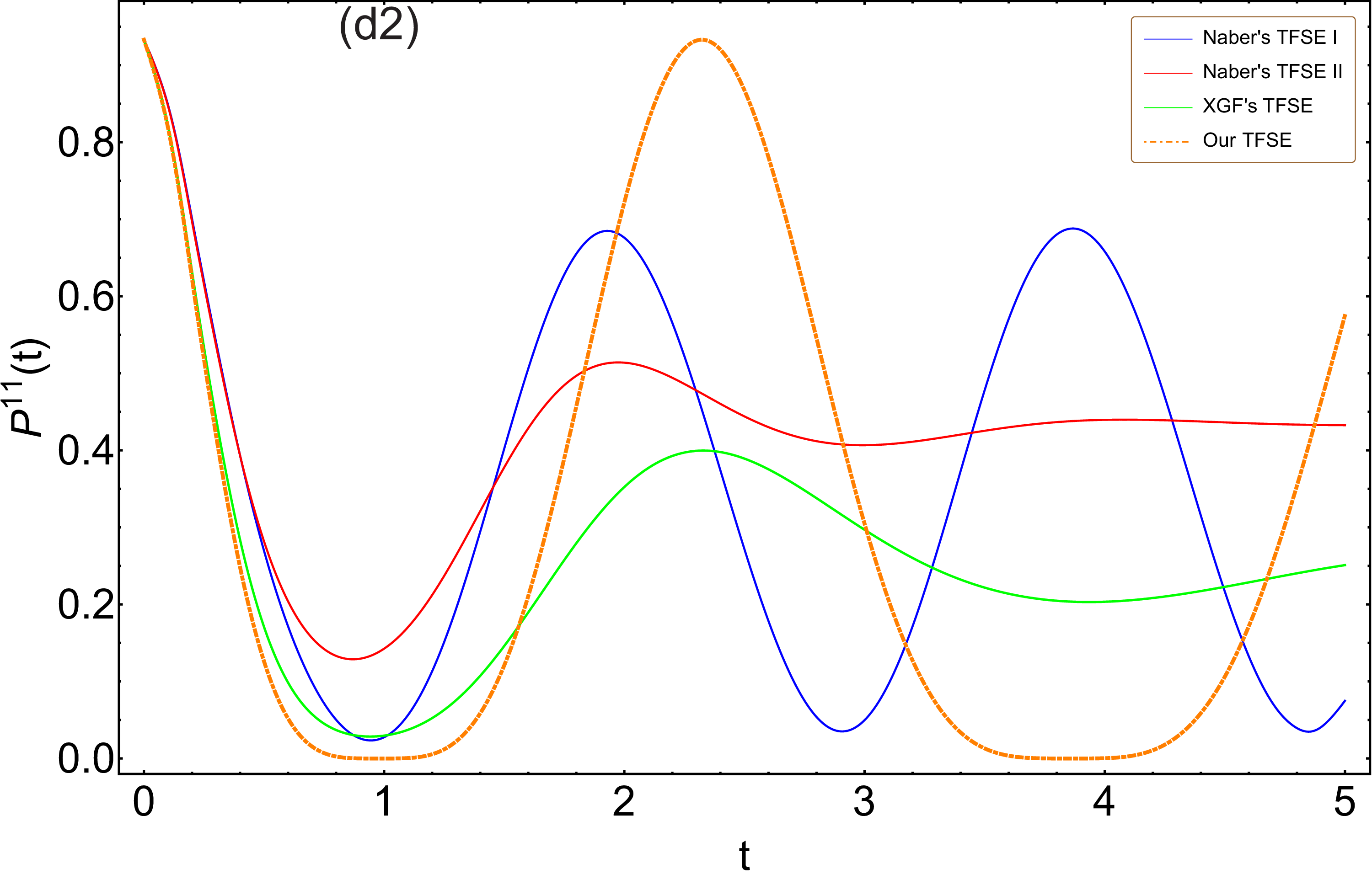}}
    \subfigure{\label{Fig12(e2)}
    \includegraphics[width=0.32\linewidth]{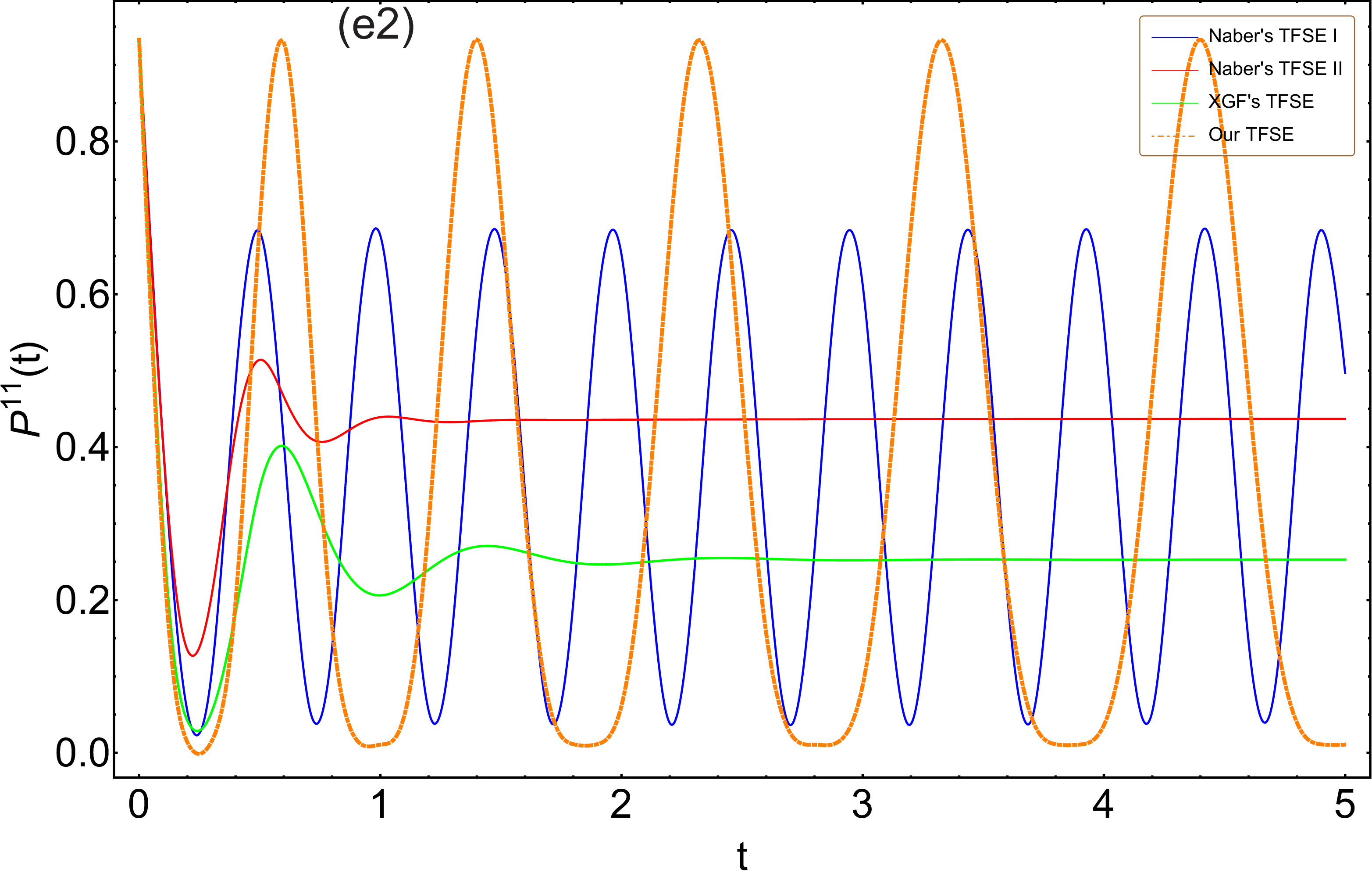}}
    \subfigure{\label{Fig12(f2)}
    \includegraphics[width=0.32\linewidth]{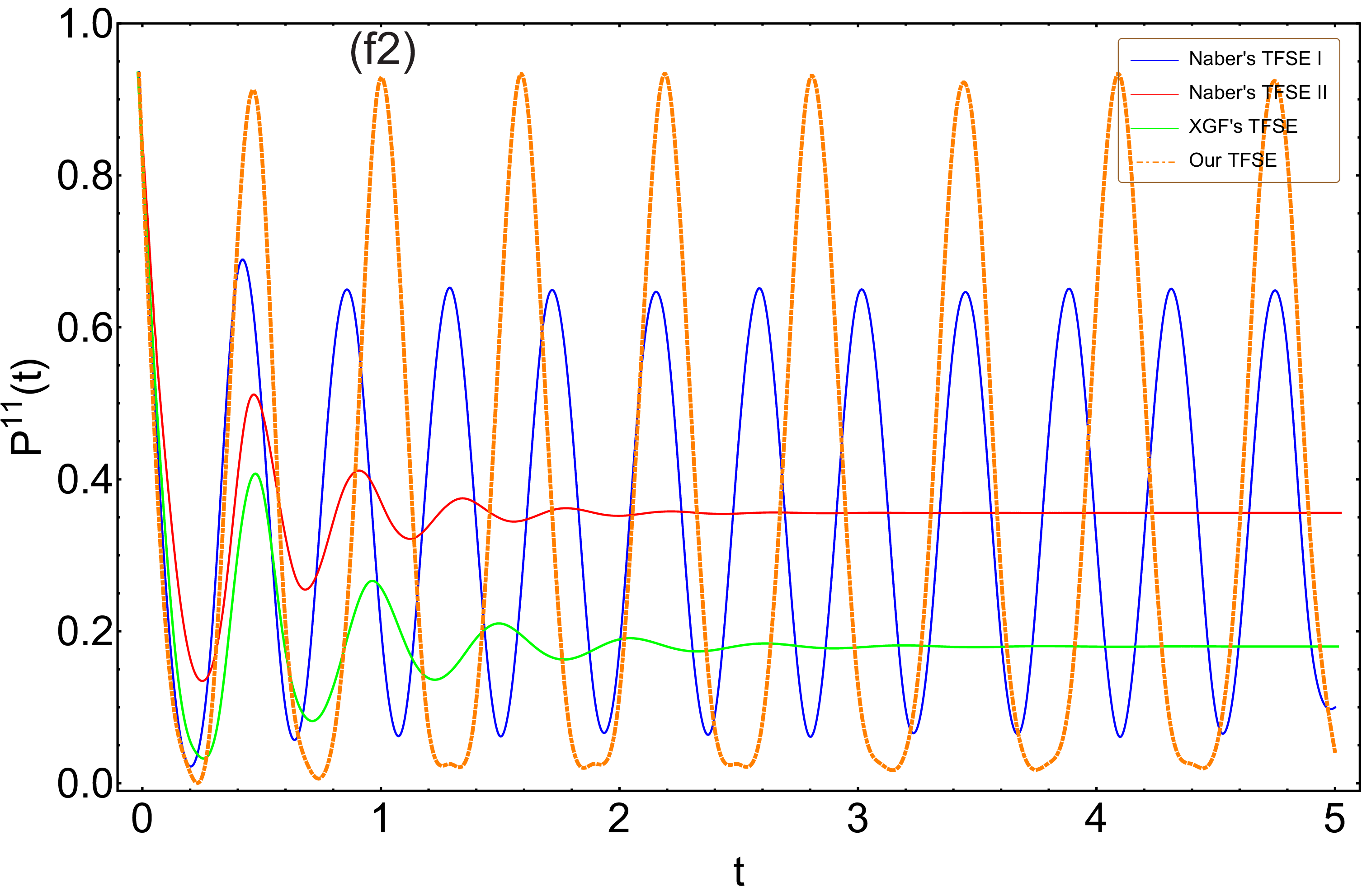}}\\
\caption{The time evolution of $P^{\left|{1}\right\rangle}_{\gamma=1,2,3,4}(t)$ and $P^{\left|{11}\right\rangle}_{\gamma=1,2,3,4}(t)$ is plotted for $\lambda=0, 0.05, 0.1, 0.2, 0.6, 1$. The settings of other parameters are $\beta=0.8$, $n=40$, and $C_0=0.5$.}
\label{Fig12}
\end{figure}

\begin{figure}[htbp]
\centering
    \includegraphics[width=0.32\linewidth]{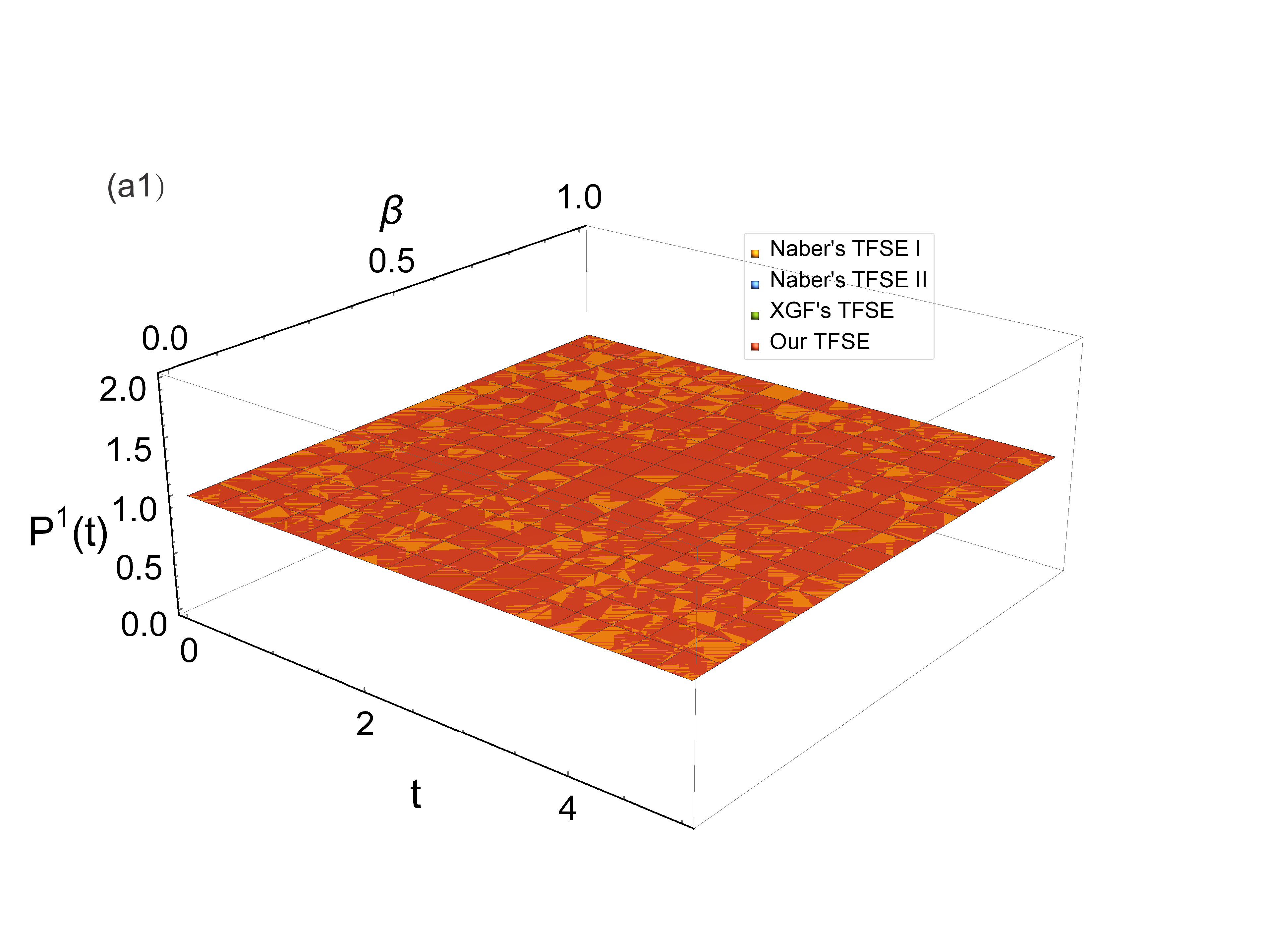}
    \includegraphics[width=0.32\linewidth]{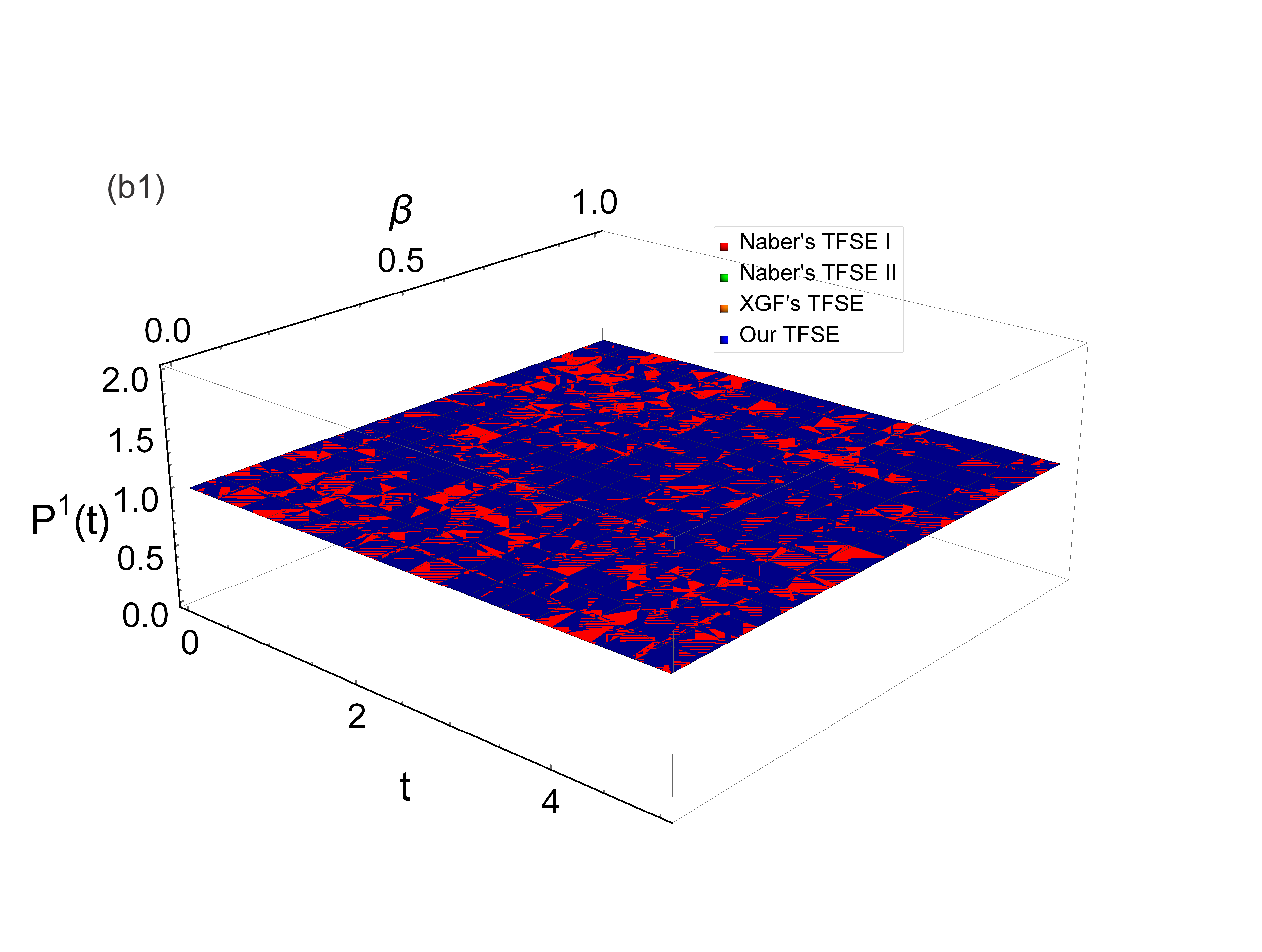}
    \includegraphics[width=0.32\linewidth]{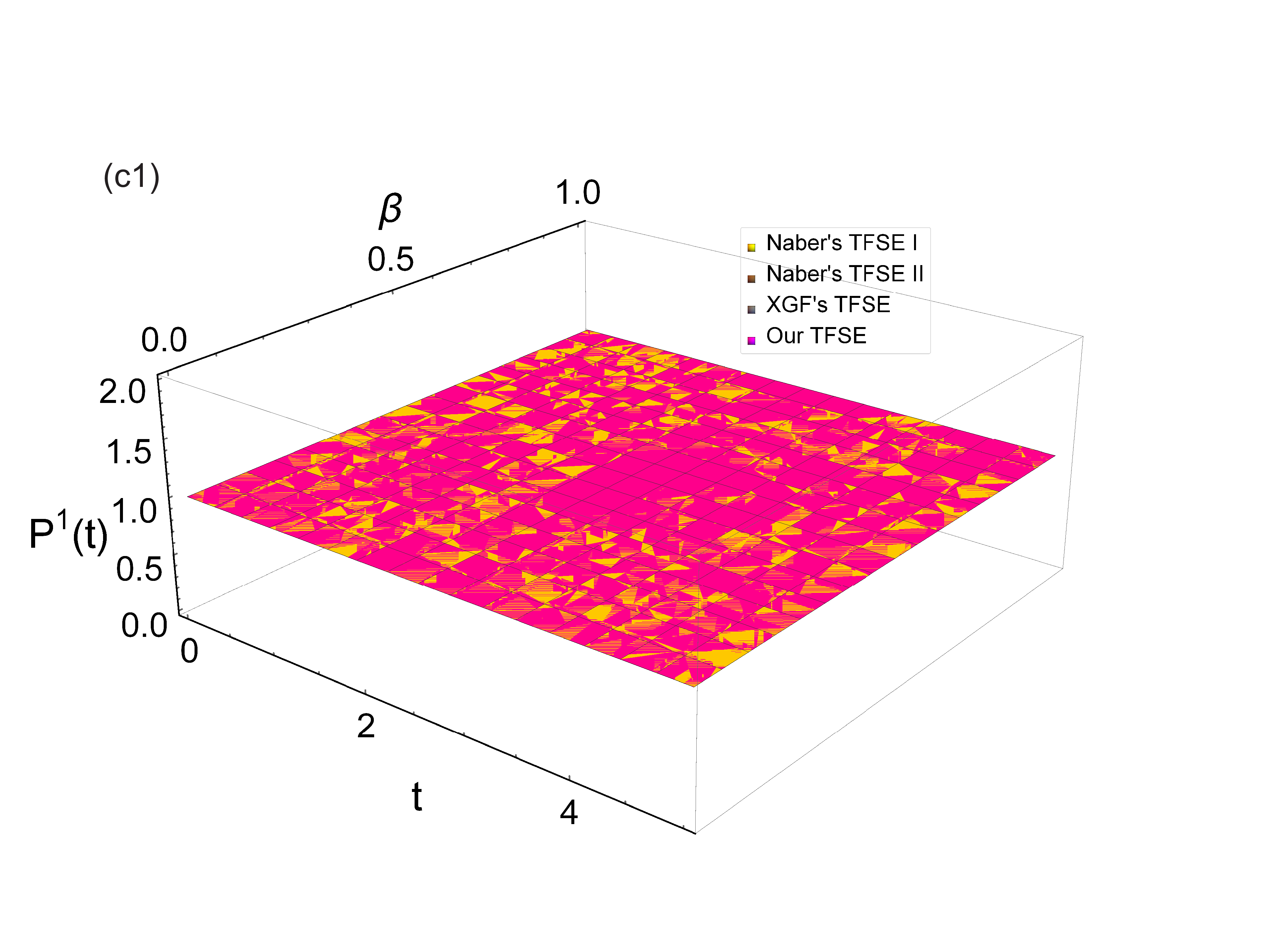}\\
    \includegraphics[width=0.32\linewidth]{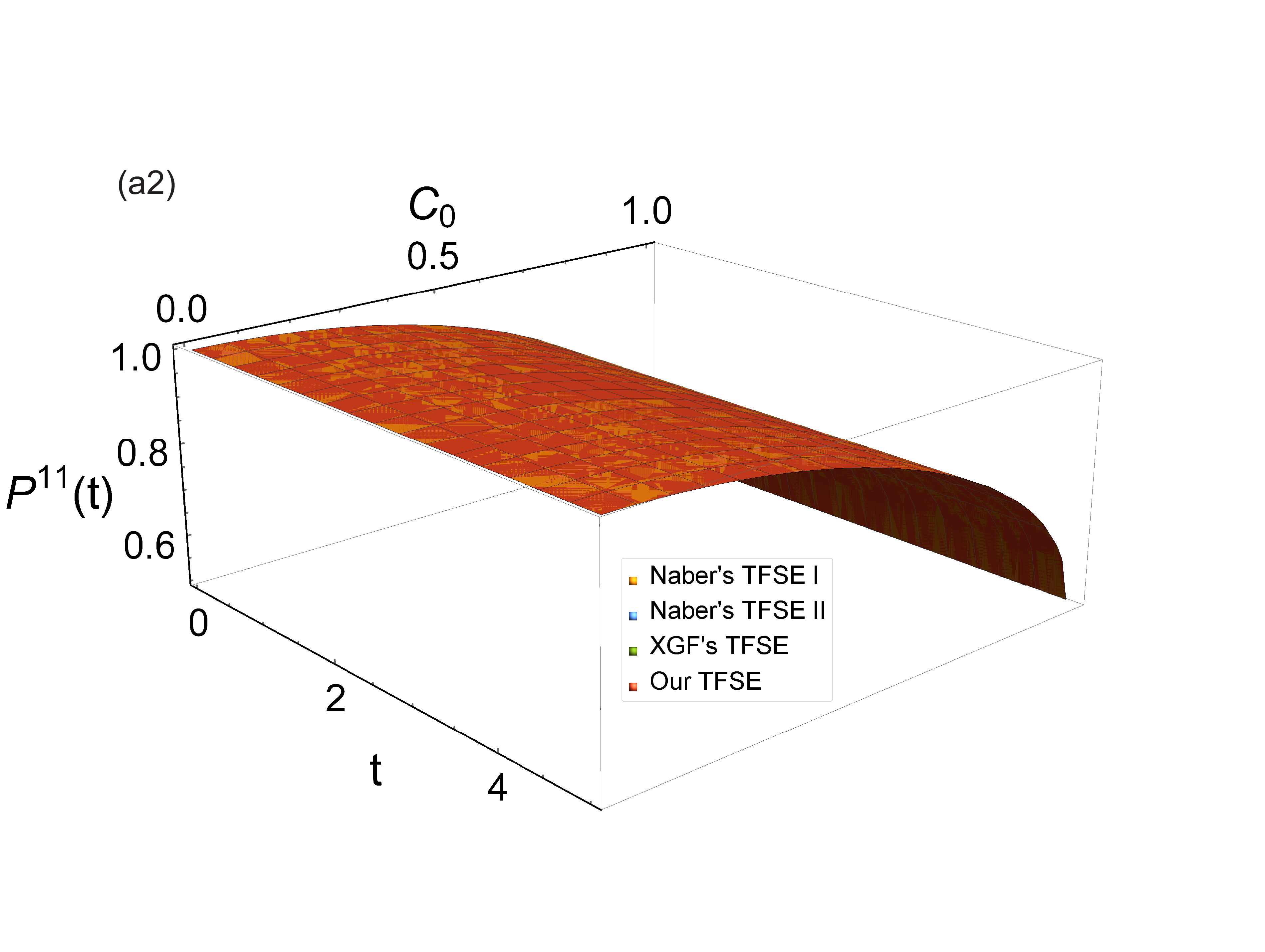}
    \includegraphics[width=0.32\linewidth]{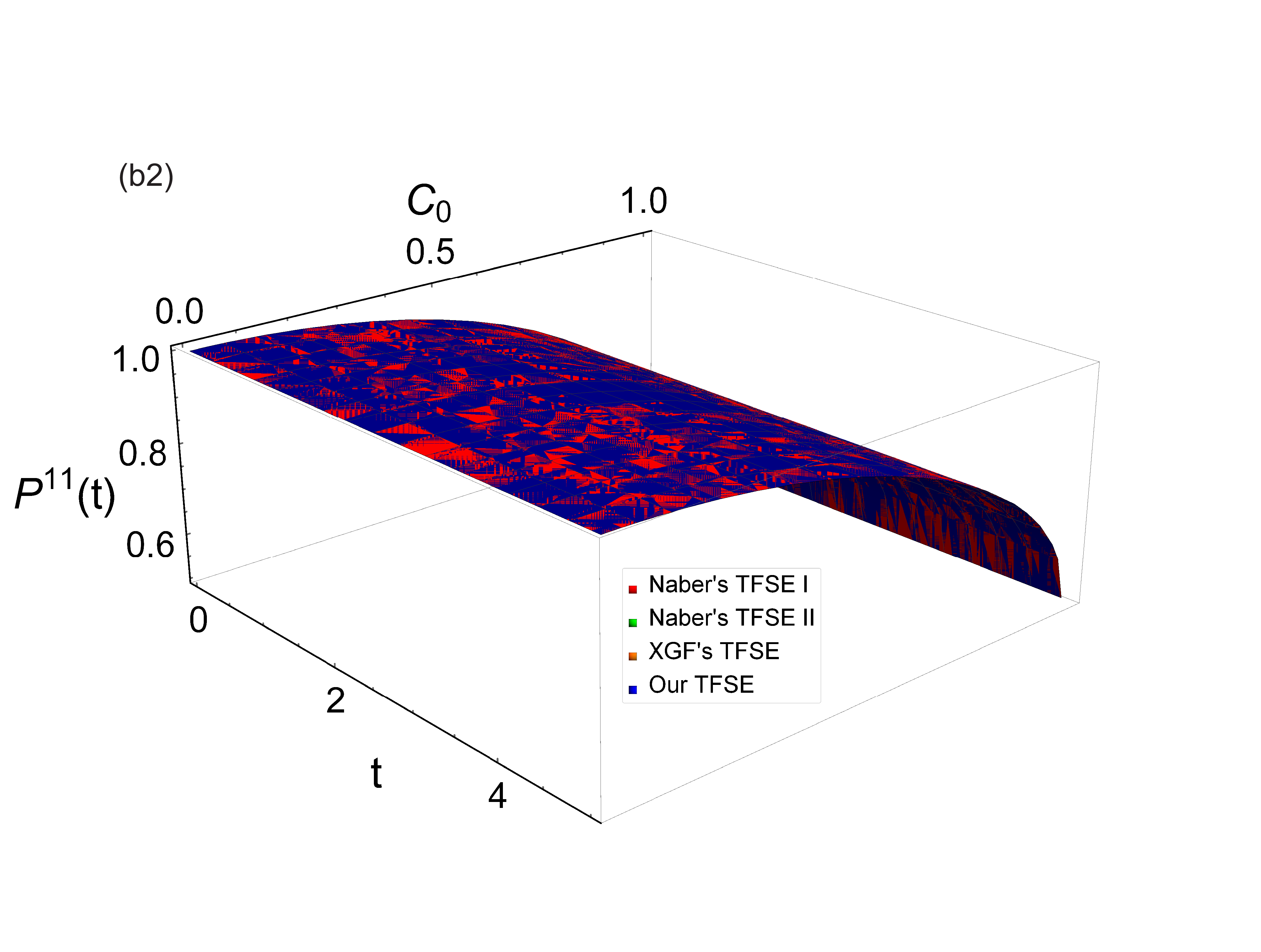}
    \includegraphics[width=0.32\linewidth]{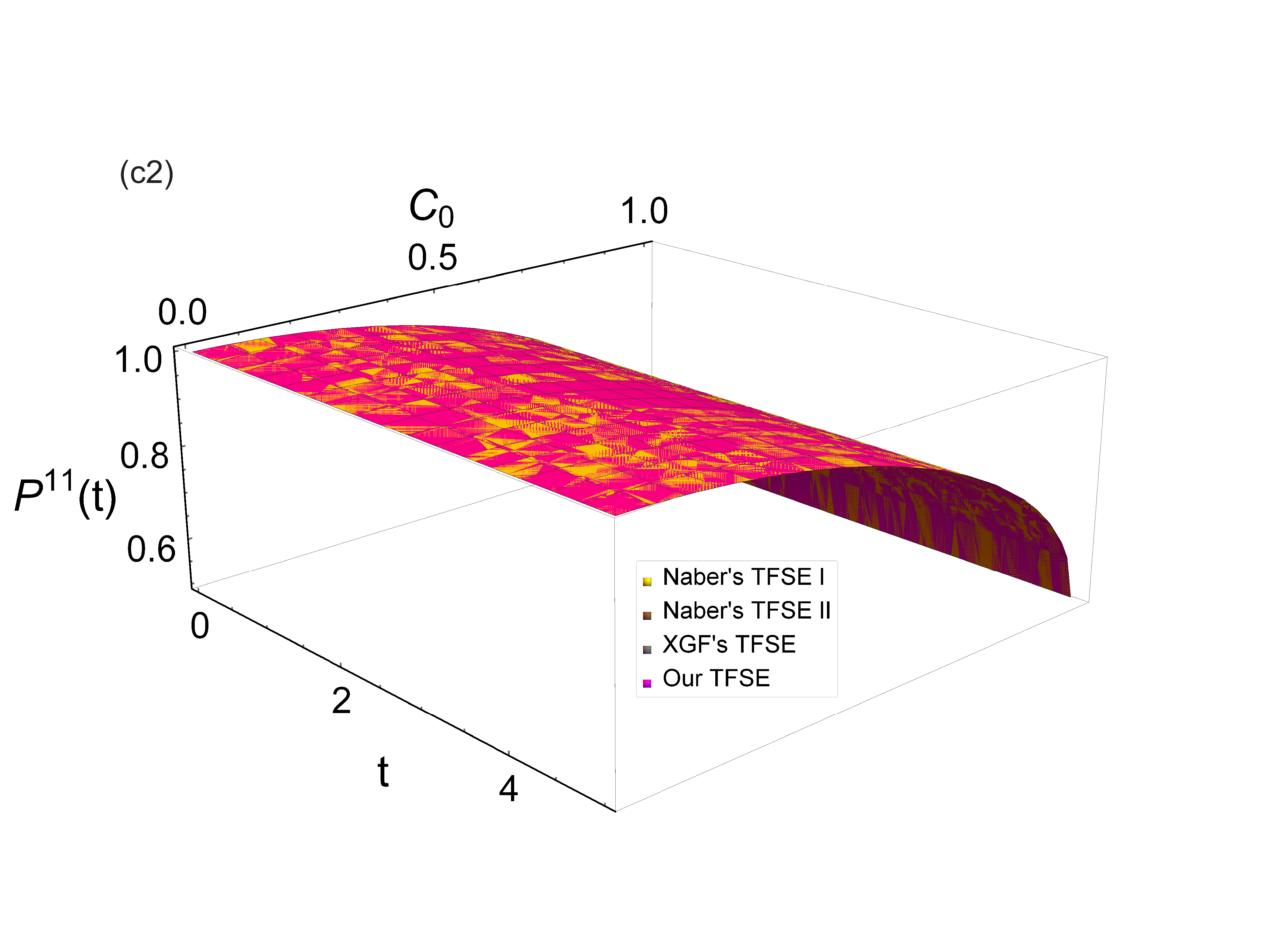}\\
\caption{In the context of $\lambda=0$, the time evolution of $P^{\left|{1}\right\rangle}_{\gamma=1,2,3,4}(t)$ and $P^{\left|{11}\right\rangle}_{\gamma=1,2,3,4}(t)$ is drawn for the cases (a1)-(c1) $n=0$; $n=10$; $n=50$; (a2)-(c2) $n=0$, $\beta=0.1$; $n=10$, $\beta=0.5$; $n=50$, $\beta=1$.}
\label{Fig13}
\end{figure}

In Fig. \ref{Fig12}, we plot the time behavior of $P^{\left|{1}\right\rangle}_{\gamma=1,2,3,4}(t)$ and $P^{\left|{11}\right\rangle}_{\gamma=1,2,3,4}(t)$ for $\lambda=0$, $0.05$, $0.1$, $0.2$, $0.6$, $1$ assuming $\beta=0.8$, $n=40$, and $C_0=0.5$. Careful observations of these subfigures yield the following meaningful results:

(a) In Figs. \ref{Fig12}(a1) and \ref{Fig12}(a2) for $\lambda=0$, $P^{\left|{1}\right\rangle}_{\gamma=1,2,3,4}(t)$ equal one and $P^{\left|{11}\right\rangle}_{\gamma=1,2,3,4}(t)$ equal some values subject to $C_0$. In such cases for the four TFSEs, the non-Markovian dynamics does not take place. To proceed with the generalization of the above results, we now attend to the joint influence of $\beta$, $n$, $C_0$ on the time behavior of all probabilities for $\lambda=0$ in Fig. \ref{Fig13}. $P^{\left|{1}\right\rangle}_{\gamma=1,2,3,4}(t)$ in Figs. \ref{Fig13}(a1) to \ref{Fig13}(c1) are one and independent of $\beta$, $n$. $P^{\left|{11}\right\rangle}_{\gamma=1,2,3,4}(t)$ in Figs. \ref{Fig13}(a2) to \ref{Fig13}(c2) decrease monotonically from one to $0.5$ as $C_0$ increases regardless of $\beta$, $n$.

(b) Between Figs. \ref{Fig12}(b) and \ref{Fig12}(c), when small $\lambda$ is larger than a certain critical value, the transition point of $P(t)$ from a monotonic function of $t$ to a non-monotonic one is just the point at which the memoryless environment turns into the memory one. Therefore, by adjusting $\lambda$ we can control the crossover from the Markovian to non-Markovian processes and vice versa. This may provide a method to manipulate the non-Markovian dynamics of time-fractional open quantum systems, which is valid for the four TFSEs.

(c) From Figs. \ref{Fig12}(d) to \ref{Fig12}(f), the time behavior of all probabilities exhibits stronger non-Markovian oscillations as small $\lambda$ increases. In other words, the four TFSEs can reproduce the non-Markovian features of time-fractional open quantum dynamics when small $\lambda$ exceeds a certain critical value.

\section{Conclusion}
\label{sec:5}
In conclusion, we present applications of three popular TFSEs, i.e., Naber's TFSE I, Naber's TFSE II, and XGF's TFSE, to different open quantum system models and further propose a new TFSE aimed to describe the non-Markovian features of time-fractional open quantum dynamics more effectively. Based on a basic open system model of a qubit coupled resonantly to a dissipative environment, we obtain the exact solutions to the three TFSEs for TFSQOSs and show that the three TFSEs underperform as follows: (i) in the respective frameworks of the three TFSEs, the total probability for finding the system in a single-qubit state is not one over time at $\beta\in\left({0,1}\right)$, which numerically verifies the previous results \cite{Naber2004,Achar2013,Xiang2019}. This implies that time-fractional quantum mechanics violates quantum mechanical probability conservation. (ii) the latter two TFSEs can not describe the non-Markovian evolution of the system at some of $\beta$. To address this, we introduce a well-performed TFSE by constructing a new ACT and employing the Co-FD, where for all $\beta$ the total probability for the system equals one at all times, and the time evolution of the system exhibits the non-Markovian features. Besides, based an open system model of two isolated qubits each locally interacting with its own dissipative environment, we obtain the exact solutions to the four TFSEs for TFTQOSs and demonstrate that our TFSE still provides the two advantages mentioned above over the other three TFSEs.

Particularly, the procedure for obtaining the exact time-fractional dynamics of multiqubit each locally interacting with its own dissipative environment is offered. And more notably, most of our results are independent of the specific form of the spectral distribution, thus they can be used directly in different physical settings. On the other hand, our treatment can be easily extended to investigate the entanglement dynamics of TFTQOSs by considering different initial conditions, such as the Werner state \cite{Werner1989}. Our approach may contribute to characterizing and analyzing the time-fractional dynamical evolution performances of quantum systems exposed to real environments.

The non-Markovian oscillations in time-fractional open quantum dynamics describe a feedback of energy and/or information from the open system to its environment, namely the non-Markovianity, and its measure is an interesting question that will be considered next. Nowadays, many useful measures of the non-Markovianity in standard open quantum dynamics have been presented. Breuer et al. \cite{Breuer2009} constructed a general measure for the non-Markovianity based on the trace distance between two initial states of open quantum systems. The entanglement was also introduced to evaluate the degree of the non-Markovian behavior of a quantum process in noise \cite{Rivas2010,Addis2016}. Recently, Chen et al. \cite{Chen2022} found that the non-Markovianity could be measured by combining the quantum mutual information with the quantum Fisher information flow for a multi-channel open quantum dynamics in a fully-controllable experiment platform. Therefore, the non-Markovianity measure for time-fractional open quantum dynamics may be another option to assess the properties of the TFSEs.

\section*{Declaration of competing interest}
The authors declare that they have no known competing financial interests or personal relationships that could have appeared to influence the work reported in this paper.

\section*{CRediT authorship contribution statement}
\textbf{Dongmei Wei}: Conceptualization, Methodology, Formal analysis, Software, Data curation, Writing - Original draft, Writing - Review \& Editing, Visualization. \textbf{Hailing Liu}: Methodology, Formal analysis, Investigation, Writing - Review \& Editing. \textbf{Yongmei Li}: Methodology, Investigation, Data curation. \textbf{Sujuan Qin}: Supervision, Investigation, Project administration. \textbf{Qiaoyan Wen}: Project administration. \textbf{Fei Gao}: Supervision, Investigation, Project administration, Writing - Review \& Editing.

\section*{Acknowledgements}
This work is supported by National Natural Science Foundation of China (Grant Nos. 62371069, 62372048, 62272056).

\section*{Availability of data and materials }
The authors confirm that the data available for non-commercial using.

\end{document}